\documentclass[aps,twocolumn,showpacs,twoside]{revtex4}

\usepackage{graphicx}
\usepackage{longtable}
\usepackage{amsmath}
\usepackage{bm}

\newcommand{\nc}{\newcommand}
\nc{\rnc}{\renewcommand}

\nc{\beq}{\begin{equation}}
\nc{\eeq}{\end{equation}}
\nc{\bea}{\begin{eqnarray}}
\nc{\eea}{\end{eqnarray}}
\nc{\bse}{\begin{subequations}}
\nc{\ese}{\end{subequations}}
\nc{\ba}{\begin{array}}
\nc{\ea}{\end{array}}
\nc{\bpi}{\begin{picture}}
\nc{\epi}{\end{picture}}
\nc{\nn}{\nonumber}
\nc{\btr}{\begin{tabular}}
\nc{\etr}{\end{tabular}}
\nc{\btb}{\begin{table}}
\nc{\etb}{\end{table}}
\nc{\btbs}{\begin{table*}}
\nc{\etbs}{\end{table*}}
\nc{\blt}{\begin{longtable*}}
\nc{\elt}{\end{longtable*}}

\nc{\al}{\alpha}
\nc{\be}{\beta}
\nc{\ga}{\gamma}
\nc{\de}{\delta}
\nc{\ep}{\epsilon}
\nc{\ka}{\kappa}
\nc{\la}{\lambda}
\nc{\om}{\omega}
\nc{\ze}{\zeta}
\nc{\De}{\Delta}
\nc{\Ga}{\Gamma}
\nc{\Si}{\Sigma}

\nc{\ds}{\displaystyle}
\nc{\ts}{\textstyle}
\nc{\scs}{\scriptstyle}
\nc{\sss}{\scriptscriptstyle}
\nc{\p}{\partial}
\nc{\f}[2]{\frac{#1}{#2}}
\nc{\od}{{\cal O}}
\nc{\ra}{\rightarrow}
\nc{\lra}{\longrightarrow}
\nc{\uh}{\hat{u}}
\nc{\boldnabla}{\bm{\nabla}}
\nc{\boldx}{\mbox{\boldmath$x$}}
\nc{\rb}{r_\text{B}}
\nc{\onehalf}{{\ts\f{1}{2}}}
\nc{\threehalf}{{\ts\f{3}{2}}}
\nc{\R}{{\cal R}}
\nc{\msbar}{\mbox{$\overline{\text{MS}}$}}
\nc{\rms}{r_{\sss\overline{\text{MS}}}}
\nc{\mubar}{\bar{\mu}}
\nc{\ur}{u_r}
\nc{\gae}{{\ga_\text{E}}}

\nc{\eprint}[2]{#1/#2}
\nc{\ibid}[3]{ibid.\ {\bf #1}, #2 (#3)}
\nc{\APNY}[3]{Ann.\ Phys.\ (N.Y.) \textbf{#1}, #2 (#3)}
\nc{\CPC}[3]{Comp.\ Phys.\ Comm.\ \textbf{#1}, #2 (#3)}
\nc{\EPJB}[3]{Eur.\ Phys.\ J.\ B \textbf{#1}, #2 (#3)}
\nc{\EPL}[3]{Europhys.\ Lett.\ \textbf{#1}, #2 (#3)}
\nc{\JChP}[3]{J.\ Chem.\ Phys.\ \textbf{#1}, #2 (#3)}
\nc{\JCpP}[3]{J.\ Comput.\ Phys.\ \textbf{#1}, #2 (#3)}
\nc{\JHEP}[3]{JHEP \textbf{#1}, #2 (#3)}
\nc{\JMP}[3]{J.\ Math.\ Phys.\ \textbf{#1}, #2 (#3)}
\nc{\JPB}[3]{J.\ Phys.\ B \textbf{#1}, #2 (#3)}
\nc{\LP}[3]{Laser Phys.\ \textbf{#1}, #2 (#3)}
\nc{\MPLB}[3]{Mod.\ Phys.\ Lett.\ B \textbf{#1}, #2 (#3)}
\nc{\MUPB}[3]{Moscow Univ.\ Phys.\ Bull.\ \textbf{#1}, #2 (#3)}
\nc{\NCL}[3]{Nuovo Cimento Lett.\ \textbf{#1}, #2 (#3)}
\nc{\NPB}[3]{Nucl.\ Phys. B \textbf{#1}, #2 (#3)}
\nc{\PA}[3]{Physica A \textbf{#1}, #2 (#3)}
\nc{\PLA}[3]{Phys.\ Lett.\ A \textbf{#1}, #2 (#3)}
\nc{\PLB}[3]{Phys.\ Lett.\ B \textbf{#1}, #2 (#3)}
\nc{\PR}[3]{Phys.\ Rev.\ \textbf{#1}, #2 (#3)}
\nc{\PRL}[3]{Phys.\ Rev.\ Lett.\ \textbf{#1}, #2 (#3)}
\nc{\PRA}[3]{Phys.\ Rev.\ A \textbf{#1}, #2 (#3)}
\nc{\PRB}[3]{Phys.\ Rev.\ B \textbf{#1}, #2 (#3)}
\nc{\PRC}[3]{Phys.\ Rev.\ C \textbf{#1}, #2 (#3)}
\nc{\PRD}[3]{Phys.\ Rev.\ D \textbf{#1}, #2 (#3)}
\nc{\PRE}[3]{Phys.\ Rev.\ E \textbf{#1}, #2 (#3)}
\nc{\PREP}[3]{Phys.\ Rep.\ \textbf{#1}, #2 (#3)}
\nc{\RMP}[3]{Rev.\ Mod.\ Phys.\ \textbf{#1}, #2 (#3)}
\nc{\Sc}[3]{Science \textbf{#1}, #2 (#3)}
\nc{\SPJETP}[3]{Sov.\ Phys.\ JETP \textbf{#1}, #2 (#3)}
\nc{\TMF}[3]{Teor.\ Mat.\ Fiz.\ \textbf{#1}, #2 (#3)}
\nc{\ZPB}[3]{Z.\ Phys.\ B Condens.\ Matter \textbf{#1}, #2 (#3)}

\begin{document}

\bibliographystyle{apsrev}

\title{
Non-universal Critical Quantities from Variational Perturbation Theory\\
and Their Application to the BEC Temperature Shift}

\author{Boris Kastening}
\email{boris.kastening@physik.fu-berlin.de}
\affiliation{Institut f\"ur Theoretische Physik\\
Freie Universit\"at Berlin\\ Arnimallee 14\\ D-14195 Berlin\\ Germany}

\date{June 1, 2004}

\begin{abstract}
For an O($N$) symmetric scalar field theory with Euclidean action
$\int d^3x\left[\f{1}{2}|\boldnabla\phi|^2
+\f{1}{2}r\phi^2+\f{1}{4!}u\phi^4\right]$,
where $\phi=(\phi_1,\ldots,\phi_N)$ is a vector of $N$ real field
components, variational perturbation theory through seven loops is
employed for $N=0,1,2,3,4$ to compute the renormalized value of
$r/(N{+}2)u^2$ at the phase transition.
Its exact large-$N$ limit is determined as well.
We also extend an earlier computation of the interaction-induced
shift $\De\langle\phi^2\rangle/Nu$ for $N=1,2,4$ to $N=0,3$.
For $N=2$, the results for the two quantities are used to compute the
second-order shift of the condensation temperature of a dilute Bose gas,
both in the homogenous case and for the wide limit of a harmonic trap.
Our results are in agreement with earlier Monte Carlo simulations
for $N=1,2,4$.
The appendix contains previously unpublished numerical seven-loop data
provided to us by B.~Nickel.
\end{abstract}


\maketitle

\section{Introduction}
Since the first experimental realizations of Bose-Ein\-stein
condensation (BEC) in dilute atomic gases \cite{expbec}, the
preparation of such quantum gases has been achieved in many
laboratories around the globe.
With the continuous improvement of experimental control over
these gases, it appears to be only a matter of time until precision
measurements of physical quantities such as the condensation
temperature, the number of bosons, and the density profile of
the gas will be possible.
The relations between these quantities were explored in many
theoretical papers, often contradicting each other.
However, after an intense investigation of the subject using
Monte-Carlo simulations and various methods of field theory,
most theorists agree by now that the shift $\De T_c\equiv T_c-T_0$
of the condensation temperature away from its ideal gas value
\beq
\label{t0n0}
T_0=\f{2\pi}{m}\left[\f{n}{\ze(\threehalf)}\right]^{2/3},
\eeq
where $n$ is the particle number density and $m$ the mass of the bosons,
and where we work throughout in units where $k_B=\hbar=1$, has, for
a dilute homogenous Bose gas, the form \cite{BaBlHoLaVa1,BaBlZi,HoBaBlLa}
\beq
\label{deltathom}
\f{\De T_c}{T_0}=\bar{c}_1an^{1/3}
+[\bar{c}_2'\ln(an^{1/3})+\bar{c}_2''](an^{1/3})^2+\cdots.
\eeq
Here $a$ is the $s$-wave scattering length corresponding to the
two-particle interaction potential of the bosons and $\bar{c}_1$,
$\bar{c}_2'$ and $\bar{c}_2''$ are constants.
For a comparison of several approaches to compute $\De T_c$ for the
homogenous gas, the reader is referred to the recent review \cite{An}.

While the shift above is expressed in terms of the number density
of bosons, the shift of $T_c$ for a gas in the wide limit of a
harmonic trap is conventionally expressed in terms of the total number
$N_b$ of bosons in the trap.
With the size of the unperturbed ground state given by
\beq
l_\text{ho}=\f{1}{\sqrt{m(\om_x\om_y\om_z)^{1/3}}},
\eeq
where $\om_x$, $\om_y$, $\om_z$ are the frequencies characterizing
the harmonic trap, the transition temperature may be written as
\beq
T_0=\left(\f{N_b}{\ze(3)}\right)^{1/3}\f{1}{ml_\text{ho}^2},
\eeq
so that the corresponding thermal wavelength $\la_0=\sqrt{2\pi/mT_0}$
becomes
\beq
\la_0=\sqrt{2\pi}\left(\f{N_b}{\ze(3)}\right)^{-1/6}l_\text{ho}.
\eeq
The equivalent of the expansion (\ref{deltathom}) is then
\cite{GiPiSt,ArTo}
\beq
\label{deltattrap}
\f{\De T_c}{T_0}
=
\breve{c}_1\f{a}{\la_0}
+\left(\breve{c}_2'\ln\f{a}{\la_0}+\breve{c}_2''\right)
\left(\f{a}{\la_0}\right)^2+\cdots
\eeq
with constants $\breve{c}_1$, $\breve{c}_2'$ and $\breve{c}_2''$.
The result (\ref{deltattrap}) is strictly valid only in the limit of
an infinitely wide trap.
In this case, however, the gas is locally homogenous and the result
(\ref{deltathom}) is applicable as well, if $n$ is taken to be the
central density \cite{ArTo,ArMoTo}.
The fact that $\De T_c$ is different in both cases is no contradiction,
since different quantities are kept fixed, namely $n$ in the homogenous
case and $N_b$ in the case of a wide trap.
In particular, $\bar{c}_1>0$, while $\breve{c}_1<0$.
Keeping $T_c$ fixed instead, this means that, for small $a$, more particles
have to be put into the trap as compared to the ideal gas, but at the
same time the central density is reduced.

The starting point for deriving the expansions (\ref{deltathom}) and
(\ref{deltattrap}) is to describe the gas of bosons by a non-relativistic
$3+1$-dimensional field theory.
Being interested in static quantities at finite temperature, one
works in the imaginary-time formalism in the grand canonical ensemble.
The corresponding Euclidean action is
\bea
S_{3+1}
&\!=\!&
\int_0^\be d\tau\int d^3x\bigg[
\psi^*\left(\f{\p}{\p\tau}-\f{1}{2m}\boldnabla^2-\mu+V(\boldx)\right)\psi
\nn\\
&&{}\quad\quad\quad\quad\quad\quad
+\f{2\pi a}{m}(\psi^*\psi)^2\bigg],
\eea
with $V(\boldx)=0$ for the homogenous case and
$V(\boldx)=\f{1}{2}m(\om_x^2x^2+\om_y^2y^2+\om_z^2z^2)$ for the harmonic
trap.
The complex field $\psi$ is periodic in the imaginary-time direction,
$\psi(\boldx,0)=\psi(\boldx,\be)$.
The two-body interaction potential is parameterized by the $s$-wave
scattering length $a$.
It has been argued \cite{ArTo,ArMoTo} that the effect of interactions
between more than two bosons and details of the two-body interaction
potential beyond the $s$-wave scattering length enter only at higher
orders in $n^{1/3}$ and $\la_0^{-1}$ than explicitely given in
(\ref{deltathom}) and (\ref{deltattrap}).

The effect of a wide trapping potential can be accommodated by
writing
\beq
\label{nb}
N_b=\int d^3x\,n(T,\mu-V(\boldx)),
\eeq
where the density function $n(T,\mu)$ holds for the homogenous gas and,
through the order needed here, may be obtained from a perturbative
calculation \cite{ArTo,HuYaLu}.
The relation (\ref{nb}) allows one to restrict all further field-theoretic
considerations to the homogenous case \cite{ArTo}.

Due to the imaginary-time periodicity of the field $\psi(\boldx,\tau)$,
it may be decomposed into imaginary-time frequency modes with Matsubara
frequencies $\om_j=2\pi j/\be$.
Subsequently, the nonzero Matsubara frequencies are integrated
out, leaving a three-dimensional field theory for the zero-Matsubara
modes $\psi_0$.
Conventionally, this theory is written as
\beq
\label{s3}
S_3=\int d^3x\left[\f{1}{2}(\boldnabla\phi_a)^2
+\f{\rb}{2}\phi_a^2+\f{u}{24}(\phi_a^2)^2+f_B\right],
\eeq
where $\phi_a$ are $N$ real field components, i.e., $a=1,\ldots,N$ and
$N=2$ is the case interesting for BEC.
In general, $S_3$ contains a hierarchy of infinitely many terms, but for
the orders given in (\ref{deltathom}) and (\ref{deltattrap}), it is
sufficient to consider the terms in (\ref{s3}).
The relations between the parameters and fields in $S_{3+1}$ and those
of the effective field theory given by $S_3$ may be determined by a
perturbative matching calculation and are provided in \cite{ArTo,ArMoTo}
through the order needed for the purposes here.

Finally, the coefficients in (\ref{deltathom}) are given by \cite{ArMoTo}
\bse
\label{cbar}
\bea
\label{c1bar}
\bar{c}_1
&=&
-\f{4(4\pi)^3}{\ze(\threehalf)^{4/3}}{\;\ka_2},
\\
\label{c2barp}
\bar{c}_2'
&=&
-\f{16(4\pi)\ze(\onehalf)}{3\ze(\threehalf)^{5/3}}
\approx19.7518,
\\
\label{c2barpp}
\bar{c}_2''
&=&
\f{16(4\pi)\ze(\onehalf)}{9\ze(\threehalf)^{5/3}}\ln[\ze(\threehalf)]
+\f{28(4\pi)^6}{\ze(\threehalf)^{8/3}}{\ka_2}^2
\nn\\
\lefteqn{+\f{8(4\pi)}{3\ze(\threehalf)^{5/3}}
\bigg\{
\f{\ze(\onehalf)^2\ln2}{\sqrt{\pi}}+2{K_2}-\sqrt{\pi}}~~~~~~~~~~~
\nn\\
\lefteqn{-[\ln2+3\ln(4\pi){+}1{-}36(4\pi)^2{R_2}
{-}24(4\pi)^2{\ka_2}]\ze(\onehalf)
\bigg\},}~~~~~~~~~
\nn\\
\eea
\ese
with perturbatively defined $K_2$ given in Appendix \ref{pertconsts}
and non-perturbative quantities $\ka_2$ and $R_2$, to which we return
below.
The coefficients in (\ref{deltattrap}), on the other hand, are given by
\cite{ArTo}
\bse
\label{cbreve}
\bea
\label{c1breve}
\breve{c}_1
&=&
\f{2}{3\ze(3)}\!\!\left[\sum_{i,j=1}^\infty
\f{1}{i^{3/2}j^{3/2}(i+j)^{1/2}}{-}2\ze(2)\ze(\threehalf)\right]
\nn\\
&\approx&
-3.426032,
\\
\label{c2brevep}
\breve{c}_2'
&=&
-\f{8(4\pi)\ze(2)}{3\ze(3)}\approx-45.8566,
\\
\label{c2brevepp}
\breve{c}_2''
&=&
C_2-\f{4(4\pi)\ze(2)}{3\ze(3)}[1-\ln2+3\ln(4\pi)
\nn\\
&&~~~~~~~~~~~~~~~~~~~~~
-4K_1-36(4\pi)^2R_2],
\eea
\ese
with perturbatively defined $K_1$ and $C_2$ given in Appendix
\ref{pertconsts} and the same $\ka_2$ and $R_2$ as in Eqs.~(\ref{cbar}).

The quantities $\ka_2$ and $R_2$ remain to be computed within the
three-dimensional theory.
Although they are well-defined, their generalization from two to
$N$ real field components is not unique.
We exploit this to define them in such a way that both their $N\ra0$
and $N\ra\infty$ limits exist.

Let $\ka_N$ be the critical limit of the interaction-induced shift
$\De\langle\phi^2\rangle/Nu$ and let $R_N$ be the critical limit of
$\rms(u,r)/(N+2)u^2$, where $\rms(u,r)$ is the renormalized version of
the bare quantity defined by $\rb(u,r)=r+\Si(0,u,\rb(u,r))$ with the
self-energy $\Si(k,u,\rb)$.
Here, renormalization refers to using the modified minimal
subtraction scheme (\msbar) after applying dimensional regularization (DR).
For $R_N$, we still have to specify the renormalization scale, which will
be done below in Sec.~\ref{defs}.
Both $\ka_N$ and $R_N$ receive contributions from all length scales,
which makes them non-universal critical quantities.
This is particularly obvious for $R_N$ because of its renormalization
scale dependence.
Application of such quantities to physical situations is restricted to
cases where the short-distance physics described by the effective field
theory is perturbative at a scale where the effective theory is still
applicable \cite{ArMoTo}.

Our motivation to generalize $\ka_2$ and $R_2$ to $N$ real field
components derives not only from the possibility to compare to the
$N=1,4$ Monte Carlo (MC) data of Sun \cite{Su}, but also because our
results may be applicable to physical situations different from BEC.
We remind the reader that $N=0,1,2,3$ correspond to the universality
classes of dilute polymer solutions, the Ising model, the
\textit{XY} model, and the Heisenberg model, respectively.

We like to comment on the use of the phrase ``universal.''
In the context of critical phenomena, quantities are considered
``universal'' if they are common to all members of a given universality
class.
They do not depend on the microscopic Hamiltonian used to compute them,
as long as the model under consideration is in the correct universality
class.
Universal quantities are typically critical exponents and critical
amplitude ratios.
In the context of BEC, on the other hand, ``universal'' refers to
quantities that depend only on the $s$-wave scattering length $a$ of
the two-body interaction potential and on no further details of the
potential such as the effective range $r_\text{s}$ or interaction
terms involving more than two particles.
Consequently, the quantities $\ka_N$ and $R_N$ are non-universal from
the critical phenomena viewpoint, but
$\bar{c}_1,\bar{c}_2',\bar{c}_2'',\breve{c}_1,\breve{c}_2'$, and
$\breve{c}_2''$ are universal BEC coefficients.

The remainder of this work is structured as follows.
In Sec.~\ref{defs}, we give a more detailed definition of the
non-perturbative quantities $\ka_N$ and $R_N$ and write down
their loop expansions.
In Sec.~\ref{largeN}, we determine the large-$N$ limit $R_\infty$.
In Sec.~\ref{pertseries} we define the perturbative series for
$R_N$ and provide perturbative coefficients for $N=0,1,2,3,4,\infty$ through
seven loops.
In Sec.~\ref{resummation}, we resum the perturbative series using
Kleinert's variational perturbation theory (VPT) and obtain seven-loop
estimates of $R_N$ for $N=0,1,2,3,4,\infty$.
Previous results for $\ka_N$ with $N=1,2,4$ are extended to $N=0,3$.
The results for $\ka_2$ and $R_2$ are translated into values for
$\bar{c}_2''$ and $\breve{c}_2''$, while the translation of our
VPT results for $\ka_2$ was already given in \cite{Ka6,Ka7,KaLPHYS03}.
We close with a discussion of our findings in Sec.~\ref{discussion}.

\section{\boldmath Definition of $\ka_N$ and $R_N$}
\label{defs}
As in \cite{Ka6,Ka7,KaLPHYS03} we use a renormalization scheme where
the bare parameter $\rb$ is traded for the renormalized quantity
$r\equiv\rb-\Si(0)$ with the self-energy $\Si(p)$.
Full and free propagator are then
\beq
\label{gp}
G(p)=\f{1}{p^2+r-[\Si(p)-\Si(0)]},\quad
G_0(p)=\f{1}{p^2+r},
\eeq
respectively, and the critical limit is identified as $r\ra0$.
This scheme renders finite all diagrams in the expansion (\ref{kappaNur})
below and all but the first two diagrams in the expansion (\ref{rb})
further below.
This is essential for numerical evaluations, since it allows to work
without regulator for all but these two diagrams.
The two divergent diagrams will be regulated by DR and the only remaining
divergent diagram renormalized by \msbar.
The corresponding Feynman rules are $\de_{ab}G_0(p)$ for internal
lines and $-u(\de_{ab}\de_{cd}+\de_{ac}\de_{bd}+\de_{ad}\de_{bc})/3$
for vertices, where indices run from $1$ to $N$.
The integration measure for the loop momenta is
$\int_p\equiv\mu^\ep\int d^Dp/(2\pi)^D$ with $D=3-\ep$ and the
renormalization scale $\mu$, whose introduction keeps the dimension of
physical quantities at their $D=3$ values even for $D\neq3$.
For convergent diagrams, we may set $D=3$ from the outset, of course.

$\ka_N$, relevant for the first-order shift of $T_c$ of a homogenous
Bose gas and also needed as one of the ingredients for the shift in
second order for both the homogenous gas and the case of a wide trap,
has been determined before.
We define it by $\ka_N\equiv\lim_{\ur\ra\infty}\ka_N(\ur)$, where
\bea
\label{kappaNur}
\ka_N(\ur)
&\equiv&
\f{\De\langle\phi^2\rangle}{Nu}
=\f{1}{u}\int_p[G(p)-G_0(p)]
\nn\\
&=&
\f{1}{Nu}\left[
{\cal R}
\rule[-9pt]{0pt}{24pt}
\bpi(34,0)
\put(17,3){\circle{24}}
\put(17,3){\oval(24,8)}
\put(5,3){\circle*{2}}
\put(29,3){\circle*{2}}
\put(17,-9){\makebox(0,0){$\times$}}
\epi
+
{\cal R}
\rule[-9pt]{0pt}{24pt}
\bpi(34,0)
\put(17,3){\circle{24}}
\put(6.6,-3){\line(1,0){20.8}}
\put(6.6,-3){\line(3,5){10.4}}
\put(27.4,-3){\line(-3,5){10.4}}
\put(6.6,-3){\circle*{2}}
\put(27.4,-3){\circle*{2}}
\put(17,15){\circle*{2}}
\put(17,-9){\makebox(0,0){$\times$}}
\epi
+\cdots\right]
\eea
with
\beq
\label{ur}
\ur\equiv\f{(N+2)u}{\sqrt{r}},
\eeq
and where the operator $\R$ recursively removes all zero-momentum
parts of any self-energy subdiagram and originates in the subtraction of
$\Si(0)$ from $\Si(p)$ in (\ref{gp}) \cite{Ka6,Ka7,KaLPHYS03}.
The cross in the diagrams is an insertion that merely separates two
propagators.
Power counting shows that $\ka_N(\ur)$ depends on $u$ and $r$ only
through the combination in (\ref{ur}).
The inclusion of $N+2$ in the definition of $\ur$ leaves the perturbative
coefficients of $\ka_N(\ur)$ finite for both $N=0$ and $N\ra\infty$,
when $\ka_N(\ur)$ is expressed as a power series in $\ur$.

Since we have computed $\ka_N$ \cite{c1kaN} for $N=1,2,4$ in the
framework of VPT through seven loops before \cite{Ka7,KaLPHYS03}, we
merely quote in Table~\ref{kappantab} our results and those other results
that appear to be most reliable (for detailed arguments supporting this
conclusion, see \cite{Ka7} and references therein), namely
those from MC simulations \cite{Su,KaPrSv,ArMoc1,ArMoMC}.
Our results for $N=0,3$ are new, though.
\btb
\caption{\label{kappantab}
$\ka_N$ for $N=1,2,4$ from MC simulations, for $N=0,1,2,3,4$ from
VPT through seven loops, and exact for $N\ra\infty$.
For obtaining the $N=0$ result from VPT, we have set $\om=0.805$ and
$\eta=0.029$ and thus $\om'=0.8817$, while for $N=3$, we have set
$\om=0.785$ and $\eta=0.037$ and thus $\om'=0.800$ (see, e.g.,
\cite{phi4book}).
For $N=1,2,4$, the reader is referred to \cite{Ka7}.}
\begin{center}
\btr{|c|c|l|}
\hline
$N$ & $\ka_N$ from MC & \multicolumn{1}{c|}{$\ka_N$ from VPT} \\\hline\hline
$0$ & \textemdash & $-(3.66\pm0.39)\times10^{-4}$ \\\hline
$1$ & $-(4.94\pm0.41)\times10^{-4}$ \cite{Su}
    & $-(4.86\pm0.45)\times10^{-4}$ \cite{Ka7} \\\hline
$2$ & 
\btr{l}
$-(5.85\pm0.23)\times10^{-4}$ \cite{KaPrSv} \\
$-(5.99\pm0.09)\times10^{-4}$ \cite{ArMoMC} 
\etr
& $-(5.75\pm0.49)\times10^{-4}$ \cite{Ka7}\\\hline
$3$ & \textemdash & $-(6.46\pm0.48)\times10^{-4}$ \\\hline
$4$ & $-(7.23\pm0.45)\times10^{-4}$ \cite{Su}
    & $-(6.99\pm0.48)\times10^{-4}$ \cite{Ka7} \\\hline
$\infty$ &
\multicolumn{2}{c|}{$-1/[6(4\pi)^2]\approx-1.05543\times10^{-3}$
\cite{BaBlZi}}
\\\hline
\etr
\end{center}
\etb

The other non-perturbative quantity needed for the second-order
contributions (\ref{c2barpp}) and (\ref{c2brevepp}) is the shift
of the chemical potential due to the interaction \cite{ArTo,ArMoTo}.
Within the three-dimensional theory (\ref{s3}), this amounts to
computing the $r\ra0$ limit of
\bea
\label{rb}
\lefteqn{\rb(u,r)=r+\Si(0,u,\rb(u,r))}
\nn\\
&=&\!
r+
\R
\rule[-8pt]{0pt}{22pt}
\bpi(22,12)
\put(11,3){\circle{12}}
\put(11,-3){\circle*{2}}
\put(5,-3){\line(1,0){12}}
\epi
+\R
\rule[-10pt]{0pt}{26pt}
\bpi(38,12)
\put(19,3){\circle{16}}
\put(5,3){\line(1,0){28}}
\put(11,3){\circle*{2}}
\put(27,3){\circle*{2}}
\epi
+\R
\rule[-14pt]{0pt}{34pt}
\bpi(34,12)
\put(17,3){\circle{24}}
\put(17,3){\oval(24,8)}
\put(5,3){\circle*{2}}
\put(29,3){\circle*{2}}
\put(17,-9){\circle*{2}}
\put(10,-9){\line(1,0){14}}
\epi
+\R
\rule[-8pt]{0pt}{22pt}
\bpi(46,12)
\put(5,-3){\line(1,0){36}}
\put(11,-3){\circle*{2}}
\put(35,-3){\circle*{2}}
\put(23,9){\circle*{2}}
\put(23,-3){\oval(24,24)[t]}
\put(11,9){\oval(24,24)[br]}
\put(35,9){\oval(24,24)[bl]}
\epi
\nn\\
&&{}\!
+\cdots
\eea
at zero external momentum.
This quantity, however, is ultraviolet (UV) divergent.
The theory (\ref{s3}) is superrenormalizable, and only the one-loop
diagram and the two-loop ``sunset'' diagram in the expansion (\ref{rb})
are divergent.

Following \cite{ArTo,ArMoTo,Su}, we use DR in $D=3-\ep$ dimensions
together with \msbar\ to renormalize $r_B$.
When using other schemes, the definition of the perturbative quantities
in (\ref{c2barpp}) and (\ref{c2brevepp}) would have to be changed as well,
leading to the same values of $\bar{c}_2''$ and $\breve{c}_2''$.
The one-loop diagram in (\ref{rb}) is finite in DR and consequently
needs no subtraction.
The only divergent diagram is then the two-loop ``sunset'' diagram,
which is computed in Appendix \ref{onetwoloop}.
Using the results derived there, we define the renormalized quantity
$\rms(u,r)$ by
\beq
\label{rmsdef}
\rb(u,r)=\rms(u,r)+\f{(N+2)u^2}{36(4\pi)^2\ep}.
\eeq
$\rms(u,r)$ then remains finite as $\ep\ra0$.
We further define $R_N=\lim_{\ur\ra\infty}R_N(\ur)$, where
\beq
\label{rndef}
R_N(\ur)\equiv\left.\lim_{\ep\ra0}\f{\rms(u,r)}{(N+2)u^2}
\right|_{\mubar=(N+2)u/12}
\eeq
with $\ur$ from (\ref{ur}), and where $\mubar$ is the renormalization
scale in the \msbar\ scheme, see also Appendix \ref{onetwoloop}.
The relation with the quantity
$R\equiv\lim_{\ep\ra0}\left.\rms(u,0)/u^2\right|_{\mubar=u/3}$
determined for $N=2$ in \cite{ArMoMC} and for $N=1,4$ in \cite{Su} is
\beq
R_N=\f{R}{N+2}+\f{1}{18(4\pi)^2}\ln\f{N+2}{4},
\eeq
so that, for the BEC case $N=2$,
\beq
R=4R_2.
\eeq
Our definition leads to finite values for both $R_0$ and $R_\infty$.
The exact value of the latter will be computed next.

\section{\boldmath$1/N$-Expansion for $\ka_N$ and $R_N$}
\label{largeN}
Consider expansions of $\ka_N$ and $R_N$ in powers of $N^{-1}$, where
$u$ is considered to be proportional to $N^{-1}$, so that $\ur$ is of
order $N^0$.
For the purposes of this section, each diagram represents only its
leading order part in powers of $N$.

Denote the leading order (LO), next-to-leading order (NLO),
etc.\ contributions to $\ka_N$ in a $1/N$-expansion by $\ka_N^{(0)}$,
$\ka_N^{(1)}$, etc., respectively, where $\ka_N^{(k)}\propto N^{-k}$.
In LO, only one type of diagrams contributes and, in
general dimension $D=3-\ep$, we have
\bea
\label{kaN0}
\lefteqn{\ka_\infty=\ka_N^{(0)}}
\nn\\
&=&
\f{1}{Nu}\lim_{r\ra0}\left[{\cal R}
\rule[-9pt]{0pt}{24pt}
\bpi(34,0)
\put(17,3){\circle{24}}
\put(17,3){\oval(24,8)}
\put(5,3){\circle*{2}}
\put(29,3){\circle*{2}}
\put(17,-9){\makebox(0,0){$\times$}}
\epi
+
{\cal R}
\rule[-9pt]{0pt}{24pt}
\bpi(34,0)
\put(17,3){\circle{24}}
\put(6.6,-3){\line(1,0){20.8}}
\put(6.6,-3){\line(3,5){10.4}}
\put(27.4,-3){\line(-3,5){10.4}}
\put(6.6,-3){\circle*{2}}
\put(27.4,-3){\circle*{2}}
\put(17,15){\circle*{2}}
\put(17,-9){\makebox(0,0){$\times$}}
\epi
+{\cal R}
\rule[-9pt]{0pt}{24pt}
\bpi(34,0)
\put(17,3){\circle{24}}
\put(8.5,-5.5){\line(1,0){17}}
\put(8.5,11.5){\line(1,0){17}}
\put(8.5,-5.5){\line(0,1){17}}
\put(25.5,-5.5){\line(0,1){17}}
\put(25.5,-5.5){\circle*{2}}
\put(25.5,11.5){\circle*{2}}
\put(8.5,11.5){\circle*{2}}
\put(8.5,-5.5){\circle*{2}}
\put(17,-9){\makebox(0,0){$\times$}}
\epi
+\cdots\right]
\nn\\
&=&
-\f{\pi\ep\ga_\ep\mu^{2\ep}}{3(1+\ep)\sin\left(\f{2\pi\ep}{1+\ep}\right)}
\left(\f{N\ga_\ep\mu^\ep u}{6}\right)^{-2\ep/(1+\ep)}
\f{S_D}{(2\pi)^D},
\nn\\
\eea
where $S_D=2\pi^{D/2}/\Ga(D/2)$ is the surface of a unit sphere in
$D$ dimensions.
See Appendix \ref{largeNdiags} for intermediate steps.
Taking the limit $\ep\ra0$, we get the result
\beq
\ka_N=-\f{1}{6(4\pi)^2}+\od(N^{-1})
\eeq
of \cite{BaBlZi}, determined also in \cite{ArToN}.
We have rederived the intermediate result (\ref{kaN0}) of \cite{BaBlZi}
in our conventions since it turns out to be useful also for computing the
large-$N$ limit of $R_N$.

Now consider $\rms(u,0)$ as defined by (\ref{rb}) and (\ref{rmsdef}).
In its $1/N$-expansion, denote the LO, NLO, etc.\ contributions to
$\rms(u,0)$ by $\rms^{(0)}(u,0)$, $\rms^{(1)}(u,0)$, etc.
The first two terms on the right hand side of (\ref{rb}) are the only
contributions to $\rms(u,r)$ that are of order $N^0$.
Since, observing (\ref{oneloop}), they both vanish in DR as $r\ra0$,
we get $\rms^{(k)}(u,0)\propto N^{-k-1}$, up to logarithmic corrections
of $\rms^{(0)}(u,0)$ and $\rms^{(1)}(u,0)$, due to the UV divergence of
the sunset diagram.

In LO, $\rms(u,0)$ receives contributions from two classes of diagrams
and the counterterm defined in (\ref{rmsdef}).
The first contribution is
\bea
\label{rms0a}
\rms^{(0a)}(u,0)
&=&
\lim_{r\ra0}\bigg[
\R
\rule[-10pt]{0pt}{26pt}
\bpi(38,12)
\put(19,3){\circle{16}}
\put(5,3){\line(1,0){28}}
\put(11,3){\circle*{2}}
\put(27,3){\circle*{2}}
\epi
+\R
\rule[-8pt]{0pt}{22pt}
\bpi(46,12)
\put(5,-3){\line(1,0){36}}
\put(11,-3){\circle*{2}}
\put(35,-3){\circle*{2}}
\put(23,9){\circle*{2}}
\put(23,-3){\oval(24,24)[t]}
\put(11,9){\oval(24,24)[br]}
\put(35,9){\oval(24,24)[bl]}
\epi
\nn\\
&&~~~~~~~
+\R
\rule[-8pt]{0pt}{22pt}
\bpi(62,12)
\put(5,-3){\line(1,0){52}}
\put(11,-3){\circle*{2}}
\put(51,-3){\circle*{2}}
\put(23,9){\circle*{2}}
\put(39,9){\circle*{2}}
\put(23,-3){\oval(24,24)[tl]}
\put(11,9){\oval(24,24)[br]}
\put(39,-3){\oval(24,24)[tr]}
\put(51,9){\oval(24,24)[bl]}
\qbezier(23,9)(31,1)(39,9)
\qbezier(23,9)(31,17)(39,9)
\epi
+\cdots\bigg]
\nn\\
&=&
-\f{Nu^2}{6\ep}\ka_N^{(0)},
\eea
with intermediate steps provided in Appendix \ref{largeNdiags}.

The computation of the second contribution is even more closely related
to that of $\ka_N^{(0)}$,
\bea
\label{rms0b}
\lefteqn{\rms^{(0b)}(u,0)}
\nn\\
&=&
\lim_{r\ra0}\left[\R
\rule[-14pt]{0pt}{34pt}
\bpi(34,12)
\put(17,3){\circle{24}}
\put(17,3){\oval(24,8)}
\put(5,3){\circle*{2}}
\put(29,3){\circle*{2}}
\put(17,-9){\circle*{2}}
\put(10,-9){\line(1,0){14}}
\epi
+{\cal R}
\rule[-14pt]{0pt}{34pt}
\bpi(34,0)
\put(17,3){\circle{24}}
\put(6.6,-3){\line(1,0){20.8}}
\put(6.6,-3){\line(3,5){10.4}}
\put(27.4,-3){\line(-3,5){10.4}}
\put(6.6,-3){\circle*{2}}
\put(27.4,-3){\circle*{2}}
\put(17,15){\circle*{2}}
\put(17,-9){\circle*{2}}
\put(10,-9){\line(1,0){14}}
\epi
+{\cal R}
\rule[-14pt]{0pt}{34pt}
\bpi(34,0)
\put(17,3){\circle{24}}
\put(8.5,-5.5){\line(1,0){17}}
\put(8.5,11.5){\line(1,0){17}}
\put(8.5,-5.5){\line(0,1){17}}
\put(25.5,-5.5){\line(0,1){17}}
\put(25.5,-5.5){\circle*{2}}
\put(25.5,11.5){\circle*{2}}
\put(8.5,11.5){\circle*{2}}
\put(8.5,-5.5){\circle*{2}}
\put(17,-9){\circle*{2}}
\put(10,-9){\line(1,0){14}}
\epi
+\cdots\right]
\nn\\
&=&
-\f{Nu^2}{6}\ka_N^{(0)},
\eea
and we arrive at
\bea
\lefteqn{\rms^{(0a)}(u,0)+\rms^{(0b)}(u,0)=-\f{Nu^2}{6\ep}(1+\ep)\ka_N^{(0)}}
\nn\\
&=&
\f{\pi\mu^\ep u}{3\sin\left(\f{2\pi\ep}{1+\ep}\right)}
\left(\f{N\ga_\ep\mu^\ep u}{6}\right)^{(1-\ep)/(1+\ep)}\f{S_D}{(2\pi)^D}.~~~~
\eea
Subtracting from this the counterterm in (\ref{rmsdef}) and taking
the limit $\ep\ra0$, we obtain
\bea
\label{rms0}
\rms(u,0)
&=&
\f{Nu^2}{18(4\pi)^2}\left(-\ln\f{Nu}{\mubar}+1+4\ln2+\ln3\right)
\nn\\
&&{}
+\od(N^{-2}\ln N).
\eea
Setting $\mubar=(N+2)u/12$, we arrive at
\beq
R_N=\f{1+2\ln2}{18(4\pi)^2}+\od(N^{-1}),
\eeq
from which we obtain the exact large-$N$ limit
\beq
\label{rninf}
R_\infty=\f{1+2\ln2}{18(4\pi)^2}
\approx8.39521\times10^{-4}.
\eeq

\section{\boldmath Perturbative Series for $R_N$}
\label{pertseries}
The perturbative series corresponding to the loop expansion
(\ref{kappaNur}) of $\ka_N(\ur)$ and its resummation with VPT
was discussed before in \cite{Ka6,Ka7,KaLPHYS03}.
Note, however, that the use of $N+2$ (instead of $N$ as in
\cite{Ka6,Ka7,KaLPHYS03}) in the definition (\ref{ur}) of $\ur$
is crucial for defining a perturbative expansion for $\ka_0(\ur)$.

Here we focus on the perturbative series corresponding to the
loop expansion of $R_N(\ur)$, defined by (\ref{rb})--(\ref{rndef}).
We have constructed the relevant diagrams through seven loops using
the recursive methods of \cite{recrel}.
The numbers of diagrams in some low loop orders is listed in
Table~\ref{nd} and the diagrams are collected in Table~\ref{diagrams}.
\btb
\caption{\label{nd}Numbers of diagrams for $R_N$ in low loop orders $l$.}
\btr{|c||c|c|c|c|c|c|c|c|c|}
\hline
$l$   & 1 & 2 & 3 & 4 &  5 &  6 &   7 &    8 &    9 \\\hline
$n_l$ & 1 & 1 & 2 & 5 & 16 & 62 & 265 & 1387 & 8038 \\
\hline
\etr
\etb
A convenient representation of each diagram for symbolic manipulations
by computer code was defined in \cite{MurNi} and is listed in
Table~\ref{encodings} and explained in Appendix~\ref{encods}.
The corresponding weights and group factors can be found in
Tables~\ref{allresults} and \ref{groupfactors}, respectively.
The integrals corresponding to the diagrams were computed by Nickel
and Murray and were used for the determination of critical exponents
for $N=0,1,2,3$ in \cite{MurNi}.
They were communicated to us by Nickel and are listed in
Table~\ref{allresults}.

Through six loops, the numerical results for the integrals corresponding
to the relevant diagrams were published in \cite{MutNi}, while the
seven-loop results have, to the best of our knowledge, never been published
(except for the small fraction of diagrams needed for the computation
of $\ka_N$ \cite{Ka7}).
We therefore provide the numerical results for the diagrams through
seven loops in Table~\ref{allresults}.

Any subtracted $l$-loop diagram $\R D_n$ represents the product
$\R D_n=(-u)^lw_ng_nI_n$, where $w_n$ is the combinatorial weight of
the diagram, $g_n$ its group factor from the O($N$) symmetry and
$I_n$ the value of the corresponding integral.
The only divergent diagrams have one and two loops, respectively,
and are treated in Sec.~\ref{defs} and in Appendix~\ref{onetwoloop}.
In DR, the one-loop diagram is in fact convergent.

Consider the case of zero external momentum, which is all we need for
$R_N$.
Dimensional analysis reveals that any $l$-loop integral $I_n$ convergent
in DR is proportional to $r^{1-l/2}$ for $D=3$.
Therefore, $I_n$ for any $l$-loop diagram with $l\neq2$ may be
obtained from the $r$-derivative of $I_n$ provided in
Table~\ref{allresults} by
\beq
I_n=\f{r^{1-l/2}}{1-l/2}\left.\f{\p I_n}{\p r}\right|_{r=1}.
\eeq
We conclude that the contribution from any $l$-loop integral with $l\neq2$
to $R_N(\ur)$ is proportional to $\ur^{l-2}$.
Combining (\ref{rb})--(\ref{rndef}), (\ref{oneloop}), and
(\ref{sunsetres}), we obtain
\beq
\label{rnexp}
R_N(\ur)=b_2'\ln\ur+\sum_{l=0}^\infty b_l\ur^{l-2}
\eeq
with
\bse
\label{bcoeffs}
\bea
b_0&=&N+2,
\\
b_1&=&\f{N+2}{6(4\pi)},
\\
b_2&=&\f{1-4\ln6}{36(4\pi)^2},
\\
b_2'&=&\f{1}{18(4\pi)^2}.
\eea
\ese
The higher-order perturbative coefficients $b_l$ with $l>2$ have to be
computed from the data in Tables~\ref{allresults} and \ref{groupfactors}.
They are listed in Table~\ref{pertcoeffs} through $l=7$ for
$N=0,1,2,3,4,\infty$.

\nc{\pertcoeffscaption}
{Perturbative coefficients $b_l$ of the expansion (\ref{rnexp}) from
three through seven loops for $N=0,1,2,3,4,\infty$.
The lower-loop coefficients $b_0$, $b_1$, $b_2$, and $b_2'$ are provided
in analytical form in Eqs.~(\ref{bcoeffs}).}
\btbs
\begin{center}
\caption{\label{pertcoeffs}\pertcoeffscaption}
\btr{|c|l|l|l|l|l|l|}
\hline
& \multicolumn{1}{c|}{$N=0$}
& \multicolumn{1}{c|}{$N=1$}
& \multicolumn{1}{c|}{$N=2$}
& \multicolumn{1}{c|}{$N=3$}
& \multicolumn{1}{c|}{$N=4$}
& \multicolumn{1}{c|}{$N\rightarrow\infty$}
\\\hline
$b_{3}$
& $-8.91161\times10^{-6}$
& $-6.51591\times10^{-6}$
& $-5.31807\times10^{-6}$
& $-4.59936\times10^{-6}$
& $-4.12022\times10^{-6}$
& $-1.72453\times10^{-6}$ \\
$b_{4}$
& $\phantom{+}1.49970\times10^{-7}$
& $\phantom{+}8.09764\times10^{-8}$
& $\phantom{+}5.42095\times10^{-8}$
& $\phantom{+}4.06228\times10^{-8}$
& $\phantom{+}3.25956\times10^{-8}$
& $\phantom{+}4.82748\times10^{-9}$ \\
$b_{5}$
& $-4.00122\times10^{-9}$
& $-1.59256\times10^{-9}$
& $-8.72249\times10^{-10}$
& $-5.64780\times10^{-10}$
& $-4.04786\times10^{-10}$
& $-1.89600\times10^{-11}$ \\
$b_{6}$
& $\phantom{+}1.35890\times10^{-10}$
& $\phantom{+}3.96903\times10^{-11}$
& $\phantom{+}1.77165\times10^{-11}$
& $\phantom{+}9.87917\times10^{-12}$
& $\phantom{+}6.30620\times10^{-12}$
& $\phantom{+}8.60806\times10^{-14}$ \\
$b_{7}$
& $-5.42433\times10^{-12}$
& $-1.15654\times10^{-12}$
& $-4.19012\times10^{-13}$
& $-2.00553\times10^{-13}$
& $-1.13703\times10^{-13}$
& $-4.23973\times10^{-16}$ \\
\hline
\etr
\end{center}
\etbs
Note that the coefficients $b_0$ and $b_1$ diverge as $N\ra\infty$.
Since they multiply negative powers of $\ur$, this is not a problem
if we take the limit $\ur\ra\infty$ before letting $N\ra\infty$.
This means that the terms involving $b_0$ and $b_1$ have to be dropped
when resumming the perturbative series for $N\ra\infty$, see also the
following section.

\section{Resummation and Results}
\label{resummation}
The loop expansions of $\ka_N(\ur)$ and $R_N(\ur)$ suffer from
infrared divergences as $r\ra0$, i.e., as $\ur\ra\infty$.
This corresponds to the well-known fact that the description of
long-distance physics by perturbation theory breaks down at second-order
phase transitions.

The resummation of $\ka_N(\ur)$ through seven loops has been carried
out in \cite{Ka7,KaLPHYS03} for $N=1,2,4$, giving the results reported
in Table~\ref{kappantab}.
We have added results for $N=0$ and $N=3$.

Here we carry out the corresponding resummation of $R_N(\ur)$, given
by the expansion (\ref{rnexp}).
There are two relevant changes as compared to the resummation of
$\ka_N(\ur)$.
One is the appearance of a logarithm of $\ur$.
This turns out not to be an obstacle and VPT can be carried out as before.
The other is the appearance of negative powers of $\ur$ with nonzero
coefficients $b_0$ and $b_1$.
These terms vanish in the limit $\ur\ra\infty$ and therefore do not
contribute to $R_N$.
However, such terms influence resummation.
For the computation of $\ka_N$, it was argued in \cite{Ka6,Ka7,KaLPHYS03}
that it is most natural to work with a scheme where such negative powers
are absent, but the same scheme leads to the nonzero coefficients $b_0$
and $b_1$ in (\ref{rnexp}).
Below, we will in fact employ the ambiguity of not including one or both
of these negative-power terms in our resummation procedure to estimate
the precision of our result.
For large $N$, one should drop the coefficients $b_0$ and $b_1$, since
they are the only ones to grow linearly with $N$ and therefore make the
perturbative series unnatural and consequently resummation with VPT tends
not to work well if they are included. 

The interactions cause the phase transition to be second order
with critical exponents of the O(2) universality class.
The leading class of corrections of a physical quantity that remains
finite in the critical limit are integer powers of $t^{\om\nu}$
\cite{phi4book,We,ZiPeVi}, where $t\equiv(T-T_c)/T_c$,
$\nu$ is the critical exponent of the correlation length, and
$\om=\be'(g^*)$ in a renormalization group approach.
Since, in our renormalization scheme, the propagator obeys
$G(p=0)=1/r\propto t^{-\ga}$, the leading corrections are integer
powers of $u_r^{-\om'}$ with
\beq
\label{omp}
\om'=\f{2\om}{2-\eta}.
\eeq
Here we have employed the universal scaling relation $\ga=\nu(2-\eta)$,
where $\eta$ is the anomalous dimension of the critical propagator,
i.e., $G(r=0)\propto1/p^{2-\eta}$ in the small-$p$ limit.
If we neglect any other powers, we have
\beq
\label{larggeur}
R_N(u_r)=\sum_{m=0}^\infty f_mu_r^{-m\om'}.
\eeq
Thus an expansion which correctly describes the leading corrections
to scaling \cite{We} has the form (\ref{larggeur}) with $\om'$ from
(\ref{omp}).

The ansatz (\ref{larggeur}) does not account for so-called confluent
singularities which cause the true large-$u_r$ expansion to also contain
other negative powers of $u_r$, which are subleading at least compared
to $u_r^{-\om'}$.
We can expect methods that can correctly accommodate the leading
behavior in (\ref{larggeur}) to converge faster to the
true result than methods having the wrong leading behavior, such as, e.g.,
Pad\'{e} approximants or the linear delta expansion (LDE; see \cite{HaKl}
for a general criticism of the application of the LDE in the context
of field theory).
On the other hand, convergence will be slowed by the fact that we
do not make an ansatz reflecting the full power structure in $u_r$,
but the expansion (\ref{larggeur}) will try to mimic the neglected
subleading powers.

The alternating signs of the $b_l$ displayed in Table~\ref{pertcoeffs}
suggest that the perturbative series for $R_N(u_r)$ is Borel summable.
In the context of critical phenomena, such series have been successfully
resummed using Kleinert's VPT (see \cite{Kl2,Kl3,Kl4} and Chapters 5 and
19 of the textbooks \cite{pibook} and \cite{phi4book}, respectively;
improving perturbation theory by a variational principle goes back at
least to \cite{Yu}).
Accurate critical exponents \cite{Kl3,Kl4,phi4book} and amplitude ratios
\cite{KlvdB} have been obtained.
For a truncated partial sum through $u_r^{L-2}$ of (\ref{rnexp}),
the method requires replacing
\bea
u_r
\!&\ra&\!
t\uh
\left\{1+t\left[\left(\f{\uh}{u_r}\right)^{\om'}-1\right]
\right\}^{-1/\om'}
\eea
(note that this is an identity for $t=1$),
reexpanding the resulting expression in $t$ through $t^{L-2}$, setting
$t=1$, and then optimizing in $\uh$, where optimizing is done in accordance
with the principle of minimal sensitivity (PMS) \cite{Ste} and in practice
means finding appropriate stationary or turning points.
While to the best of our knowledge, this has so far been done only for
powers of $\ur$, the procedure goes straightforwardly through for the
logarithmic term in (\ref{rnexp}) as well, if $\ln t$ is treated as $t^0$
for the purpose of power counting. 
Before optimizing, we may take the limit $\ur\ra\infty$, since we are
only interested in $R_N=\lim_{\ur\ra\infty}R_N(\ur)$.
The result is a function $f_0^{(L)}(\uh,\om')$, whose value optimized
in $\uh$ is our $L$-loop VPT estimate $f_0^{(L)}$ for $f_0$ in the
expansion (\ref{larggeur}), i.e., for $R_N$.

There are different ways of fixing the exponent $\om'$.
One may take it from other sources, since it is a universal critical
exponent describing the approach to the critical point.
One may selfconsistently determine it by setting to zero an appropriate
logarithmic derivative of the quantity to be determined \cite{phi4book,Kl3}.
These two approaches have been followed to determine $\ka_N$ in
\cite{Ka6,Ka7,KaLPHYS03}.
It turns out, however, that, for $R_N$, even when including up to
seven-loop perturbative coefficients, these approaches do not lead
to satisfactory plateaus when plotting $f_0^{(L)}(\uh,\om')$ as a
function of $\uh$.
However, the existence of such plateaus that become wider and flatter
as the number of involved perturbative coefficients rises, is an
indication that the PMS criterion works and therefore for the
applicability of VPT to the problem at hand.
We have therefore adopted an alternative approach \cite{HaKl}, where
$\om'$ is varied such that satisfactory plateaus develop.
In practice, this means finding $\om'$ so that not only the first,
but also the second or third derivative with respect to $\uh$ vanishes.
As a check, it is mandatory to inspect the resulting plots of
$f_0^{(L)}(\uh,\om')$ as a function of $\uh$ for the resulting
plateaus.

\begin{figure}[ht]
\begin{center}
\bpi(230,140)(0,-2)
\put(0,0){\includegraphics[width=8cm,angle=0]{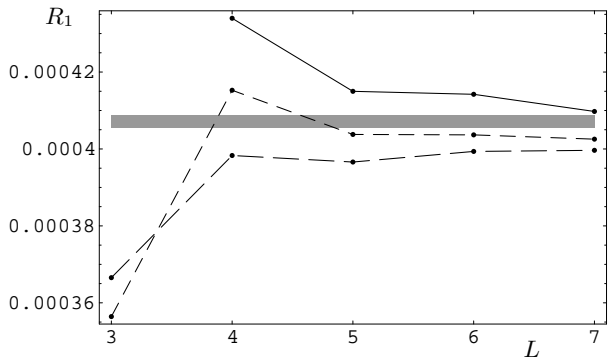}}
\put(196,-2){$L$}
\put(15,124){$R_1$}
\epi
\end{center}
\vspace{-15pt}
\caption{\label{r1fig}
$R_1$ from VPT as a function of the number of loops $L$.
Perturbative coefficients starting at zero loops (long dashes),
one loop (short dashes) and two loops (solid) are used.
The horizontal bar is the MC result including its error bar from
Table~\ref{rntab}.}
\end{figure}

\begin{figure}[ht]
\begin{center}
\bpi(230,140)(0,-2)
\put(0,0){\includegraphics[width=8cm,angle=0]{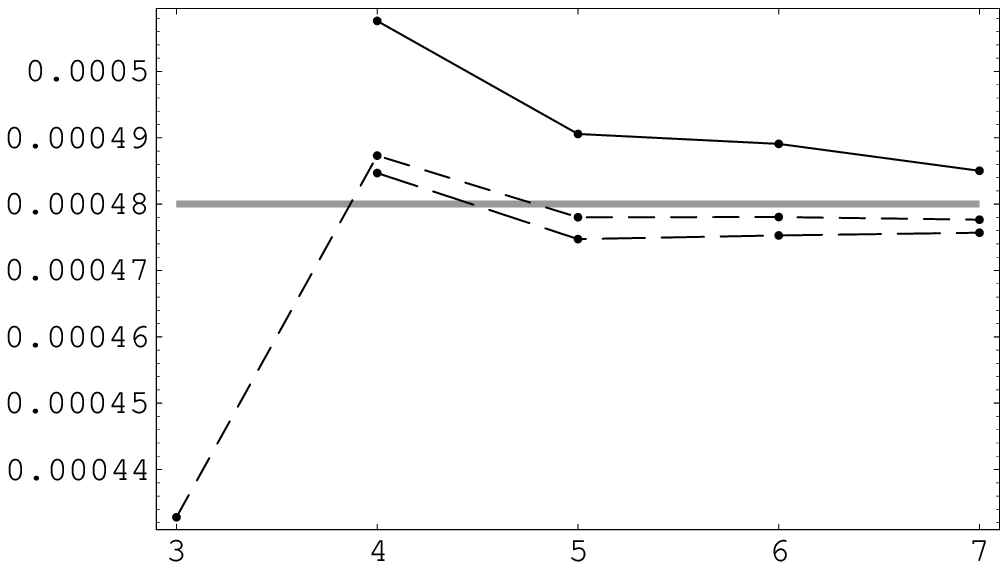}}
\put(196,-2){$L$}
\put(15,124){$R_2$}
\epi
\end{center}
\vspace{-15pt}
\caption{\label{r2fig}
Same as in Fig.~\ref{r1fig}, but for $R_2$.}
\end{figure}

\begin{figure}[ht]
\begin{center}
\bpi(230,140)(0,-2)
\put(0,0){\includegraphics[width=8cm,angle=0]{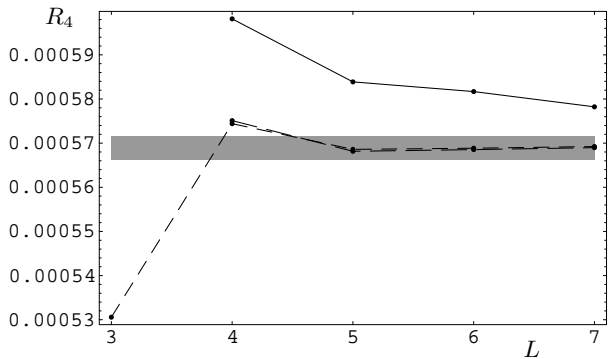}}
\put(196,-2){$L$}
\put(15,124){$R_4$}
\epi
\end{center}
\vspace{-15pt}
\caption{\label{r4fig}
Same as in Fig.~\ref{r1fig}, but for $R_4$.}
\end{figure}

As mentioned above, the zero- and one-loop contributions to $R_N(\ur)$
ultimately do not contribute to $R_N$.
However, they influence resummation if only a finite number of
perturbative coefficients are available.
We have used the variance of the results when omitting the first or
the first two orders in (\ref{rnexp}) to estimate the error of our
estimates.
The corresponding estimates for $R_N$ as a function of the number of
loops for $N=0,1,2,3,4,\infty$ can be found in Table~\ref{vptres}.
\nc{\vptrescaption}{
Results from VPT for $10^4R_N$ for $N=0,1,2,3,4,\infty$.
The perturbative coefficients involved are $b_{l_{\min}}$ through
$b_{l_{\max}}$ in each case.
Missing entries indicate that the PMS criterion does not provide a
satisfactory solution or no solution at all.
The data of this table for $N=1,2,4$ are, together with the corresponding
MC data, visualized in Figs.~\ref{r1fig},\ref{r2fig},\ref{r4fig},
respectively.}
\btb
\begin{center}
\caption{\label{vptres}\vptrescaption}
\btr{|c|c|l|l|l|l|l|}
\hline
\multicolumn{2}{|c|}{$10^4R_N$} & \multicolumn{5}{|c|}{$l_{\max}$}\\
\hline
$N$ & $l_{\min}$ & \multicolumn{1}{c|}{$3$}
& \multicolumn{1}{c|}{$4$}
& \multicolumn{1}{c|}{$5$}
& \multicolumn{1}{c|}{$6$}
& \multicolumn{1}{c|}{$7$}
\\\hline
  & 0 & $2.52263$  & $3.23565$  & $2.79559$  & $2.83351$  & $2.83130$ \\
0 & 1 & $2.40522$  & $3.02281$  & $2.88227$  & $2.88454$  & $2.86700$ \\
  & 2 &  & $3.19514$  & $2.98071$  & $2.98519$  & $2.93592$ \\
\hline
  & 0 & $3.66540$  & $3.98298$  & $3.96602$  & $3.99360$  & $3.99648$ \\
1 & 1 & $3.56421$  & $4.15265$  & $4.03773$  & $4.03662$  & $4.02549$ \\
  & 2 &  & $4.33980$  & $4.14986$  & $4.14212$  & $4.09751$ \\
\hline
  & 0 &  & $4.84689$  & $4.74733$  & $4.75283$  & $4.75693$ \\
2 & 1 & $4.32805$  & $4.87315$  & $4.78025$  & $4.78056$  & $4.77655$ \\
  & 2 &  & $5.07615$  & $4.90577$  & $4.89084$  & $4.85027$ \\
\hline
  & 0 & $5.15826$  & $5.37860$  & $5.28793$  & $5.29284$  & $5.29755$ \\
3 & 1 & $4.88146$  & $5.37527$  & $5.30106$  & $5.30329$  & $5.30637$ \\
  & 2 &  & $5.59508$  & $5.44013$  & $5.42087$  & $5.38362$ \\
\hline
  & 0 &  & $5.75103$  & $5.68154$  & $5.68540$  & $5.68979$ \\
4 & 1 & $5.30578$  & $5.74407$  & $5.68587$  & $5.68850$  & $5.69231$ \\
  & 2 &  & $5.98154$  & $5.83879$  & $5.81683$  & $5.78226$ \\
\hline
$\infty$ & 2 &  & $8.35299$  & $8.39432$  & $8.40716$  & $8.40705$ \\
\hline
\etr
\end{center}
\etb
In Figs.~\ref{r1fig},\ref{r2fig},\ref{r4fig}, we plot these results
for $N=1,2,4$ as a function of the number of loops together with
the corresponding MC data.
We conservatively estimate our mean result at the seven loop level by
taking the average of the most distant of the three estimates and take
their difference as the total error bar.
This results in the values given in Table~\ref{rntab}, showing agreement
with the corresponding MC values.
\btb
\caption{\label{rntab}
$R_N$ for $N=1,2,4$ from MC simulations, for $N=0,1,2,3,4$ from
VPT through seven loops, and exact for $N\ra\infty$.}
\begin{center}
\btr{|c|c|c|}
\hline
$N$ & $R_N$ from MC & $R_N$ from VPT \\\hline\hline
$0$ & \textemdash
    & $(2.884\pm0.052)\times10^{-4}$ \\\hline
$1$ & $(4.071\pm0.016)\times10^{-4}$ \cite{Su}
    & $(4.047\pm0.051)\times10^{-4}$ \\\hline
$2$ & $(4.8003\pm0.0053)\times10^{-4}$ \cite{ArMoMC} 
    & $(4.804\pm0.047)\times10^{-4}$ \\\hline
$3$ & \textemdash
    & $(5.341\pm0.043)\times10^{-4}$ \\\hline
$4$ & $(5.690\pm0.027)\times10^{-4}$ \cite{Su}
    & $(5.736\pm0.046)\times10^{-4}$ \\\hline
$\infty$ &
\multicolumn{2}{c|}{$(1+2\ln2)/[18(4\pi)^2]\approx8.39521\times10^{-4}$}
\\\hline
\etr
\end{center}
\etb
Our VPT results and the MC data may also be compared to the large-$N$
result (\ref{rninf}).
Note that only the differences between these values are relevant.
This is due to the arbitrariness in fixing the scale $\mubar$ that
effectively separates perturbative from non-perturbative physics and
causes a common arbitrary additive constant to the $R_N$.

\begin{figure}[ht]
\begin{center}
\bpi(230,140)(0,-2)
\put(0,0){\includegraphics[width=8cm,angle=0]{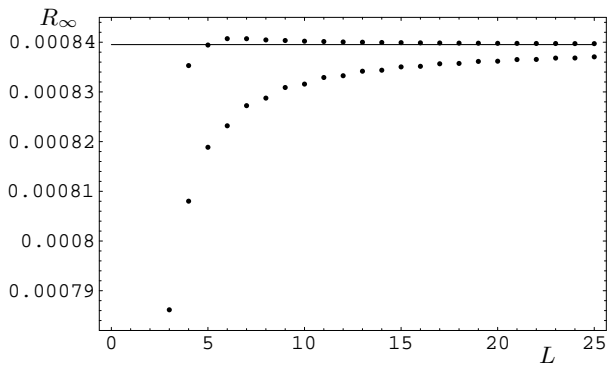}}
\put(202,-2){$L$}
\put(13,126){$R_\infty$}
\epi
\end{center}
\vspace{-15pt}
\caption{\label{largeNfig}
Exact result for $R_\infty$ (solid line) and perturbative results
resummed with VPT as a function of the number of loops $L$.
Lower dots: fixed $\om'=1$.
Upper dots: $\om'$ from the plateau method, described in the main text.
For $N\ra\infty$, the perturbative coefficients $b_l$ are easy to
numerically compute at any loop order.
Through seven loops, they are listed in Table~\ref{pertcoeffs}, while
the corresponding VPT resummed values for $R_\infty$ can be found in
Table~\ref{vptres}.}
\end{figure}

We have also applied VPT to the large-$N$ case.
As argued at the end of Sec.~\ref{pertseries}, we have to drop the one-
and two-loop terms since $b_0$ and $b_1$ diverge as $N\ra\infty$.
In Fig.~\ref{largeNfig}, we plot the results using fixed $\om'=1$
(since $\om=1$ and $\eta=0$ in this limit, see, e.g., \cite{phi4book})
and for the method of tuning $\om'$ as described above.
The second method has faster apparent convergence than the one with
$\om'=1$.
However, it approaches the limiting value monotonously only for $L\geq6$.
This shows the potential danger of making any extrapolations towards
higher loop orders based on the VPT results through seven loops for the
perturbative results for $N<\infty$.

Using (\ref{c2barpp}) and (\ref{c2brevepp}), our VPT results for $\ka_2$
given in Table~\ref{kappantab} and for $R_2$ given in Table~\ref{rntab}
translate into
\bse
\label{c2vptres}
\bea
\bar{c}_2''  &=&74.6\pm2.3,
\\
\breve{c}_2''&=&-155.0\pm0.7,
\eea
\ese
which should be compared to the MC results \cite{ArTo,ArMoTo}
\bse
\label{c2mcres}
\bea
\bar{c}_2''  &=&75.7\pm0.4,
\\
\breve{c}_2''&=&-155.0\pm0.1,
\eea
\ese
following from the MC data for $\ka_2$ and $R_2$ given in
Tables~\ref{kappantab} and \ref{rntab}, respectively.

Recently, an earlier attempt \cite{SCPiRaSe} to compute $\bar{c}_2''$
using the LDE has been improved by using a generalization of the LDE
that involves more free parameters \cite{KnNePi}.
These seem, however, to be introduced in a somewhat ad hoc manner, which
makes it difficult to assess the validity of the method.
Also, there is a small disagreement between our perturbative coefficients
and those used there \cite{ldeerror}.
The results  at five-loop order, the highest order
considered for $\bar{c}_2''$ in \cite{KnNePi}, lie between $73.5$ and
$79.3$ and are therefore in good agreement with the MC data and our VPT
results.

\section{Summary}
\label{discussion}
We have applied VPT through seven loops to the computation of two
non-perturbative non-universal quantities $\ka_N$ and $R_N$ of critical
O($N$) symmetric $\phi^4$ theory for $N=0,1,2,3,4$.
The results for $\ka_N$ and $R_N$ are listed in Tables~\ref{kappantab}
and \ref{rntab}, respectively.
While for $\ka_N$ this was just an extension of earlier work
\cite{Ka6,Ka7,KaLPHYS03} to $N=0,3$, the computation of $R_N$ with VPT
is new.
Our results are in agreement with the apparently most reliable
other results available obtained by MC simulations \cite{ArMoMC,Su}.
For comparison, the MC results are also listed in Tables~\ref{kappantab}
and \ref{rntab}.
They are available only for $N=1,2,4$, though.

In Tables~\ref{allresults} and \ref{groupfactors}, we provide all
perturbative input data that allow for an extension of our results to
other $N$, for checks of the input data and for other uses of these data.
E.g., they have been used before to compute critical exponents for
$N=0,1,2,3$ \cite{MurNi}.
The data, provided to us by Nickel, have, to the best of our knowledge,
not previously been published.

In a computation similar to the one that lead to the large-$N$ limit
$\ka_\infty$ \cite{BaBlZi}, we have computed the exact large-$N$
limit $R_\infty$.
The result is given in (\ref{rninf}).

Employing the findings of \cite{ArTo,ArMoTo}, our values $\ka_2$ and $R_2$
may be translated into the coefficients $\bar{c}_2''$ and $\breve{c}_2''$
of the second-order shifts (\ref{deltathom}) and (\ref{deltattrap})
of the BEC temperature for dilute homogenous and trapped Bose gases,
respectively.
This leads to the values in (\ref{c2vptres}), which are in good agreement
with the presumably most reliable other results (\ref{c2mcres})
\cite{ArTo,ArMoTo}, which follow from the MC data of \cite{ArMoMC}
listed in Tables~\ref{kappantab} and \ref{rntab}.

\begin{acknowledgments}
The author wishes to express his deep gratitude to B.~Nickel for
providing the results of his seven-loop calculations, without which
this work would not have been possible.
He thanks H.~Kleinert for discussions and him as well as J.O.~Andersen
and B.~Tom\'a\v{s}ik for a careful reading of the manuscript.
\end{acknowledgments}

\appendix

\section{\boldmath Perturbative Quantities contributing to
$\bar{c}_2''$ and $\breve{c}_2''$}
\label{pertconsts}
The perturbative constants needed in (\ref{c2barpp}) and (\ref{c2brevepp})
are given by \cite{ArMoTo}
\begin{widetext}
\beq
K_1
=
\f{1}{4\pi}\int_0^1\f{dt}{t}
\left[\text{Li}_{1/2}(t)^2-\f{\pi t}{1-t}\right]
\approx-0.6305682,
\eeq
\bea
K_2
&=&
\f{1}{4\pi}\int_0^1\f{ds}{s}\int_0^1\f{dt}{t}
\bigg\{\text{Li}_{1/2}(s)\text{Li}_{1/2}(t)\text{Li}_{-1/2}(st)
-\f{\sqrt{\pi}}{2}st(1-st)^{-3/2}
\left[\sqrt{\f{\pi}{1-s}}+\ze(\onehalf)\right]
\left[\sqrt{\f{\pi}{1-t}}+\ze(\onehalf)\right]\bigg\}
\nn\\
&\approx&
-0.135083,
\eea
and \cite{ArTo}
\bea
C_2
&=&
\f{5}{2}{\breve{c}_1}^2+\f{4}{3\ze(3)}\bigg[
\ze(\threehalf)^2+2\ze(\threehalf)\sum_{i,j=1}^\infty
\f{(i+j)^{1/2}-i^{1/2}-j^{1/2}}{i^{3/2}j^{3/2}}
\nn\\
&&~~~~~~~~~~~~~~~~~
-2\ba{c}\scs\infty\\\sum\\\scs\!\!\!\!\!\!i,j,k=1\!\!\!\!\!\ea
\f{1}{(ij)^{3/2}k^{1/2}}
\left(\f{1}{(i{+}j{+}k)^{1/2}}
{+}\f{ij}{(i{+}k)(j{+}k)(i{+}j{+}k)^{1/2}}{-}\f{1}{k^{1/2}}
\right)
\bigg]
\nn\\
&\approx&
21.4
\eea
\end{widetext}
with $\breve{c}_1$ from (\ref{c1breve}).

\section{Perturbative One- and Two-Loop Diagrams}
\label{onetwoloop}
Here we compute the first two diagrams in (\ref{rb}).
The one-loop diagram in (\ref{rb}), called \texttt{2-M1} in
Table~\ref{diagrams}, is finite in DR for $D\ra3$.
With the integration measure $\int_p\equiv\mu^\ep\int d^Dp/(2\pi)^D$
with $D=3-\ep$, it is given by
\bea
\label{oneloop}
\R
\rule[-8pt]{0pt}{22pt}
\bpi(22,12)
\put(11,3){\circle{12}}
\put(11,-3){\circle*{2}}
\put(5,-3){\line(1,0){12}}
\epi
&=&
\rule[-8pt]{0pt}{22pt}
\bpi(22,12)
\put(11,3){\circle{12}}
\put(11,-3){\circle*{2}}
\put(5,-3){\line(1,0){12}}
\epi
=\f{N+2}{6}(-u)\int_p\f{1}{p^2+r}
\nn\\
&=&
-\f{N+2}{6}
\f{\mu^\ep u\Ga(\f{\ep}{2}-\f{1}{2})}{(4\pi)^{(3-\ep)/2}r^{\ep/2-1/2}}
\nn\\
&=&
\f{(N+2)u\sqrt{r}}{6(4\pi)}+\od(\ep).
\eea

The second, ``sunset,'' diagram in (\ref{rb}), called \texttt{3-S2} in
Table~\ref{diagrams}, is the only diagram we need that is UV divergent
in DR for $D\ra3$.
At zero external momentum it is given by
\bea
\label{sunsetdef}
\lefteqn{\R
\rule[-10pt]{0pt}{26pt}
\bpi(38,12)
\put(19,3){\circle{16}}
\put(5,3){\line(1,0){28}}
\put(11,3){\circle*{2}}
\put(27,3){\circle*{2}}
\epi
=
\rule[-10pt]{0pt}{26pt}
\bpi(38,12)
\put(19,3){\circle{16}}
\put(5,3){\line(1,0){28}}
\put(11,3){\circle*{2}}
\put(27,3){\circle*{2}}
\epi}
\nn\\
&=&
\f{N+2}{18}(-u)^2\int_{pq}\f{1}{(p^2+r)(q^2+r)[(p+q)^2+r]}
\nn\\
&=&
\f{(N+2)u^2}{18}(I_{2-1}^a+I_{2-1}^b),
\eea
where
\beq
I_{2-1}^a=\int_{pq}\f{1}{p^2(q^2+r)[(p+q)^2+r]}
\eeq
is divergent as $D\ra3$ and will be computed in DR, and
\beq
I_{2-1}^b=-r\int_{pq}\f{1}{p^2(p^2+r)(q^2+r)[(p+q)^2+r]}
\eeq
is convergent for $D=3$.
For the computation of both $I_{2-1}^a$ and $I_{2-1}^b$ it is useful
to introduce Feynman parameters.
$I_{2-1}^a$ can easily be computed in closed form in arbitrary dimension,
\bea
\label{i21a}
I_{2-1}^a
&=&
-\f{1}{(4\pi)^3}\left(\f{r}{4\pi\mu^2}\right)^{-\ep}
\f{\Ga(\f{\ep+1}{2})\Ga(\f{\ep-1}{2})}{\ep}
\nn\\
&=&
\f{1}{2(4\pi)^2}\left[\f{1}{\ep}-\ln\f{r}{\mubar^2}-2\ln2+1+\od(\ep)\right],~~
\eea
where $\mubar$ is defined by $4\pi\mu^2=e^\gae\mubar^2$ and the use of
$\mubar$ instead of $\mu$ amounts by definition to working in \msbar\
instead of minimal subtraction (MS).
$I_{2-1}^b$ is evaluated for $D=3$ with the result
\beq
\label{i21b}
I_{2-1}^b
=-\f{r}{(2\pi)^3}\int_0^\infty dp\f{\arctan\f{p}{2\sqrt{r}}}{p(p^2+r)}
=\f{1}{(4\pi)^2}\ln\f{2}{3}.
\eeq
Adding (\ref{i21a}) and (\ref{i21b}), we get
\beq
\label{sunsetres}
\R\!
\rule[-10pt]{0pt}{26pt}
\bpi(38,12)
\put(19,3){\circle{16}}
\put(5,3){\line(1,0){28}}
\put(11,3){\circle*{2}}
\put(27,3){\circle*{2}}
\epi
\!=\f{(N+2)u^2}{36(4\pi)^2}
\left[\f{1}{\ep}-\ln\f{r}{\mubar^2}-2\ln3+1+\od(\ep)\right].
\eeq

\section{\boldmath Large-$N$ Expansion}
\label{largeNdiags}
Here we fill in the intermediate steps in deriving the results
(\ref{kaN0}) and (\ref{rms0a}).
The diagrams in (\ref{kaN0}) represent a geometric series, such that
\bea
\label{kappaN0}
\ka_N^{(0)}
&=&
\f{Nu}{18}\int_k\left[\De'(k)-\f{1}{k^2}\int_p\f{1}{p^4}\right]
\f{\De(k)}{1+\f{Nu}{6}\De(k)}
\nn\\
&=&
\f{1}{3}\int_k
\f{\De'(k)}{1+\left[\f{Nu}{6}\De(k)\right]^{-1}},
\eea
where
\beq
\label{oneloopbubble}
\De(k)\equiv\int_p\f{1}{(k+p)^2p^2}=\f{\ga_\ep\mu^\ep}{k^{1+\ep}}
\eeq
and
\beq
\label{oneloopbubbleprimem0}
\De'(k)\equiv\int_p\f{1}{(k+p)^2p^4}=\f{\ep\ga_\ep\mu^\ep}{k^{3+\ep}}
\eeq
with
\beq
\label{gammaeps}
\ga_\ep=\f{\Ga(\f{1}{2}-\f{\ep}{2})^2\Ga(\f{1}{2}+\f{\ep}{2})}
{(4\pi)^{(3-\ep)/2}\Ga(1-\ep)},
\eeq
as can easily be shown, e.g.\ using Feynman parameters.
In (\ref{kappaN0}) we have used that the second term in the square
brackets does not contribute, since in DR the integral over an arbitrary
power $s$ of $p$ vanishes, 
\beq
\label{drrule}
\int_pp^s=0.
\eeq
Changing the integration variable according to
\beq
\label{kchange}
k\ra\left(\f{N\ga_\ep\mu^\ep u}{6}\right)^{1/(1+\ep)}k
\eeq
gives
\beq
\label{kaN0res}
\ka_N^{(0)}=\f{\ep\ga_\ep\mu^\ep}{3}
\left(\f{N\ga_\ep\mu^\ep u}{6}\right)^{-2\ep/(1+\ep)}
\int_k\f{1}{k^{3+\ep}(1+k^{1+\ep})}.
\eeq
Together with the identity
\beq
\int_0^\infty dk\f{k^a}{1+k^b}=\f{\pi}{b\sin\f{(1+a)\pi}{b}},
\eeq
this leads immediately to the result (\ref{kaN0}).

The diagrams in (\ref{rms0a}) also represent a geometric series, such that
\bea
\lefteqn{\rms^{(0a)}(u,0)
=\f{N(-u)^2}{18}\int_k\f{\De(k)}
{k^2\left[1+\f{Nu}{6}\De(k)\right]}}
\nn\\
&=&
\f{u}{3}\int_k\f{1}
{k^2\left\{1+\left[\f{Nu}{6}\De(k)\right]^{-1}\right\}}
\nn\\
&=&
\f{u}{3}\left(\f{N\ga_\ep\mu^\ep u}{6}\right)^{(1-\ep)/(1+\ep)}
\int_k\f{1}{k^2(1+k^{1+\ep})}.~~~~
\eea
In the last step, we have changed the integration variable according to
(\ref{kchange}).
Due to (\ref{drrule}), we have
\beq
\int_k\f{1}{k^2(1+k^{1+\ep})}=-\int_k\f{1}{k^{3+\ep}(1+k^{1+\ep})}
\eeq
and therefore, comparison with (\ref{kaN0res}) leads to (\ref{rms0a}).

\section{Symbolic Representation of Diagrams}
\label{encods}
In Table~\ref{encodings}, we provide the representation of diagrams
as defined in \cite{NiMeBa}.
While the pictorial representation of Table~\ref{diagrams} is best
for inspection by the human eye, the representation given in
Table~\ref{encodings} is useful for the symbolic manipulation of
diagrams by computer code.
Here we explain the rules for this representation along the lines
of \cite{NiMeBa}.

Label the $n$ vertices of a diagram by integers $0$ through $n-1$
and construct the sequence
\beq
\label{encodeq}
\begin{minipage}{7cm}
/vertices connected to $0$/ vertices connected to $1$ excluding $0$/
vertices connected to $2$ excluding $0,1$/\ldots/vertices connected to $m$
excluding $0,1,\ldots,m-1$/\ldots/vertices connected to $n-1$ excluding
$0,1,\ldots,n-2$/
\end{minipage}
\eeq
External lines are regarded as terminating on a vertex labeled E
and are included in (\ref{encodeq}).
By convention, any line connecting a vertex to itself is listed only
once.
To any allowed sequence (\ref{encodeq}) corresponds only one diagram,
but to make the sequence unique requires additional rules.

Since the diagrams in which we are interested here have only vertices
with four legs, it turns out that the slashes may be omitted from
(\ref{encodeq}) and that the labels E may be replaced by zeros
without compromising the uniqueness of the diagram.
If $n_{\max}$ is the maximal number of vertices in any diagram to be
considered, choose this as the radix of the integers labeling the
vertices in every diagram.
Assuming this has been done, we apply the following rules.
\begin{enumerate}
\item
\label{step1}
When listing from left to right the vertices connected to any
one particular vertex, list E, if present, first and then the remaining
integers in ascending order.
Apply this rule to all vertices in turn; the resulting sequence
is a unique description for the particular vertex labeling assigned
to the diagram under consideration.
\item
\label{step2}
Form these unique sequences for all $n!$ relabelings of the true
vertices of the diagram.
\item
\label{step3}
Interprete each resulting sequence as a number and choose that sequence
from step \ref{step2} that results in the smallest number as the unique
descriptor for the diagram under consideration.
\end{enumerate}
For better readability, the virtual vertex collecting the external lines
may now be labeled by E again and the slashes may be reinstated, so that
the format (\ref{encodeq}) is recovered.
As noted in \cite{NiMeBa}, this is essentially the algorithm given
by Nagle \cite{Na}.
The diagrams in Tables~\ref{diagrams}, \ref{encodings}, and \ref{allresults}
are ordered by the numbers that result in step \ref{step3} above.



\nc{\diagramscaption}{Diagrams through seven loops
A suitable representation for processing by computer code as defined
in \cite{NiMeBa} may be found in Table~\ref{encodings}.
We do not provide a graphical representation of the zero-loop term
$k^2+r$, called \texttt{1-S0} in Table~\ref{encodings}.}
\blt[c]{l}
\caption{\label{diagrams}\diagramscaption}
\\
\btr{c}
\makebox(40,40){}
\etr
\btr{c}
\makebox(40,40){\includegraphics[width=0.45cm,angle=0]{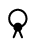}}\\
\texttt{2-M1}
\etr
\btr{c}
\makebox(40,40){\includegraphics[width=0.65cm,angle=0]{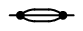}}\\
\texttt{3-S2}
\etr
\btr{c}
\makebox(40,40){\includegraphics[width=0.85cm,angle=0]{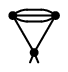}}\\
\texttt{4-M3}
\etr
\btr{c}
\makebox(40,40){\includegraphics[width=0.85cm,angle=0]{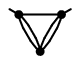}}\\
\texttt{5-S3}
\etr
\btr{c}
\makebox(40,40){\includegraphics[width=1.05cm,angle=0]{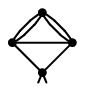}}\\
\texttt{6-M4}
\etr
\btr{c}
\makebox(40,40){\includegraphics[width=1.05cm,angle=0]{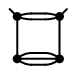}}\\
\texttt{7-S4}
\etr
\btr{c}
\makebox(40,40){\includegraphics[width=1.05cm,angle=0]{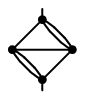}}\\
\texttt{8-S4}
\etr
\btr{c}
\makebox(40,40){\includegraphics[width=1.05cm,angle=0]{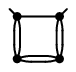}}\\
\texttt{9-S4}
\etr
\btr{c}
\makebox(40,40){\includegraphics[width=1.05cm,angle=0]{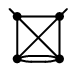}}\\
\texttt{10-S4}
\etr
\\
\btr{c}
\makebox(40,40){\includegraphics[width=1.25cm,angle=0]{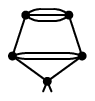}}\\
\texttt{11-M5}
\etr
\btr{c}
\makebox(40,40){\includegraphics[width=1.25cm,angle=0]{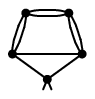}}\\
\texttt{12-M5}
\etr
\btr{c}
\makebox(40,40){\includegraphics[width=1.25cm,angle=0]{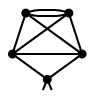}}\\
\texttt{13-M5}
\etr
\btr{c}
\makebox(40,40){\includegraphics[width=1.25cm,angle=0]{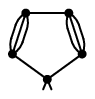}}\\
\texttt{14-M5}
\etr
\btr{c}
\makebox(40,40){\includegraphics[width=1.25cm,angle=0]{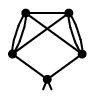}}\\
\texttt{15-M5}
\etr
\btr{c}
\makebox(40,40){\includegraphics[width=1.25cm,angle=0]{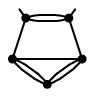}}\\
\texttt{16-S5}
\etr
\btr{c}
\makebox(40,40){\includegraphics[width=1.25cm,angle=0]{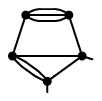}}\\
\texttt{17-S5}
\etr
\btr{c}
\makebox(40,40){\includegraphics[width=1.25cm,angle=0]{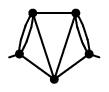}}\\
\texttt{18-S5}
\etr
\btr{c}
\makebox(40,40){\includegraphics[width=1.25cm,angle=0]{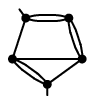}}\\
\texttt{19-S5}
\etr
\btr{c}
\makebox(40,40){\includegraphics[width=1.25cm,angle=0]{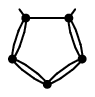}}\\
\texttt{20-S5}
\etr
\\
\btr{c}
\makebox(40,40){\includegraphics[width=1.25cm,angle=0]{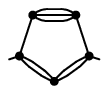}}\\
\texttt{21-S5}
\etr
\btr{c}
\makebox(40,40){\includegraphics[width=1.25cm,angle=0]{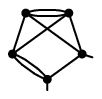}}\\
\texttt{22-S5}
\etr
\btr{c}
\makebox(40,40){\includegraphics[width=1.25cm,angle=0]{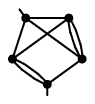}}\\
\texttt{23-S5}
\etr
\btr{c}
\makebox(40,40){\includegraphics[width=1.25cm,angle=0]{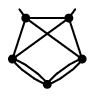}}\\
\texttt{24-S5}
\etr
\btr{c}
\makebox(40,40){\includegraphics[width=1.25cm,angle=0]{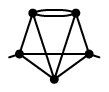}}\\
\texttt{25-S5}
\etr
\btr{c}
\makebox(40,40){\includegraphics[width=1.25cm,angle=0]{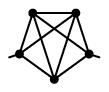}}\\
\texttt{26-S5}
\etr
\btr{c}
\makebox(40,40){\includegraphics[width=1.45cm,angle=0]{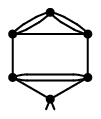}}\\
\texttt{27-M6}
\etr
\btr{c}
\makebox(40,40){\includegraphics[width=1.45cm,angle=0]{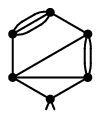}}\\
\texttt{28-M6}
\etr
\btr{c}
\makebox(40,40){\includegraphics[width=1.45cm,angle=0]{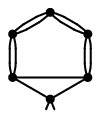}}\\
\texttt{29-M6}
\etr
\btr{c}
\makebox(40,40){\includegraphics[width=1.45cm,angle=0]{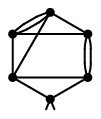}}\\
\texttt{30-M6}
\etr
\\
\btr{c}
\makebox(40,40){\includegraphics[width=1.45cm,angle=0]{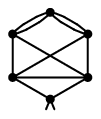}}\\
\texttt{31-M6}
\etr
\btr{c}
\makebox(40,40){\includegraphics[width=1.45cm,angle=0]{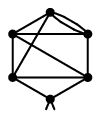}}\\
\texttt{32-M6}
\etr
\btr{c}
\makebox(40,40){\includegraphics[width=1.45cm,angle=0]{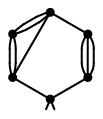}}\\
\texttt{33-M6}
\etr
\btr{c}
\makebox(40,40){\includegraphics[width=1.45cm,angle=0]{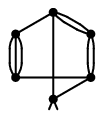}}\\
\texttt{34-M6}
\etr
\btr{c}
\makebox(40,40){\includegraphics[width=1.45cm,angle=0]{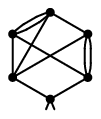}}\\
\texttt{35-M6}
\etr
\btr{c}
\makebox(40,40){\includegraphics[width=1.45cm,angle=0]{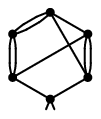}}\\
\texttt{36-M6}
\etr
\btr{c}
\makebox(40,40){\includegraphics[width=1.45cm,angle=0]{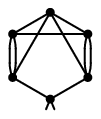}}\\
\texttt{37-M6}
\etr
\btr{c}
\makebox(40,40){\includegraphics[width=1.45cm,angle=0]{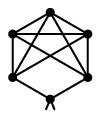}}\\
\texttt{38-M6}
\etr
\btr{c}
\makebox(40,40){\includegraphics[width=1.45cm,angle=0]{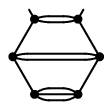}}\\
\texttt{39-S6}
\etr
\btr{c}
\makebox(40,40){\includegraphics[width=1.45cm,angle=0]{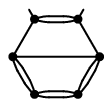}}\\
\texttt{40-S6}
\etr
\\
\btr{c}
\makebox(40,40){\includegraphics[width=1.45cm,angle=0]{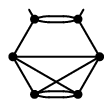}}\\
\texttt{41-S6}
\etr
\btr{c}
\makebox(40,40){\includegraphics[width=1.45cm,angle=0]{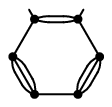}}\\
\texttt{42-S6}
\etr
\btr{c}
\makebox(40,40){\includegraphics[width=1.45cm,angle=0]{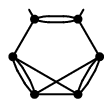}}\\
\texttt{43-S6}
\etr
\btr{c}
\makebox(40,40){\includegraphics[width=1.45cm,angle=0]{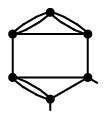}}\\
\texttt{44-S6}
\etr
\btr{c}
\makebox(40,40){\includegraphics[width=1.45cm,angle=0]{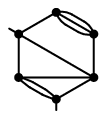}}\\
\texttt{45-S6}
\etr
\btr{c}
\makebox(40,40){\includegraphics[width=1.45cm,angle=0]{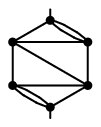}}\\
\texttt{46-S6}
\etr
\btr{c}
\makebox(40,40){\includegraphics[width=1.45cm,angle=0]{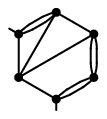}}\\
\texttt{47-S6}
\etr
\btr{c}
\makebox(40,40){\includegraphics[width=1.45cm,angle=0]{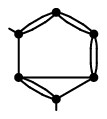}}\\
\texttt{48-S6}
\etr
\btr{c}
\makebox(40,40){\includegraphics[width=1.45cm,angle=0]{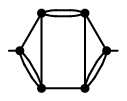}}\\
\texttt{49-S6}
\etr
\btr{c}
\makebox(40,40){\includegraphics[width=1.45cm,angle=0]{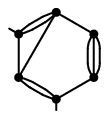}}\\
\texttt{50-S6}
\etr
\\
\btr{c}
\makebox(40,40){\includegraphics[width=1.45cm,angle=0]{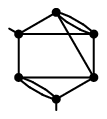}}\\
\texttt{51-S6}
\etr
\btr{c}
\makebox(40,40){\includegraphics[width=1.45cm,angle=0]{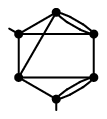}}\\
\texttt{52-S6}
\etr
\btr{c}
\makebox(40,40){\includegraphics[width=1.45cm,angle=0]{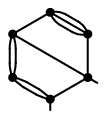}}\\
\texttt{53-S6}
\etr
\btr{c}
\makebox(40,40){\includegraphics[width=1.45cm,angle=0]{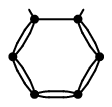}}\\
\texttt{54-S6}
\etr
\btr{c}
\makebox(40,40){\includegraphics[width=1.45cm,angle=0]{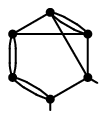}}\\
\texttt{55-S6}
\etr
\btr{c}
\makebox(40,40){\includegraphics[width=1.45cm,angle=0]{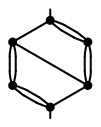}}\\
\texttt{56-S6}
\etr
\btr{c}
\makebox(40,40){\includegraphics[width=1.45cm,angle=0]{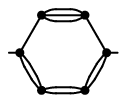}}\\
\texttt{57-S6}
\etr
\btr{c}
\makebox(40,40){\includegraphics[width=1.45cm,angle=0]{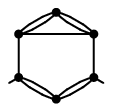}}\\
\texttt{58-S6}
\etr
\btr{c}
\makebox(40,40){\includegraphics[width=1.45cm,angle=0]{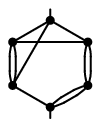}}\\
\texttt{59-S6}
\etr
\btr{c}
\makebox(40,40){\includegraphics[width=1.45cm,angle=0]{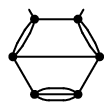}}\\
\texttt{60-S6}
\etr
\\
\btr{c}
\makebox(40,40){\includegraphics[width=1.45cm,angle=0]{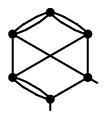}}\\
\texttt{61-S6}
\etr
\btr{c}
\makebox(40,40){\includegraphics[width=1.45cm,angle=0]{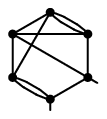}}\\
\texttt{62-S6}
\etr
\btr{c}
\makebox(40,40){\includegraphics[width=1.45cm,angle=0]{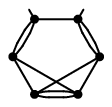}}\\
\texttt{63-S6}
\etr
\btr{c}
\makebox(40,40){\includegraphics[width=1.45cm,angle=0]{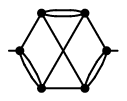}}\\
\texttt{64-S6}
\etr
\btr{c}
\makebox(40,40){\includegraphics[width=1.45cm,angle=0]{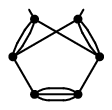}}\\
\texttt{65-S6}
\etr
\btr{c}
\makebox(40,40){\includegraphics[width=1.45cm,angle=0]{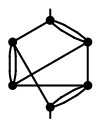}}\\
\texttt{66-S6}
\etr
\btr{c}
\makebox(40,40){\includegraphics[width=1.45cm,angle=0]{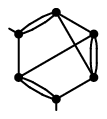}}\\
\texttt{67-S6}
\etr
\btr{c}
\makebox(40,40){\includegraphics[width=1.45cm,angle=0]{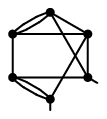}}\\
\texttt{68-S6}
\etr
\btr{c}
\makebox(40,40){\includegraphics[width=1.45cm,angle=0]{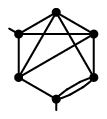}}\\
\texttt{69-S6}
\etr
\btr{c}
\makebox(40,40){\includegraphics[width=1.45cm,angle=0]{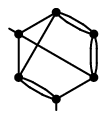}}\\
\texttt{70-S6}
\etr
\\
\btr{c}
\makebox(40,40){\includegraphics[width=1.45cm,angle=0]{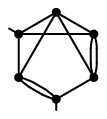}}\\
\texttt{71-S6}
\etr
\btr{c}
\makebox(40,40){\includegraphics[width=1.45cm,angle=0]{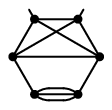}}\\
\texttt{72-S6}
\etr
\btr{c}
\makebox(40,40){\includegraphics[width=1.45cm,angle=0]{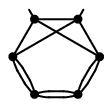}}\\
\texttt{73-S6}
\etr
\btr{c}
\makebox(40,40){\includegraphics[width=1.45cm,angle=0]{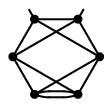}}\\
\texttt{74-S6}
\etr
\btr{c}
\makebox(40,40){\includegraphics[width=1.45cm,angle=0]{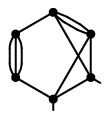}}\\
\texttt{75-S6}
\etr
\btr{c}
\makebox(40,40){\includegraphics[width=1.45cm,angle=0]{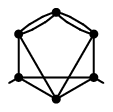}}\\
\texttt{76-S6}
\etr
\btr{c}
\makebox(40,40){\includegraphics[width=1.45cm,angle=0]{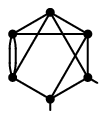}}\\
\texttt{77-S6}
\etr
\btr{c}
\makebox(40,40){\includegraphics[width=1.45cm,angle=0]{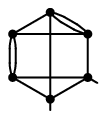}}\\
\texttt{78-S6}
\etr
\btr{c}
\makebox(40,40){\includegraphics[width=1.45cm,angle=0]{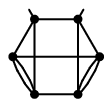}}\\
\texttt{79-S6}
\etr
\btr{c}
\makebox(40,40){\includegraphics[width=1.45cm,angle=0]{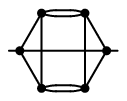}}\\
\texttt{80-S6}
\etr
\\
\btr{c}
\makebox(40,40){\includegraphics[width=1.45cm,angle=0]{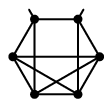}}\\
\texttt{81-S6}
\etr
\btr{c}
\makebox(40,40){\includegraphics[width=1.45cm,angle=0]{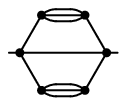}}\\
\texttt{82-S6}
\etr
\btr{c}
\makebox(40,40){\includegraphics[width=1.45cm,angle=0]{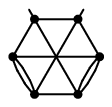}}\\
\texttt{83-S6}
\etr
\btr{c}
\makebox(40,40){\includegraphics[width=1.45cm,angle=0]{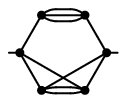}}\\
\texttt{84-S6}
\etr
\btr{c}
\makebox(40,40){\includegraphics[width=1.45cm,angle=0]{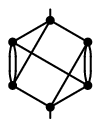}}\\
\texttt{85-S6}
\etr
\btr{c}
\makebox(40,40){\includegraphics[width=1.45cm,angle=0]{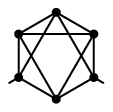}}\\
\texttt{86-S6}
\etr
\btr{c}
\makebox(40,40){\includegraphics[width=1.45cm,angle=0]{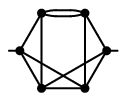}}\\
\texttt{87-S6}
\etr
\btr{c}
\makebox(40,40){\includegraphics[width=1.45cm,angle=0]{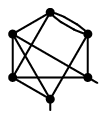}}\\
\texttt{88-S6}
\etr
\btr{c}
\makebox(40,40){\includegraphics[width=1.65cm,angle=0]{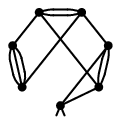}}\\
\texttt{89-M7}
\etr
\btr{c}
\makebox(40,40){\includegraphics[width=1.65cm,angle=0]{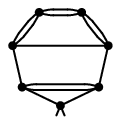}}\\
\texttt{90-M7}
\etr
\\
\btr{c}
\makebox(40,40){\includegraphics[width=1.65cm,angle=0]{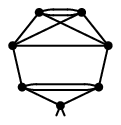}}\\
\texttt{91-M7}
\etr
\btr{c}
\makebox(40,40){\includegraphics[width=1.65cm,angle=0]{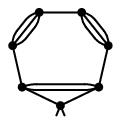}}\\
\texttt{92-M7}
\etr
\btr{c}
\makebox(40,40){\includegraphics[width=1.65cm,angle=0]{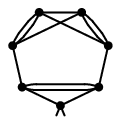}}\\
\texttt{93-M7}
\etr
\btr{c}
\makebox(40,40){\includegraphics[width=1.65cm,angle=0]{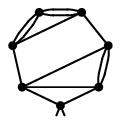}}\\
\texttt{94-M7}
\etr
\btr{c}
\makebox(40,40){\includegraphics[width=1.65cm,angle=0]{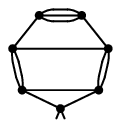}}\\
\texttt{95-M7}
\etr
\btr{c}
\makebox(40,40){\includegraphics[width=1.65cm,angle=0]{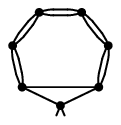}}\\
\texttt{96-M7}
\etr
\btr{c}
\makebox(40,40){\includegraphics[width=1.65cm,angle=0]{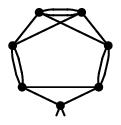}}\\
\texttt{97-M7}
\etr
\btr{c}
\makebox(40,40){\includegraphics[width=1.65cm,angle=0]{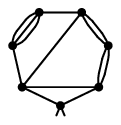}}\\
\texttt{98-M7}
\etr
\btr{c}
\makebox(40,40){\includegraphics[width=1.65cm,angle=0]{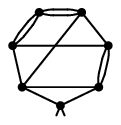}}\\
\texttt{99-M7}
\etr
\btr{c}
\makebox(40,40){\includegraphics[width=1.65cm,angle=0]{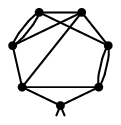}}\\
\texttt{100-M7}
\etr
\\
\btr{c}
\makebox(40,40){\includegraphics[width=1.65cm,angle=0]{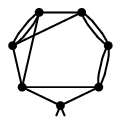}}\\
\texttt{101-M7}
\etr
\btr{c}
\makebox(40,40){\includegraphics[width=1.65cm,angle=0]{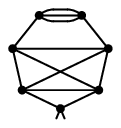}}\\
\texttt{102-M7}
\etr
\btr{c}
\makebox(40,40){\includegraphics[width=1.65cm,angle=0]{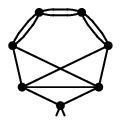}}\\
\texttt{103-M7}
\etr
\btr{c}
\makebox(40,40){\includegraphics[width=1.65cm,angle=0]{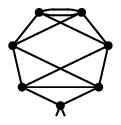}}\\
\texttt{104-M7}
\etr
\btr{c}
\makebox(40,40){\includegraphics[width=1.65cm,angle=0]{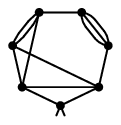}}\\
\texttt{105-M7}
\etr
\btr{c}
\makebox(40,40){\includegraphics[width=1.65cm,angle=0]{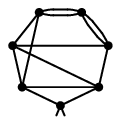}}\\
\texttt{106-M7}
\etr
\btr{c}
\makebox(40,40){\includegraphics[width=1.65cm,angle=0]{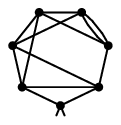}}\\
\texttt{107-M7}
\etr
\btr{c}
\makebox(40,40){\includegraphics[width=1.65cm,angle=0]{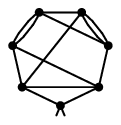}}\\
\texttt{108-M7}
\etr
\btr{c}
\makebox(40,40){\includegraphics[width=1.65cm,angle=0]{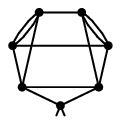}}\\
\texttt{109-M7}
\etr
\btr{c}
\makebox(40,40){\includegraphics[width=1.65cm,angle=0]{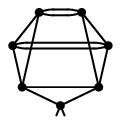}}\\
\texttt{110-M7}
\etr
\\
\btr{c}
\makebox(40,40){\includegraphics[width=1.65cm,angle=0]{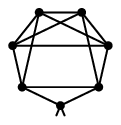}}\\
\texttt{111-M7}
\etr
\btr{c}
\makebox(40,40){\includegraphics[width=1.65cm,angle=0]{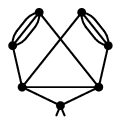}}\\
\texttt{112-M7}
\etr
\btr{c}
\makebox(40,40){\includegraphics[width=1.65cm,angle=0]{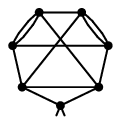}}\\
\texttt{113-M7}
\etr
\btr{c}
\makebox(40,40){\includegraphics[width=1.65cm,angle=0]{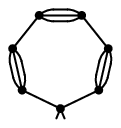}}\\
\texttt{114-M7}
\etr
\btr{c}
\makebox(40,40){\includegraphics[width=1.65cm,angle=0]{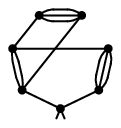}}\\
\texttt{115-M7}
\etr
\btr{c}
\makebox(40,40){\includegraphics[width=1.65cm,angle=0]{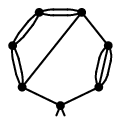}}\\
\texttt{116-M7}
\etr
\btr{c}
\makebox(40,40){\includegraphics[width=1.65cm,angle=0]{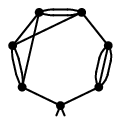}}\\
\texttt{117-M7}
\etr
\btr{c}
\makebox(40,40){\includegraphics[width=1.65cm,angle=0]{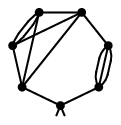}}\\
\texttt{118-M7}
\etr
\btr{c}
\makebox(40,40){\includegraphics[width=1.65cm,angle=0]{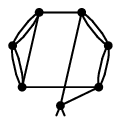}}\\
\texttt{119-M7}
\etr
\btr{c}
\makebox(40,40){\includegraphics[width=1.65cm,angle=0]{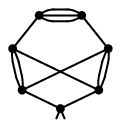}}\\
\texttt{120-M7}
\etr
\\
\btr{c}
\makebox(40,40){\includegraphics[width=1.65cm,angle=0]{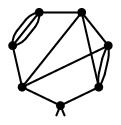}}\\
\texttt{121-M7}
\etr
\btr{c}
\makebox(40,40){\includegraphics[width=1.65cm,angle=0]{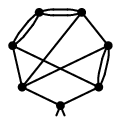}}\\
\texttt{122-M7}
\etr
\btr{c}
\makebox(40,40){\includegraphics[width=1.65cm,angle=0]{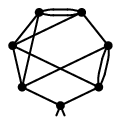}}\\
\texttt{123-M7}
\etr
\btr{c}
\makebox(40,40){\includegraphics[width=1.65cm,angle=0]{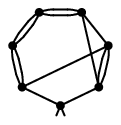}}\\
\texttt{124-M7}
\etr
\btr{c}
\makebox(40,40){\includegraphics[width=1.65cm,angle=0]{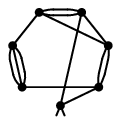}}\\
\texttt{125-M7}
\etr
\btr{c}
\makebox(40,40){\includegraphics[width=1.65cm,angle=0]{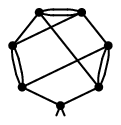}}\\
\texttt{126-M7}
\etr
\btr{c}
\makebox(40,40){\includegraphics[width=1.65cm,angle=0]{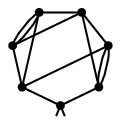}}\\
\texttt{127-M7}
\etr
\btr{c}
\makebox(40,40){\includegraphics[width=1.65cm,angle=0]{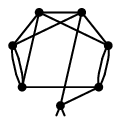}}\\
\texttt{128-M7}
\etr
\btr{c}
\makebox(40,40){\includegraphics[width=1.65cm,angle=0]{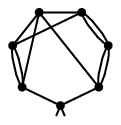}}\\
\texttt{129-M7}
\etr
\btr{c}
\makebox(40,40){\includegraphics[width=1.65cm,angle=0]{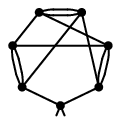}}\\
\texttt{130-M7}
\etr
\\
\btr{c}
\makebox(40,40){\includegraphics[width=1.65cm,angle=0]{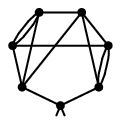}}\\
\texttt{131-M7}
\etr
\btr{c}
\makebox(40,40){\includegraphics[width=1.65cm,angle=0]{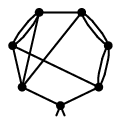}}\\
\texttt{132-M7}
\etr
\btr{c}
\makebox(40,40){\includegraphics[width=1.65cm,angle=0]{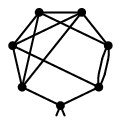}}\\
\texttt{133-M7}
\etr
\btr{c}
\makebox(40,40){\includegraphics[width=1.65cm,angle=0]{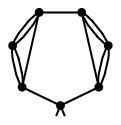}}\\
\texttt{134-M7}
\etr
\btr{c}
\makebox(40,40){\includegraphics[width=1.65cm,angle=0]{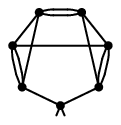}}\\
\texttt{135-M7}
\etr
\btr{c}
\makebox(40,40){\includegraphics[width=1.65cm,angle=0]{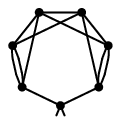}}\\
\texttt{136-M7}
\etr
\btr{c}
\makebox(40,40){\includegraphics[width=1.65cm,angle=0]{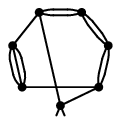}}\\
\texttt{137-M7}
\etr
\btr{c}
\makebox(40,40){\includegraphics[width=1.65cm,angle=0]{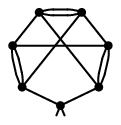}}\\
\texttt{138-M7}
\etr
\btr{c}
\makebox(40,40){\includegraphics[width=1.65cm,angle=0]{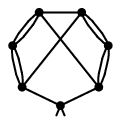}}\\
\texttt{139-M7}
\etr
\btr{c}
\makebox(40,40){\includegraphics[width=1.65cm,angle=0]{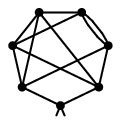}}\\
\texttt{140-M7}
\etr
\\
\btr{c}
\makebox(40,40){\includegraphics[width=1.65cm,angle=0]{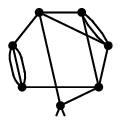}}\\
\texttt{141-M7}
\etr
\btr{c}
\makebox(40,40){\includegraphics[width=1.65cm,angle=0]{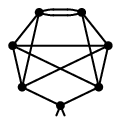}}\\
\texttt{142-M7}
\etr
\btr{c}
\makebox(40,40){\includegraphics[width=1.65cm,angle=0]{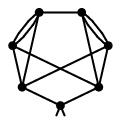}}\\
\texttt{143-M7}
\etr
\btr{c}
\makebox(40,40){\includegraphics[width=1.65cm,angle=0]{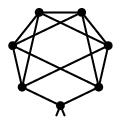}}\\
\texttt{144-M7}
\etr
\btr{c}
\makebox(40,40){\includegraphics[width=1.65cm,angle=0]{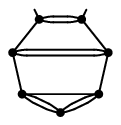}}\\
\texttt{145-S7}
\etr
\btr{c}
\makebox(40,40){\includegraphics[width=1.65cm,angle=0]{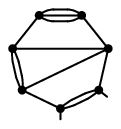}}\\
\texttt{146-S7}
\etr
\btr{c}
\makebox(40,40){\includegraphics[width=1.65cm,angle=0]{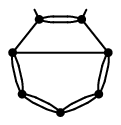}}\\
\texttt{147-S7}
\etr
\btr{c}
\makebox(40,40){\includegraphics[width=1.65cm,angle=0]{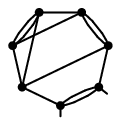}}\\
\texttt{148-S7}
\etr
\btr{c}
\makebox(40,40){\includegraphics[width=1.65cm,angle=0]{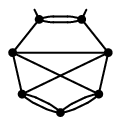}}\\
\texttt{149-S7}
\etr
\btr{c}
\makebox(40,40){\includegraphics[width=1.65cm,angle=0]{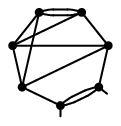}}\\
\texttt{150-S7}
\etr
\\
\btr{c}
\makebox(40,40){\includegraphics[width=1.65cm,angle=0]{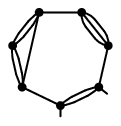}}\\
\texttt{151-S7}
\etr
\btr{c}
\makebox(40,40){\includegraphics[width=1.65cm,angle=0]{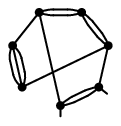}}\\
\texttt{152-S7}
\etr
\btr{c}
\makebox(40,40){\includegraphics[width=1.65cm,angle=0]{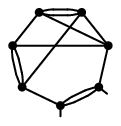}}\\
\texttt{153-S7}
\etr
\btr{c}
\makebox(40,40){\includegraphics[width=1.65cm,angle=0]{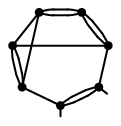}}\\
\texttt{154-S7}
\etr
\btr{c}
\makebox(40,40){\includegraphics[width=1.65cm,angle=0]{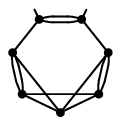}}\\
\texttt{155-S7}
\etr
\btr{c}
\makebox(40,40){\includegraphics[width=1.65cm,angle=0]{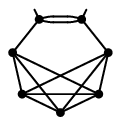}}\\
\texttt{156-S7}
\etr
\btr{c}
\makebox(40,40){\includegraphics[width=1.65cm,angle=0]{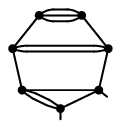}}\\
\texttt{157-S7}
\etr
\btr{c}
\makebox(40,40){\includegraphics[width=1.65cm,angle=0]{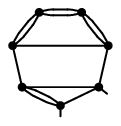}}\\
\texttt{158-S7}
\etr
\btr{c}
\makebox(40,40){\includegraphics[width=1.65cm,angle=0]{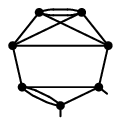}}\\
\texttt{159-S7}
\etr
\btr{c}
\makebox(40,40){\includegraphics[width=1.65cm,angle=0]{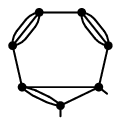}}\\
\texttt{160-S7}
\etr
\\
\btr{c}
\makebox(40,40){\includegraphics[width=1.65cm,angle=0]{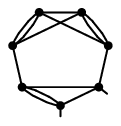}}\\
\texttt{161-S7}
\etr
\btr{c}
\makebox(40,40){\includegraphics[width=1.65cm,angle=0]{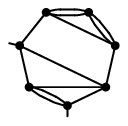}}\\
\texttt{162-S7}
\etr
\btr{c}
\makebox(40,40){\includegraphics[width=1.65cm,angle=0]{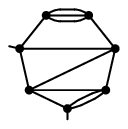}}\\
\texttt{163-S7}
\etr
\btr{c}
\makebox(40,40){\includegraphics[width=1.65cm,angle=0]{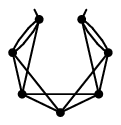}}\\
\texttt{164-S7}
\etr
\btr{c}
\makebox(40,40){\includegraphics[width=1.65cm,angle=0]{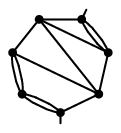}}\\
\texttt{165-S7}
\etr
\btr{c}
\makebox(40,40){\includegraphics[width=1.65cm,angle=0]{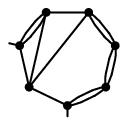}}\\
\texttt{166-S7}
\etr
\btr{c}
\makebox(40,40){\includegraphics[width=1.65cm,angle=0]{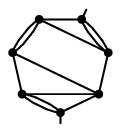}}\\
\texttt{167-S7}
\etr
\btr{c}
\makebox(40,40){\includegraphics[width=1.65cm,angle=0]{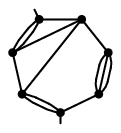}}\\
\texttt{168-S7}
\etr
\btr{c}
\makebox(40,40){\includegraphics[width=1.65cm,angle=0]{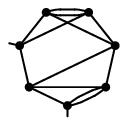}}\\
\texttt{169-S7}
\etr
\btr{c}
\makebox(40,40){\includegraphics[width=1.65cm,angle=0]{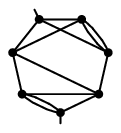}}\\
\texttt{170-S7}
\etr
\\
\btr{c}
\makebox(40,40){\includegraphics[width=1.65cm,angle=0]{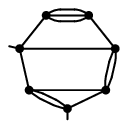}}\\
\texttt{171-S7}
\etr
\btr{c}
\makebox(40,40){\includegraphics[width=1.65cm,angle=0]{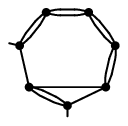}}\\
\texttt{172-S7}
\etr
\btr{c}
\makebox(40,40){\includegraphics[width=1.65cm,angle=0]{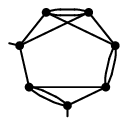}}\\
\texttt{173-S7}
\etr
\btr{c}
\makebox(40,40){\includegraphics[width=1.65cm,angle=0]{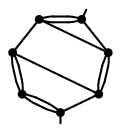}}\\
\texttt{174-S7}
\etr
\btr{c}
\makebox(40,40){\includegraphics[width=1.65cm,angle=0]{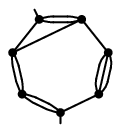}}\\
\texttt{175-S7}
\etr
\btr{c}
\makebox(40,40){\includegraphics[width=1.65cm,angle=0]{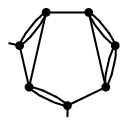}}\\
\texttt{176-S7}
\etr
\btr{c}
\makebox(40,40){\includegraphics[width=1.65cm,angle=0]{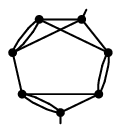}}\\
\texttt{177-S7}
\etr
\btr{c}
\makebox(40,40){\includegraphics[width=1.65cm,angle=0]{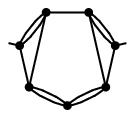}}\\
\texttt{178-S7}
\etr
\btr{c}
\makebox(40,40){\includegraphics[width=1.65cm,angle=0]{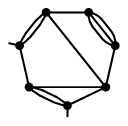}}\\
\texttt{179-S7}
\etr
\btr{c}
\makebox(40,40){\includegraphics[width=1.65cm,angle=0]{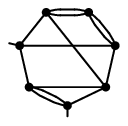}}\\
\texttt{180-S7}
\etr
\\
\btr{c}
\makebox(40,40){\includegraphics[width=1.65cm,angle=0]{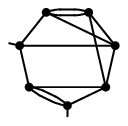}}\\
\texttt{181-S7}
\etr
\btr{c}
\makebox(40,40){\includegraphics[width=1.65cm,angle=0]{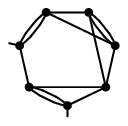}}\\
\texttt{182-S7}
\etr
\btr{c}
\makebox(40,40){\includegraphics[width=1.65cm,angle=0]{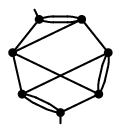}}\\
\texttt{183-S7}
\etr
\btr{c}
\makebox(40,40){\includegraphics[width=1.65cm,angle=0]{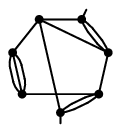}}\\
\texttt{184-S7}
\etr
\btr{c}
\makebox(40,40){\includegraphics[width=1.65cm,angle=0]{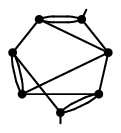}}\\
\texttt{185-S7}
\etr
\btr{c}
\makebox(40,40){\includegraphics[width=1.65cm,angle=0]{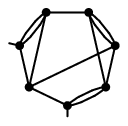}}\\
\texttt{186-S7}
\etr
\btr{c}
\makebox(40,40){\includegraphics[width=1.65cm,angle=0]{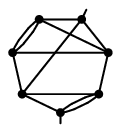}}\\
\texttt{187-S7}
\etr
\btr{c}
\makebox(40,40){\includegraphics[width=1.65cm,angle=0]{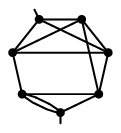}}\\
\texttt{188-S7}
\etr
\btr{c}
\makebox(40,40){\includegraphics[width=1.65cm,angle=0]{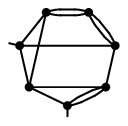}}\\
\texttt{189-S7}
\etr
\btr{c}
\makebox(40,40){\includegraphics[width=1.65cm,angle=0]{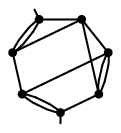}}\\
\texttt{190-S7}
\etr
\\
\btr{c}
\makebox(40,40){\includegraphics[width=1.65cm,angle=0]{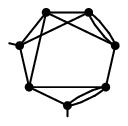}}\\
\texttt{191-S7}
\etr
\btr{c}
\makebox(40,40){\includegraphics[width=1.65cm,angle=0]{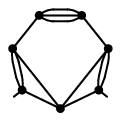}}\\
\texttt{192-S7}
\etr
\btr{c}
\makebox(40,40){\includegraphics[width=1.65cm,angle=0]{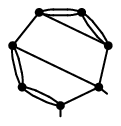}}\\
\texttt{193-S7}
\etr
\btr{c}
\makebox(40,40){\includegraphics[width=1.65cm,angle=0]{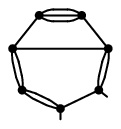}}\\
\texttt{194-S7}
\etr
\btr{c}
\makebox(40,40){\includegraphics[width=1.65cm,angle=0]{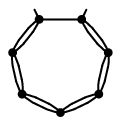}}\\
\texttt{195-S7}
\etr
\btr{c}
\makebox(40,40){\includegraphics[width=1.65cm,angle=0]{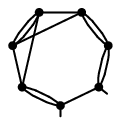}}\\
\texttt{196-S7}
\etr
\btr{c}
\makebox(40,40){\includegraphics[width=1.65cm,angle=0]{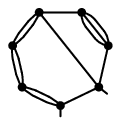}}\\
\texttt{197-S7}
\etr
\btr{c}
\makebox(40,40){\includegraphics[width=1.65cm,angle=0]{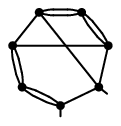}}\\
\texttt{198-S7}
\etr
\btr{c}
\makebox(40,40){\includegraphics[width=1.65cm,angle=0]{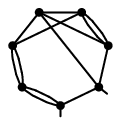}}\\
\texttt{199-S7}
\etr
\btr{c}
\makebox(40,40){\includegraphics[width=1.65cm,angle=0]{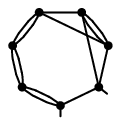}}\\
\texttt{200-S7}
\etr
\\
\btr{c}
\makebox(40,40){\includegraphics[width=1.65cm,angle=0]{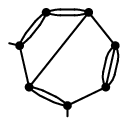}}\\
\texttt{201-S7}
\etr
\btr{c}
\makebox(40,40){\includegraphics[width=1.65cm,angle=0]{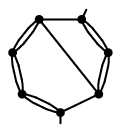}}\\
\texttt{202-S7}
\etr
\btr{c}
\makebox(40,40){\includegraphics[width=1.65cm,angle=0]{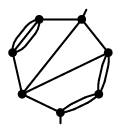}}\\
\texttt{203-S7}
\etr
\btr{c}
\makebox(40,40){\includegraphics[width=1.65cm,angle=0]{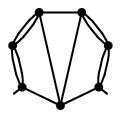}}\\
\texttt{204-S7}
\etr
\btr{c}
\makebox(40,40){\includegraphics[width=1.65cm,angle=0]{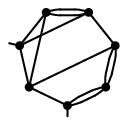}}\\
\texttt{205-S7}
\etr
\btr{c}
\makebox(40,40){\includegraphics[width=1.65cm,angle=0]{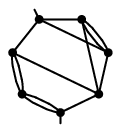}}\\
\texttt{206-S7}
\etr
\btr{c}
\makebox(40,40){\includegraphics[width=1.65cm,angle=0]{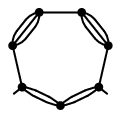}}\\
\texttt{207-S7}
\etr
\btr{c}
\makebox(40,40){\includegraphics[width=1.65cm,angle=0]{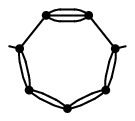}}\\
\texttt{208-S7}
\etr
\btr{c}
\makebox(40,40){\includegraphics[width=1.65cm,angle=0]{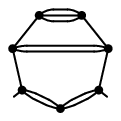}}\\
\texttt{209-S7}
\etr
\btr{c}
\makebox(40,40){\includegraphics[width=1.65cm,angle=0]{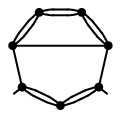}}\\
\texttt{210-S7}
\etr
\\
\btr{c}
\makebox(40,40){\includegraphics[width=1.65cm,angle=0]{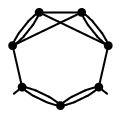}}\\
\texttt{211-S7}
\etr
\btr{c}
\makebox(40,40){\includegraphics[width=1.65cm,angle=0]{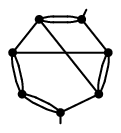}}\\
\texttt{212-S7}
\etr
\btr{c}
\makebox(40,40){\includegraphics[width=1.65cm,angle=0]{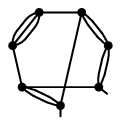}}\\
\texttt{213-S7}
\etr
\btr{c}
\makebox(40,40){\includegraphics[width=1.65cm,angle=0]{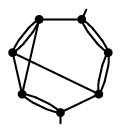}}\\
\texttt{214-S7}
\etr
\btr{c}
\makebox(40,40){\includegraphics[width=1.65cm,angle=0]{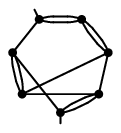}}\\
\texttt{215-S7}
\etr
\btr{c}
\makebox(40,40){\includegraphics[width=1.65cm,angle=0]{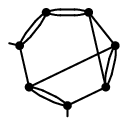}}\\
\texttt{216-S7}
\etr
\btr{c}
\makebox(40,40){\includegraphics[width=1.65cm,angle=0]{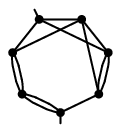}}\\
\texttt{217-S7}
\etr
\btr{c}
\makebox(40,40){\includegraphics[width=1.65cm,angle=0]{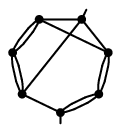}}\\
\texttt{218-S7}
\etr
\btr{c}
\makebox(40,40){\includegraphics[width=1.65cm,angle=0]{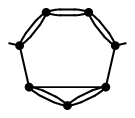}}\\
\texttt{219-S7}
\etr
\btr{c}
\makebox(40,40){\includegraphics[width=1.65cm,angle=0]{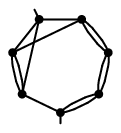}}\\
\texttt{220-S7}
\etr
\\
\btr{c}
\makebox(40,40){\includegraphics[width=1.65cm,angle=0]{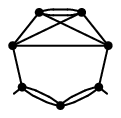}}\\
\texttt{221-S7}
\etr
\btr{c}
\makebox(40,40){\includegraphics[width=1.65cm,angle=0]{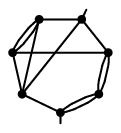}}\\
\texttt{222-S7}
\etr
\btr{c}
\makebox(40,40){\includegraphics[width=1.65cm,angle=0]{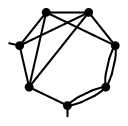}}\\
\texttt{223-S7}
\etr
\btr{c}
\makebox(40,40){\includegraphics[width=1.65cm,angle=0]{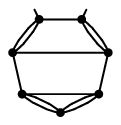}}\\
\texttt{224-S7}
\etr
\btr{c}
\makebox(40,40){\includegraphics[width=1.65cm,angle=0]{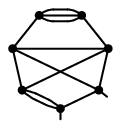}}\\
\texttt{225-S7}
\etr
\btr{c}
\makebox(40,40){\includegraphics[width=1.65cm,angle=0]{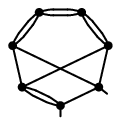}}\\
\texttt{226-S7}
\etr
\btr{c}
\makebox(40,40){\includegraphics[width=1.65cm,angle=0]{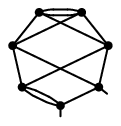}}\\
\texttt{227-S7}
\etr
\btr{c}
\makebox(40,40){\includegraphics[width=1.65cm,angle=0]{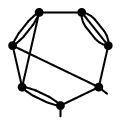}}\\
\texttt{228-S7}
\etr
\btr{c}
\makebox(40,40){\includegraphics[width=1.65cm,angle=0]{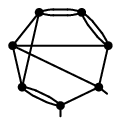}}\\
\texttt{229-S7}
\etr
\btr{c}
\makebox(40,40){\includegraphics[width=1.65cm,angle=0]{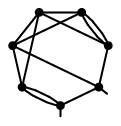}}\\
\texttt{230-S7}
\etr
\\
\btr{c}
\makebox(40,40){\includegraphics[width=1.65cm,angle=0]{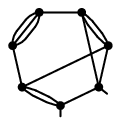}}\\
\texttt{231-S7}
\etr
\btr{c}
\makebox(40,40){\includegraphics[width=1.65cm,angle=0]{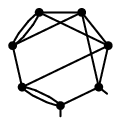}}\\
\texttt{232-S7}
\etr
\btr{c}
\makebox(40,40){\includegraphics[width=1.65cm,angle=0]{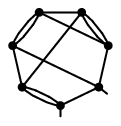}}\\
\texttt{233-S7}
\etr
\btr{c}
\makebox(40,40){\includegraphics[width=1.65cm,angle=0]{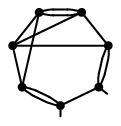}}\\
\texttt{234-S7}
\etr
\btr{c}
\makebox(40,40){\includegraphics[width=1.65cm,angle=0]{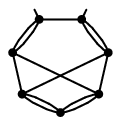}}\\
\texttt{235-S7}
\etr
\btr{c}
\makebox(40,40){\includegraphics[width=1.65cm,angle=0]{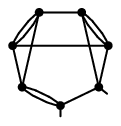}}\\
\texttt{236-S7}
\etr
\btr{c}
\makebox(40,40){\includegraphics[width=1.65cm,angle=0]{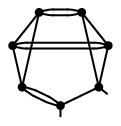}}\\
\texttt{237-S7}
\etr
\btr{c}
\makebox(40,40){\includegraphics[width=1.65cm,angle=0]{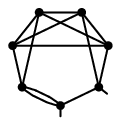}}\\
\texttt{238-S7}
\etr
\btr{c}
\makebox(40,40){\includegraphics[width=1.65cm,angle=0]{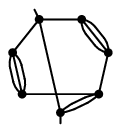}}\\
\texttt{239-S7}
\etr
\btr{c}
\makebox(40,40){\includegraphics[width=1.65cm,angle=0]{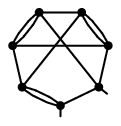}}\\
\texttt{240-S7}
\etr
\\
\btr{c}
\makebox(40,40){\includegraphics[width=1.65cm,angle=0]{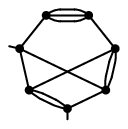}}\\
\texttt{241-S7}
\etr
\btr{c}
\makebox(40,40){\includegraphics[width=1.65cm,angle=0]{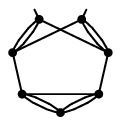}}\\
\texttt{242-S7}
\etr
\btr{c}
\makebox(40,40){\includegraphics[width=1.65cm,angle=0]{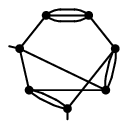}}\\
\texttt{243-S7}
\etr
\btr{c}
\makebox(40,40){\includegraphics[width=1.65cm,angle=0]{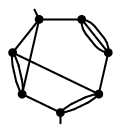}}\\
\texttt{244-S7}
\etr
\btr{c}
\makebox(40,40){\includegraphics[width=1.65cm,angle=0]{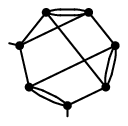}}\\
\texttt{245-S7}
\etr
\btr{c}
\makebox(40,40){\includegraphics[width=1.65cm,angle=0]{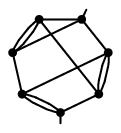}}\\
\texttt{246-S7}
\etr
\btr{c}
\makebox(40,40){\includegraphics[width=1.65cm,angle=0]{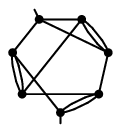}}\\
\texttt{247-S7}
\etr
\btr{c}
\makebox(40,40){\includegraphics[width=1.65cm,angle=0]{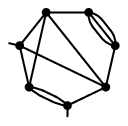}}\\
\texttt{248-S7}
\etr
\btr{c}
\makebox(40,40){\includegraphics[width=1.65cm,angle=0]{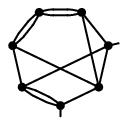}}\\
\texttt{249-S7}
\etr
\btr{c}
\makebox(40,40){\includegraphics[width=1.65cm,angle=0]{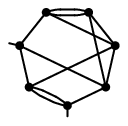}}\\
\texttt{250-S7}
\etr
\\
\btr{c}
\makebox(40,40){\includegraphics[width=1.65cm,angle=0]{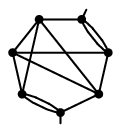}}\\
\texttt{251-S7}
\etr
\btr{c}
\makebox(40,40){\includegraphics[width=1.65cm,angle=0]{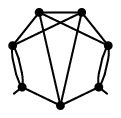}}\\
\texttt{252-S7}
\etr
\btr{c}
\makebox(40,40){\includegraphics[width=1.65cm,angle=0]{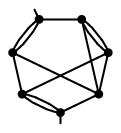}}\\
\texttt{253-S7}
\etr
\btr{c}
\makebox(40,40){\includegraphics[width=1.65cm,angle=0]{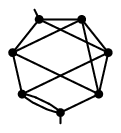}}\\
\texttt{254-S7}
\etr
\btr{c}
\makebox(40,40){\includegraphics[width=1.65cm,angle=0]{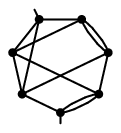}}\\
\texttt{255-S7}
\etr
\btr{c}
\makebox(40,40){\includegraphics[width=1.65cm,angle=0]{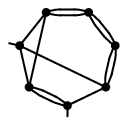}}\\
\texttt{256-S7}
\etr
\btr{c}
\makebox(40,40){\includegraphics[width=1.65cm,angle=0]{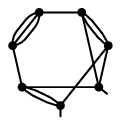}}\\
\texttt{257-S7}
\etr
\btr{c}
\makebox(40,40){\includegraphics[width=1.65cm,angle=0]{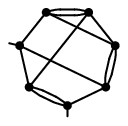}}\\
\texttt{258-S7}
\etr
\btr{c}
\makebox(40,40){\includegraphics[width=1.65cm,angle=0]{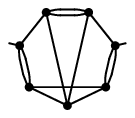}}\\
\texttt{259-S7}
\etr
\btr{c}
\makebox(40,40){\includegraphics[width=1.65cm,angle=0]{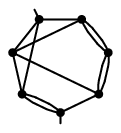}}\\
\texttt{260-S7}
\etr
\\
\btr{c}
\makebox(40,40){\includegraphics[width=1.65cm,angle=0]{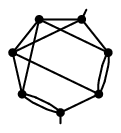}}\\
\texttt{261-S7}
\etr
\btr{c}
\makebox(40,40){\includegraphics[width=1.65cm,angle=0]{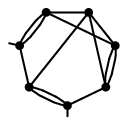}}\\
\texttt{262-S7}
\etr
\btr{c}
\makebox(40,40){\includegraphics[width=1.65cm,angle=0]{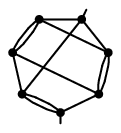}}\\
\texttt{263-S7}
\etr
\btr{c}
\makebox(40,40){\includegraphics[width=1.65cm,angle=0]{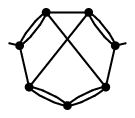}}\\
\texttt{264-S7}
\etr
\btr{c}
\makebox(40,40){\includegraphics[width=1.65cm,angle=0]{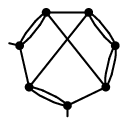}}\\
\texttt{265-S7}
\etr
\btr{c}
\makebox(40,40){\includegraphics[width=1.65cm,angle=0]{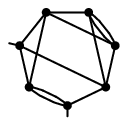}}\\
\texttt{266-S7}
\etr
\btr{c}
\makebox(40,40){\includegraphics[width=1.65cm,angle=0]{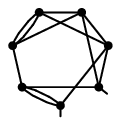}}\\
\texttt{267-S7}
\etr
\btr{c}
\makebox(40,40){\includegraphics[width=1.65cm,angle=0]{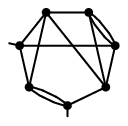}}\\
\texttt{268-S7}
\etr
\btr{c}
\makebox(40,40){\includegraphics[width=1.65cm,angle=0]{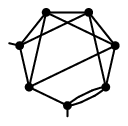}}\\
\texttt{269-S7}
\etr
\btr{c}
\makebox(40,40){\includegraphics[width=1.65cm,angle=0]{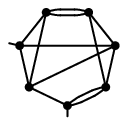}}\\
\texttt{270-S7}
\etr
\\
\btr{c}
\makebox(40,40){\includegraphics[width=1.65cm,angle=0]{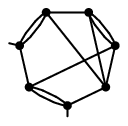}}\\
\texttt{271-S7}
\etr
\btr{c}
\makebox(40,40){\includegraphics[width=1.65cm,angle=0]{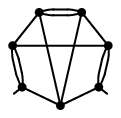}}\\
\texttt{272-S7}
\etr
\btr{c}
\makebox(40,40){\includegraphics[width=1.65cm,angle=0]{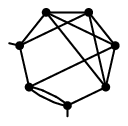}}\\
\texttt{273-S7}
\etr
\btr{c}
\makebox(40,40){\includegraphics[width=1.65cm,angle=0]{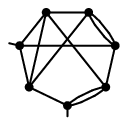}}\\
\texttt{274-S7}
\etr
\btr{c}
\makebox(40,40){\includegraphics[width=1.65cm,angle=0]{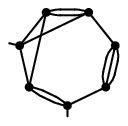}}\\
\texttt{275-S7}
\etr
\btr{c}
\makebox(40,40){\includegraphics[width=1.65cm,angle=0]{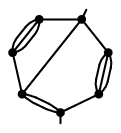}}\\
\texttt{276-S7}
\etr
\btr{c}
\makebox(40,40){\includegraphics[width=1.65cm,angle=0]{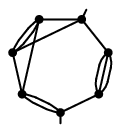}}\\
\texttt{277-S7}
\etr
\btr{c}
\makebox(40,40){\includegraphics[width=1.65cm,angle=0]{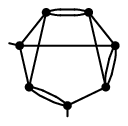}}\\
\texttt{278-S7}
\etr
\btr{c}
\makebox(40,40){\includegraphics[width=1.65cm,angle=0]{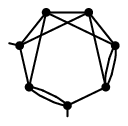}}\\
\texttt{279-S7}
\etr
\btr{c}
\makebox(40,40){\includegraphics[width=1.65cm,angle=0]{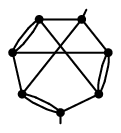}}\\
\texttt{280-S7}
\etr
\\
\btr{c}
\makebox(40,40){\includegraphics[width=1.65cm,angle=0]{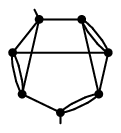}}\\
\texttt{281-S7}
\etr
\btr{c}
\makebox(40,40){\includegraphics[width=1.65cm,angle=0]{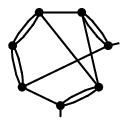}}\\
\texttt{282-S7}
\etr
\btr{c}
\makebox(40,40){\includegraphics[width=1.65cm,angle=0]{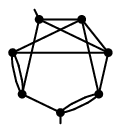}}\\
\texttt{283-S7}
\etr
\btr{c}
\makebox(40,40){\includegraphics[width=1.65cm,angle=0]{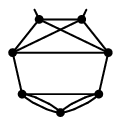}}\\
\texttt{284-S7}
\etr
\btr{c}
\makebox(40,40){\includegraphics[width=1.65cm,angle=0]{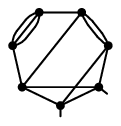}}\\
\texttt{285-S7}
\etr
\btr{c}
\makebox(40,40){\includegraphics[width=1.65cm,angle=0]{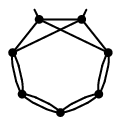}}\\
\texttt{286-S7}
\etr
\btr{c}
\makebox(40,40){\includegraphics[width=1.65cm,angle=0]{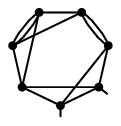}}\\
\texttt{287-S7}
\etr
\btr{c}
\makebox(40,40){\includegraphics[width=1.65cm,angle=0]{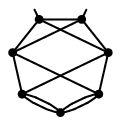}}\\
\texttt{288-S7}
\etr
\btr{c}
\makebox(40,40){\includegraphics[width=1.65cm,angle=0]{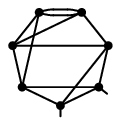}}\\
\texttt{289-S7}
\etr
\btr{c}
\makebox(40,40){\includegraphics[width=1.65cm,angle=0]{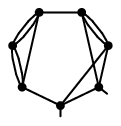}}\\
\texttt{290-S7}
\etr
\\
\btr{c}
\makebox(40,40){\includegraphics[width=1.65cm,angle=0]{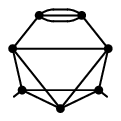}}\\
\texttt{291-S7}
\etr
\btr{c}
\makebox(40,40){\includegraphics[width=1.65cm,angle=0]{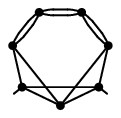}}\\
\texttt{292-S7}
\etr
\btr{c}
\makebox(40,40){\includegraphics[width=1.65cm,angle=0]{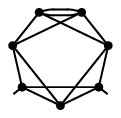}}\\
\texttt{293-S7}
\etr
\btr{c}
\makebox(40,40){\includegraphics[width=1.65cm,angle=0]{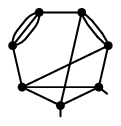}}\\
\texttt{294-S7}
\etr
\btr{c}
\makebox(40,40){\includegraphics[width=1.65cm,angle=0]{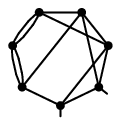}}\\
\texttt{295-S7}
\etr
\btr{c}
\makebox(40,40){\includegraphics[width=1.65cm,angle=0]{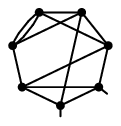}}\\
\texttt{296-S7}
\etr
\btr{c}
\makebox(40,40){\includegraphics[width=1.65cm,angle=0]{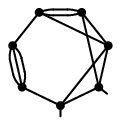}}\\
\texttt{297-S7}
\etr
\btr{c}
\makebox(40,40){\includegraphics[width=1.65cm,angle=0]{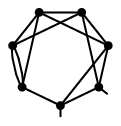}}\\
\texttt{298-S7}
\etr
\btr{c}
\makebox(40,40){\includegraphics[width=1.65cm,angle=0]{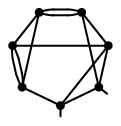}}\\
\texttt{299-S7}
\etr
\btr{c}
\makebox(40,40){\includegraphics[width=1.65cm,angle=0]{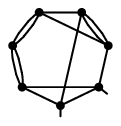}}\\
\texttt{300-S7}
\etr
\\
\btr{c}
\makebox(40,40){\includegraphics[width=1.65cm,angle=0]{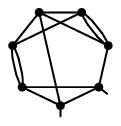}}\\
\texttt{301-S7}
\etr
\btr{c}
\makebox(40,40){\includegraphics[width=1.65cm,angle=0]{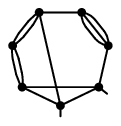}}\\
\texttt{302-S7}
\etr
\btr{c}
\makebox(40,40){\includegraphics[width=1.65cm,angle=0]{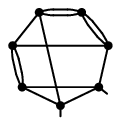}}\\
\texttt{303-S7}
\etr
\btr{c}
\makebox(40,40){\includegraphics[width=1.65cm,angle=0]{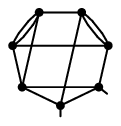}}\\
\texttt{304-S7}
\etr
\btr{c}
\makebox(40,40){\includegraphics[width=1.65cm,angle=0]{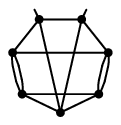}}\\
\texttt{305-S7}
\etr
\btr{c}
\makebox(40,40){\includegraphics[width=1.65cm,angle=0]{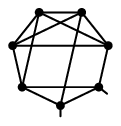}}\\
\texttt{306-S7}
\etr
\btr{c}
\makebox(40,40){\includegraphics[width=1.65cm,angle=0]{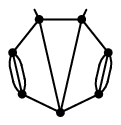}}\\
\texttt{307-S7}
\etr
\btr{c}
\makebox(40,40){\includegraphics[width=1.65cm,angle=0]{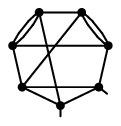}}\\
\texttt{308-S7}
\etr
\btr{c}
\makebox(40,40){\includegraphics[width=1.65cm,angle=0]{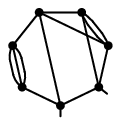}}\\
\texttt{309-S7}
\etr
\btr{c}
\makebox(40,40){\includegraphics[width=1.65cm,angle=0]{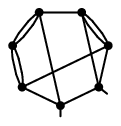}}\\
\texttt{310-S7}
\etr
\\
\btr{c}
\makebox(40,40){\includegraphics[width=1.65cm,angle=0]{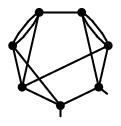}}\\
\texttt{311-S7}
\etr
\btr{c}
\makebox(40,40){\includegraphics[width=1.65cm,angle=0]{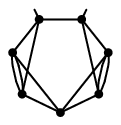}}\\
\texttt{312-S7}
\etr
\btr{c}
\makebox(40,40){\includegraphics[width=1.65cm,angle=0]{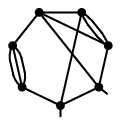}}\\
\texttt{313-S7}
\etr
\btr{c}
\makebox(40,40){\includegraphics[width=1.65cm,angle=0]{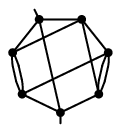}}\\
\texttt{314-S7}
\etr
\btr{c}
\makebox(40,40){\includegraphics[width=1.65cm,angle=0]{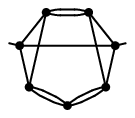}}\\
\texttt{315-S7}
\etr
\btr{c}
\makebox(40,40){\includegraphics[width=1.65cm,angle=0]{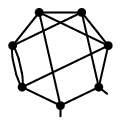}}\\
\texttt{316-S7}
\etr
\btr{c}
\makebox(40,40){\includegraphics[width=1.65cm,angle=0]{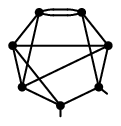}}\\
\texttt{317-S7}
\etr
\btr{c}
\makebox(40,40){\includegraphics[width=1.65cm,angle=0]{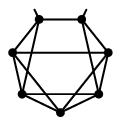}}\\
\texttt{318-S7}
\etr
\btr{c}
\makebox(40,40){\includegraphics[width=1.65cm,angle=0]{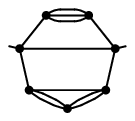}}\\
\texttt{319-S7}
\etr
\btr{c}
\makebox(40,40){\includegraphics[width=1.65cm,angle=0]{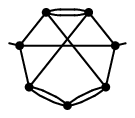}}\\
\texttt{320-S7}
\etr
\\
\btr{c}
\makebox(40,40){\includegraphics[width=1.65cm,angle=0]{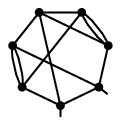}}\\
\texttt{321-S7}
\etr
\btr{c}
\makebox(40,40){\includegraphics[width=1.65cm,angle=0]{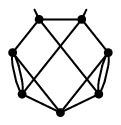}}\\
\texttt{322-S7}
\etr
\btr{c}
\makebox(40,40){\includegraphics[width=1.65cm,angle=0]{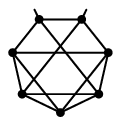}}\\
\texttt{323-S7}
\etr
\btr{c}
\makebox(40,40){\includegraphics[width=1.65cm,angle=0]{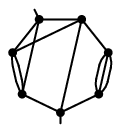}}\\
\texttt{324-S7}
\etr
\btr{c}
\makebox(40,40){\includegraphics[width=1.65cm,angle=0]{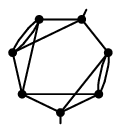}}\\
\texttt{325-S7}
\etr
\btr{c}
\makebox(40,40){\includegraphics[width=1.65cm,angle=0]{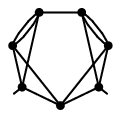}}\\
\texttt{326-S7}
\etr
\btr{c}
\makebox(40,40){\includegraphics[width=1.65cm,angle=0]{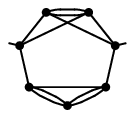}}\\
\texttt{327-S7}
\etr
\btr{c}
\makebox(40,40){\includegraphics[width=1.65cm,angle=0]{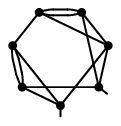}}\\
\texttt{328-S7}
\etr
\btr{c}
\makebox(40,40){\includegraphics[width=1.65cm,angle=0]{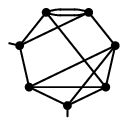}}\\
\texttt{329-S7}
\etr
\btr{c}
\makebox(40,40){\includegraphics[width=1.65cm,angle=0]{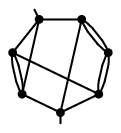}}\\
\texttt{330-S7}
\etr
\\
\btr{c}
\makebox(40,40){\includegraphics[width=1.65cm,angle=0]{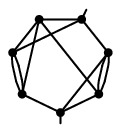}}\\
\texttt{331-S7}
\etr
\btr{c}
\makebox(40,40){\includegraphics[width=1.65cm,angle=0]{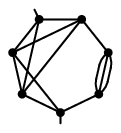}}\\
\texttt{332-S7}
\etr
\btr{c}
\makebox(40,40){\includegraphics[width=1.65cm,angle=0]{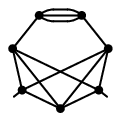}}\\
\texttt{333-S7}
\etr
\btr{c}
\makebox(40,40){\includegraphics[width=1.65cm,angle=0]{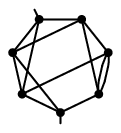}}\\
\texttt{334-S7}
\etr
\btr{c}
\makebox(40,40){\includegraphics[width=1.65cm,angle=0]{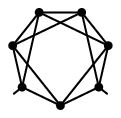}}\\
\texttt{335-S7}
\etr
\btr{c}
\makebox(40,40){\includegraphics[width=1.65cm,angle=0]{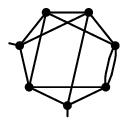}}\\
\texttt{336-S7}
\etr
\btr{c}
\makebox(40,40){\includegraphics[width=1.65cm,angle=0]{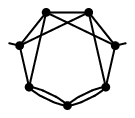}}\\
\texttt{337-S7}
\etr
\btr{c}
\makebox(40,40){\includegraphics[width=1.65cm,angle=0]{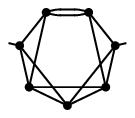}}\\
\texttt{338-S7}
\etr
\btr{c}
\makebox(40,40){\includegraphics[width=1.65cm,angle=0]{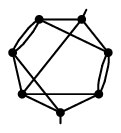}}\\
\texttt{339-S7}
\etr
\btr{c}
\makebox(40,40){\includegraphics[width=1.65cm,angle=0]{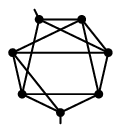}}\\
\texttt{340-S7}
\etr
\\
\btr{c}
\makebox(40,40){\includegraphics[width=1.65cm,angle=0]{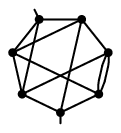}}\\
\texttt{341-S7}
\etr
\btr{c}
\makebox(40,40){\includegraphics[width=1.65cm,angle=0]{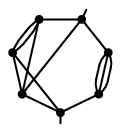}}\\
\texttt{342-S7}
\etr
\btr{c}
\makebox(40,40){\includegraphics[width=1.65cm,angle=0]{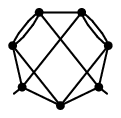}}\\
\texttt{343-S7}
\etr
\btr{c}
\makebox(40,40){\includegraphics[width=1.65cm,angle=0]{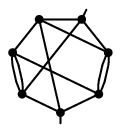}}\\
\texttt{344-S7}
\etr
\btr{c}
\makebox(40,40){\includegraphics[width=1.65cm,angle=0]{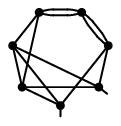}}\\
\texttt{345-S7}
\etr
\btr{c}
\makebox(40,40){\includegraphics[width=1.65cm,angle=0]{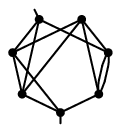}}\\
\texttt{346-S7}
\etr
\btr{c}
\makebox(40,40){\includegraphics[width=1.65cm,angle=0]{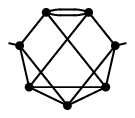}}\\
\texttt{347-S7}
\etr
\btr{c}
\makebox(40,40){\includegraphics[width=1.65cm,angle=0]{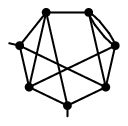}}\\
\texttt{348-S7}
\etr
\btr{c}
\makebox(40,40){\includegraphics[width=1.65cm,angle=0]{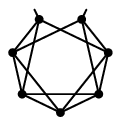}}\\
\texttt{349-S7}
\etr
\btr{c}
\makebox(40,40){\includegraphics[width=1.65cm,angle=0]{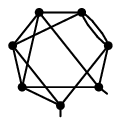}}\\
\texttt{350-S7}
\etr
\\
\btr{c}
\makebox(40,40){\includegraphics[width=1.65cm,angle=0]{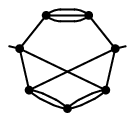}}\\
\texttt{351-S7}
\etr
\btr{c}
\makebox(40,40){\includegraphics[width=1.65cm,angle=0]{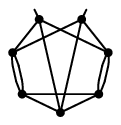}}\\
\texttt{352-S7}
\etr
\btr{c}
\makebox(40,40){\includegraphics[width=1.65cm,angle=0]{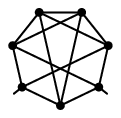}}\\
\texttt{353-S7}
\etr
\elt

\nc{\encodingscaption}
{Suitable representation of diagrams for processing by computer code
according to \cite{NiMeBa}.
Explanations are given in Appendix~\ref{encods}.}
\blt[c]{|r|l||r|l||r|l|}
\caption{\label{encodings}\encodingscaption}
\\\hline\rule{0pt}{1pt}
&&&&&\\\multicolumn{1}{|c|}{diagram} &\multicolumn{1}{|c||}{code}
& \multicolumn{1}{|c|}{diagram} &\multicolumn{1}{|c||}{code}
& \multicolumn{1}{|c|}{diagram} &\multicolumn{1}{|c|}{code}
\\\rule{0pt}{1pt}&&&&&\\\hline\texttt{1-S0} & // & 
\texttt{119-M7} & /EE12/334/335//566/66// & 
\texttt{237-S7} & /E112/34/E56/455/66/6// \\ 
\texttt{2-M1} & /EE0/ & 
\texttt{120-M7} & /EE12/334/344/5/6/666// & 
\texttt{238-S7} & /E112/34/E56/456/56/6// \\ 
\texttt{3-S2} & /E111/E/ & 
\texttt{121-M7} & /EE12/334/345/4/6/666// & 
\texttt{239-S7} & /E112/34/E56/555/666/// \\ 
\texttt{4-M3} & /EE12/222// & 
\texttt{122-M7} & /EE12/334/345/5/66/66// & 
\texttt{240-S7} & /E112/34/E56/556/566/// \\ 
\texttt{5-S3} & /E112/22/E/ & 
\texttt{123-M7} & /EE12/334/345/6/56/66// & 
\texttt{241-S7} & /E112/34/334/5/E6/666// \\ 
\texttt{6-M4} & /EE12/233/33// & 
\texttt{124-M7} & /EE12/334/355/4/66/66// & 
\texttt{242-S7} & /E112/34/335/E/566/66// \\ 
\texttt{7-S4} & /E112/E3/333// & 
\texttt{125-M7} & /EE12/334/355/5/666/6// & 
\texttt{243-S7} & /E112/34/335/4/E6/666// \\ 
\texttt{8-S4} & /E112/23/33/E/ & 
\texttt{126-M7} & /EE12/334/355/6/566/6// & 
\texttt{244-S7} & /E112/34/335/5/666/E6// \\ 
\texttt{9-S4} & /E112/33/E33// & 
\texttt{127-M7} & /EE12/334/356/4/56/66// & 
\texttt{245-S7} & /E112/34/335/6/E56/66// \\ 
\texttt{10-S4} & /E123/E23/33// & 
\texttt{128-M7} & /EE12/334/356/5/566/6// & 
\texttt{246-S7} & /E112/34/335/6/556/6/E/ \\ 
\texttt{11-M5} & /EE12/223/4/444// & 
\texttt{129-M7} & /EE12/334/455/46/6/66// & 
\texttt{247-S7} & /E112/34/335/6/566/E6// \\ 
\texttt{12-M5} & /EE12/233/44/44// & 
\texttt{130-M7} & /EE12/334/455/56/66/6// & 
\texttt{248-S7} & /E112/34/345/E4/6/666// \\ 
\texttt{13-M5} & /EE12/234/34/44// & 
\texttt{131-M7} & /EE12/334/456/45/6/66// & 
\texttt{249-S7} & /E112/34/345/E5/66/66// \\ 
\texttt{14-M5} & /EE12/333/444/4// & 
\texttt{132-M7} & /EE12/334/456/55/66/6// & 
\texttt{250-S7} & /E112/34/345/E6/56/66// \\ 
\texttt{15-M5} & /EE12/334/344/4// & 
\texttt{133-M7} & /EE12/334/456/56/56/6// & 
\texttt{251-S7} & /E112/34/345/45/6/66/E/ \\ 
\texttt{16-S5} & /E112/E3/344/44// & 
\texttt{134-M7} & /EE12/334/556/44/6/66// & 
\texttt{252-S7} & /E112/34/345/46/6/E66// \\ 
\texttt{17-S5} & /E112/23/E4/444// & 
\texttt{135-M7} & /EE12/334/556/45/66/6// & 
\texttt{253-S7} & /E112/34/345/55/66/6/E/ \\ 
\texttt{18-S5} & /E112/23/34/44/E/ & 
\texttt{136-M7} & /EE12/334/556/46/56/6// & 
\texttt{254-S7} & /E112/34/345/56/56/6/E/ \\ 
\texttt{19-S5} & /E112/23/44/E44// & 
\texttt{137-M7} & /EE12/334/556/55/666/// & 
\texttt{255-S7} & /E112/34/345/56/66/E6// \\ 
\texttt{20-S5} & /E112/33/E44/44// & 
\texttt{138-M7} & /EE12/334/556/56/566/// & 
\texttt{256-S7} & /E112/34/355/E4/66/66// \\ 
\texttt{21-S5} & /E112/33/444/E4// & 
\texttt{139-M7} & /EE12/334/556/66/556/// & 
\texttt{257-S7} & /E112/34/355/E5/666/6// \\ 
\texttt{22-S5} & /E112/34/E34/44// & 
\texttt{140-M7} & /EE12/345/345/46/6/66// & 
\texttt{258-S7} & /E112/34/355/E6/566/6// \\ 
\texttt{23-S5} & /E112/34/334/4/E/ & 
\texttt{141-M7} & /EE12/345/346/44//666// & 
\texttt{259-S7} & /E112/34/355/45/66/6/E/ \\ 
\texttt{24-S5} & /E123/E23/44/44// & 
\texttt{142-M7} & /EE12/345/346/45/6/66// & 
\texttt{260-S7} & /E112/34/355/46/E6/66// \\ 
\texttt{25-S5} & /E123/E24/34/44// & 
\texttt{143-M7} & /EE12/345/346/55/66/6// & 
\texttt{261-S7} & /E112/34/355/46/56/6/E/ \\ 
\texttt{26-S5} & /E123/234/34/4/E/ & 
\texttt{144-M7} & /EE12/345/346/56/56/6// & 
\texttt{262-S7} & /E112/34/355/46/66/E6// \\ 
\texttt{27-M6} & /EE12/223/4/455/55// & 
\texttt{145-S7} & /E112/E3/334/5/566/66// & 
\texttt{263-S7} & /E112/34/355/66/E56/6// \\ 
\texttt{28-M6} & /EE12/233/34/5/555// & 
\texttt{146-S7} & /E112/E3/344/45/6/666// & 
\texttt{264-S7} & /E112/34/355/66/556//E/ \\ 
\texttt{29-M6} & /EE12/233/44/55/55// & 
\texttt{147-S7} & /E112/E3/344/55/66/66// & 
\texttt{265-S7} & /E112/34/355/66/566/E// \\ 
\texttt{30-M6} & /EE12/233/45/45/55// & 
\texttt{148-S7} & /E112/E3/344/56/56/66// & 
\texttt{266-S7} & /E112/34/356/E4/56/66// \\ 
\texttt{31-M6} & /EE12/234/34/55/55// & 
\texttt{149-S7} & /E112/E3/345/45/66/66// & 
\texttt{267-S7} & /E112/34/356/E5/566/6// \\ 
\texttt{32-M6} & /EE12/234/35/45/55// & 
\texttt{150-S7} & /E112/E3/345/46/56/66// & 
\texttt{268-S7} & /E112/34/356/45/E6/66// \\ 
\texttt{33-M6} & /EE12/333/445/5/55// & 
\texttt{151-S7} & /E112/E3/444/556/6/66// & 
\texttt{269-S7} & /E112/34/356/45/56/6/E/ \\ 
\texttt{34-M6} & /EE12/334/335//555// & 
\texttt{152-S7} & /E112/E3/445/446//666// & 
\texttt{270-S7} & /E112/34/356/45/66/E6// \\ 
\texttt{35-M6} & /EE12/334/345/5/55// & 
\texttt{153-S7} & /E112/E3/445/456/6/66// & 
\texttt{271-S7} & /E112/34/356/55/E66/6// \\ 
\texttt{36-M6} & /EE12/334/355/4/55// & 
\texttt{154-S7} & /E112/E3/445/466/5/66// & 
\texttt{272-S7} & /E112/34/356/55/566//E/ \\ 
\texttt{37-M6} & /EE12/334/455/45/5// & 
\texttt{155-S7} & /E112/E3/445/566/56/6// & 
\texttt{273-S7} & /E112/34/356/56/E56/6// \\ 
\texttt{38-M6} & /EE12/345/345/45/5// & 
\texttt{156-S7} & /E112/E3/456/456/56/6// & 
\texttt{274-S7} & /E112/34/356/56/556//E/ \\ 
\texttt{39-S6} & /E112/E3/334/5/555// & 
\texttt{157-S7} & /E112/23/E4/445/6/666// & 
\texttt{275-S7} & /E112/34/555/E46/66/6// \\ 
\texttt{40-S6} & /E112/E3/344/55/55// & 
\texttt{158-S7} & /E112/23/E4/455/66/66// & 
\texttt{276-S7} & /E112/34/555/E56/666/// \\ 
\texttt{41-S6} & /E112/E3/345/45/55// & 
\texttt{159-S7} & /E112/23/E4/456/56/66// & 
\texttt{277-S7} & /E112/34/555/446/6/6/E/ \\ 
\texttt{42-S6} & /E112/E3/444/555/5// & 
\texttt{160-S7} & /E112/23/E4/555/666/6// & 
\texttt{278-S7} & /E112/34/556/E45/66/6// \\ 
\texttt{43-S6} & /E112/E3/445/455/5// & 
\texttt{161-S7} & /E112/23/E4/556/566/6// & 
\texttt{279-S7} & /E112/34/556/E46/56/6// \\ 
\texttt{44-S6} & /E112/23/E4/455/55// & 
\texttt{162-S7} & /E112/23/34/E5/566/66// & 
\texttt{280-S7} & /E112/34/556/E56/566/// \\ 
\texttt{45-S6} & /E112/23/34/E5/555// & 
\texttt{163-S7} & /E112/23/34/45/E6/666// & 
\texttt{281-S7} & /E112/34/556/445/6/6/E/ \\ 
\texttt{46-S6} & /E112/23/34/45/55/E/ & 
\texttt{164-S7} & /E112/23/34/45/56/66/E/ & 
\texttt{282-S7} & /E112/34/556/455/66//E/ \\ 
\texttt{47-S6} & /E112/23/34/55/E55// & 
\texttt{165-S7} & /E112/23/34/45/66/E66// & 
\texttt{283-S7} & /E112/34/556/456/56//E/ \\ 
\texttt{48-S6} & /E112/23/44/E55/55// & 
\texttt{166-S7} & /E112/23/34/55/E66/66// & 
\texttt{284-S7} & /E123/E23/34/5/566/66// \\ 
\texttt{49-S6} & /E112/23/44/455/5/E/ & 
\texttt{167-S7} & /E112/23/34/55/566/6/E/ & 
\texttt{285-S7} & /E123/E23/44/45/6/666// \\ 
\texttt{50-S6} & /E112/23/44/555/E5// & 
\texttt{168-S7} & /E112/23/34/55/666/E6// & 
\texttt{286-S7} & /E123/E23/44/55/66/66// \\ 
\texttt{51-S6} & /E112/23/45/E45/55// & 
\texttt{169-S7} & /E112/23/34/56/E56/66// & 
\texttt{287-S7} & /E123/E23/44/56/56/66// \\ 
\texttt{52-S6} & /E112/23/45/445/5/E/ & 
\texttt{170-S7} & /E112/23/34/56/556/6/E/ & 
\texttt{288-S7} & /E123/E23/45/45/66/66// \\ 
\texttt{53-S6} & /E112/33/E34/5/555// & 
\texttt{171-S7} & /E112/23/44/E45/6/666// & 
\texttt{289-S7} & /E123/E23/45/46/56/66// \\ 
\texttt{54-S6} & /E112/33/E44/55/55// & 
\texttt{172-S7} & /E112/23/44/E55/66/66// & 
\texttt{290-S7} & /E123/E24/33/5/566/66// \\ 
\texttt{55-S6} & /E112/33/E45/45/55// & 
\texttt{173-S7} & /E112/23/44/E56/56/66// & 
\texttt{291-S7} & /E123/E24/34/45/6/666// \\ 
\texttt{56-S6} & /E112/33/344/5/55/E/ & 
\texttt{174-S7} & /E112/23/44/455/6/66/E/ & 
\texttt{292-S7} & /E123/E24/34/55/66/66// \\ 
\texttt{57-S6} & /E112/33/444/55/5/E/ & 
\texttt{175-S7} & /E112/23/44/555/66/6/E/ & 
\texttt{293-S7} & /E123/E24/34/56/56/66// \\ 
\texttt{58-S6} & /E112/33/445/E5/55// & 
\texttt{176-S7} & /E112/23/44/556/E6/66// & 
\texttt{294-S7} & /E123/E24/35/44/6/666// \\ 
\texttt{59-S6} & /E112/33/445/45/5/E/ & 
\texttt{177-S7} & /E112/23/44/556/56/6/E/ & 
\texttt{295-S7} & /E123/E24/35/45/66/66// \\ 
\texttt{60-S6} & /E112/34/E33/5/555// & 
\texttt{178-S7} & /E112/23/44/556/66/E6// & 
\texttt{296-S7} & /E123/E24/35/46/56/66// \\ 
\texttt{61-S6} & /E112/34/E34/55/55// & 
\texttt{179-S7} & /E112/23/45/E44/6/666// & 
\texttt{297-S7} & /E123/E24/35/55/666/6// \\ 
\texttt{62-S6} & /E112/34/E35/45/55// & 
\texttt{180-S7} & /E112/23/45/E45/66/66// & 
\texttt{298-S7} & /E123/E24/35/56/566/6// \\ 
\texttt{63-S6} & /E112/34/E55/445/5// & 
\texttt{181-S7} & /E112/23/45/E46/56/66// & 
\texttt{299-S7} & /E123/E24/35/66/556/6// \\ 
\texttt{64-S6} & /E112/34/334/5/55/E/ & 
\texttt{182-S7} & /E112/23/45/E66/556/6// & 
\texttt{300-S7} & /E123/E24/55/446/6/66// \\ 
\texttt{65-S6} & /E112/34/335/E/555// & 
\texttt{183-S7} & /E112/23/45/445/6/66/E/ & 
\texttt{301-S7} & /E123/E24/55/456/66/6// \\ 
\texttt{66-S6} & /E112/34/335/4/55/E/ & 
\texttt{184-S7} & /E112/23/45/446/E/666// & 
\texttt{302-S7} & /E123/E24/55/556/666/// \\ 
\texttt{67-S6} & /E112/34/335/5/E55// & 
\texttt{185-S7} & /E112/23/45/446/5/66/E/ & 
\texttt{303-S7} & /E123/E24/55/566/566/// \\ 
\texttt{68-S6} & /E112/34/345/E5/55// & 
\texttt{186-S7} & /E112/23/45/446/6/E66// & 
\texttt{304-S7} & /E123/E24/56/445/6/66// \\ 
\texttt{69-S6} & /E112/34/345/45/5/E/ & 
\texttt{187-S7} & /E112/23/45/456/E6/66// & 
\texttt{305-S7} & /E123/E24/56/455/66/6// \\ 
\texttt{70-S6} & /E112/34/355/E4/55// & 
\texttt{188-S7} & /E112/23/45/456/56/6/E/ & 
\texttt{306-S7} & /E123/E24/56/456/56/6// \\ 
\texttt{71-S6} & /E112/34/355/45/E5// & 
\texttt{189-S7} & /E112/23/45/466/E5/66// & 
\texttt{307-S7} & /E123/E24/56/555/666/// \\ 
\texttt{72-S6} & /E123/E23/34/5/555// & 
\texttt{190-S7} & /E112/23/45/466/55/6/E/ & 
\texttt{308-S7} & /E123/E24/56/556/566/// \\ 
\texttt{73-S6} & /E123/E23/44/55/55// & 
\texttt{191-S7} & /E112/23/45/466/56/E6// & 
\texttt{309-S7} & /E123/E45/334/4/6/666// \\ 
\texttt{74-S6} & /E123/E23/45/45/55// & 
\texttt{192-S7} & /E112/23/45/666/E55/6// & 
\texttt{310-S7} & /E123/E45/334/5/66/66// \\ 
\texttt{75-S6} & /E123/E24/33/5/555// & 
\texttt{193-S7} & /E112/33/E34/5/566/66// & 
\texttt{311-S7} & /E123/E45/334/6/56/66// \\ 
\texttt{76-S6} & /E123/E24/34/55/55// & 
\texttt{194-S7} & /E112/33/E44/45/6/666// & 
\texttt{312-S7} & /E123/E45/336/6/556/6// \\ 
\texttt{77-S6} & /E123/E24/35/45/55// & 
\texttt{195-S7} & /E112/33/E44/55/66/66// & 
\texttt{313-S7} & /E123/E45/344/46//666// \\ 
\texttt{78-S6} & /E123/E24/55/445/5// & 
\texttt{196-S7} & /E112/33/E44/56/56/66// & 
\texttt{314-S7} & /E123/E45/344/56/6/66// \\ 
\texttt{79-S6} & /E123/E45/334/5/55// & 
\texttt{197-S7} & /E112/33/E45/44/6/666// & 
\texttt{315-S7} & /E123/E45/344/66/5/66// \\ 
\texttt{80-S6} & /E123/E45/344/55/5// & 
\texttt{198-S7} & /E112/33/E45/45/66/66// & 
\texttt{316-S7} & /E123/E45/345/46/6/66// \\ 
\texttt{81-S6} & /E123/E45/345/45/5// & 
\texttt{199-S7} & /E112/33/E45/46/56/66// & 
\texttt{317-S7} & /E123/E45/345/66/56/6// \\ 
\texttt{82-S6} & /E123/E45/444/555/// & 
\texttt{200-S7} & /E112/33/E45/66/556/6// & 
\texttt{318-S7} & /E123/E45/346/56/56/6// \\ 
\texttt{83-S6} & /E123/E45/445/455/// & 
\texttt{201-S7} & /E112/33/344/5/E6/666// & 
\texttt{319-S7} & /E123/E45/444/566//66// \\ 
\texttt{84-S6} & /E123/224/4/555/E5// & 
\texttt{202-S7} & /E112/33/344/5/66/E66// & 
\texttt{320-S7} & /E123/E45/445/466//66// \\ 
\texttt{85-S6} & /E123/224/5/445/5/E/ & 
\texttt{203-S7} & /E112/33/345/4/E6/666// & 
\texttt{321-S7} & /E123/E45/445/566/6/6// \\ 
\texttt{86-S6} & /E123/234/45/45/5/E/ & 
\texttt{204-S7} & /E112/33/345/4/66/E66// & 
\texttt{322-S7} & /E123/E45/446/556/6/6// \\ 
\texttt{87-S6} & /E123/234/45/55/E5// & 
\texttt{205-S7} & /E112/33/345/6/E56/66// & 
\texttt{323-S7} & /E123/E45/456/456/6/6// \\ 
\texttt{88-S6} & /E123/245/45/445//E/ & 
\texttt{206-S7} & /E112/33/345/6/556/6/E/ & 
\texttt{324-S7} & /E123/223/4/45/E6/666// \\ 
\texttt{89-M7} & /EE12/223/4/445/6/666// & 
\texttt{207-S7} & /E112/33/444/E5/6/666// & 
\texttt{325-S7} & /E123/223/4/56/E56/66// \\ 
\texttt{90-M7} & /EE12/223/4/455/66/66// & 
\texttt{208-S7} & /E112/33/444/55/6/66/E/ & 
\texttt{326-S7} & /E123/223/4/56/556/6/E/ \\ 
\texttt{91-M7} & /EE12/223/4/456/56/66// & 
\texttt{209-S7} & /E112/33/445/E4/6/666// & 
\texttt{327-S7} & /E123/224/4/556/E6/66// \\ 
\texttt{92-M7} & /EE12/223/4/555/666/6// & 
\texttt{210-S7} & /E112/33/445/E5/66/66// & 
\texttt{328-S7} & /E123/224/5/456/E6/66// \\ 
\texttt{93-M7} & /EE12/223/4/556/566/6// & 
\texttt{211-S7} & /E112/33/445/E6/56/66// & 
\texttt{329-S7} & /E123/224/5/456/56/6/E/ \\ 
\texttt{94-M7} & /EE12/233/34/5/566/66// & 
\texttt{212-S7} & /E112/33/445/45/6/66/E/ & 
\texttt{330-S7} & /E123/224/5/466/E5/66// \\ 
\texttt{95-M7} & /EE12/233/44/45/6/666// & 
\texttt{213-S7} & /E112/33/445/46/E/666// & 
\texttt{331-S7} & /E123/224/5/466/56/E6// \\ 
\texttt{96-M7} & /EE12/233/44/55/66/66// & 
\texttt{214-S7} & /E112/33/445/46/5/66/E/ & 
\texttt{332-S7} & /E123/234/34/5/E6/666// \\ 
\texttt{97-M7} & /EE12/233/44/56/56/66// & 
\texttt{215-S7} & /E112/33/445/46/6/E66// & 
\texttt{333-S7} & /E123/234/45/46/E/666// \\ 
\texttt{98-M7} & /EE12/233/45/44/6/666// & 
\texttt{216-S7} & /E112/33/445/56/E6/66// & 
\texttt{334-S7} & /E123/234/45/56/E6/66// \\ 
\texttt{99-M7} & /EE12/233/45/45/66/66// & 
\texttt{217-S7} & /E112/33/445/56/56/6/E/ & 
\texttt{335-S7} & /E123/234/45/56/56/6/E/ \\ 
\texttt{100-M7} & /EE12/233/45/46/56/66// & 
\texttt{218-S7} & /E112/33/445/56/66/E6// & 
\texttt{336-S7} & /E123/234/45/56/66/E6// \\ 
\texttt{101-M7} & /EE12/233/45/66/556/6// & 
\texttt{219-S7} & /E112/33/445/66/55/6/E/ & 
\texttt{337-S7} & /E123/234/45/66/E5/66// \\ 
\texttt{102-M7} & /EE12/234/34/45/6/666// & 
\texttt{220-S7} & /E112/33/445/66/56/E6// & 
\texttt{338-S7} & /E123/234/45/66/56/E6// \\ 
\texttt{103-M7} & /EE12/234/34/55/66/66// & 
\texttt{221-S7} & /E112/33/456/E4/56/66// & 
\texttt{339-S7} & /E123/234/55/66/E56/6// \\ 
\texttt{104-M7} & /EE12/234/34/56/56/66// & 
\texttt{222-S7} & /E112/33/456/45/E6/66// & 
\texttt{340-S7} & /E123/234/56/56/E56/6// \\ 
\texttt{105-M7} & /EE12/234/35/44/6/666// & 
\texttt{223-S7} & /E112/33/456/45/56/6/E/ & 
\texttt{341-S7} & /E123/234/56/56/556//E/ \\ 
\texttt{106-M7} & /EE12/234/35/45/66/66// & 
\texttt{224-S7} & /E112/34/E33/5/566/66// & 
\texttt{342-S7} & /E123/244/45/666/5/E6// \\ 
\texttt{107-M7} & /EE12/234/35/46/56/66// & 
\texttt{225-S7} & /E112/34/E34/45/6/666// & 
\texttt{343-S7} & /E123/244/56/455/6/6/E/ \\ 
\texttt{108-M7} & /EE12/234/35/66/556/6// & 
\texttt{226-S7} & /E112/34/E34/55/66/66// & 
\texttt{344-S7} & /E123/244/56/556/56//E/ \\ 
\texttt{109-M7} & /EE12/234/56/445/6/66// & 
\texttt{227-S7} & /E112/34/E34/56/56/66// & 
\texttt{345-S7} & /E123/245/45/466/E/66// \\ 
\texttt{110-M7} & /EE12/234/56/455/66/6// & 
\texttt{228-S7} & /E112/34/E35/44/6/666// & 
\texttt{346-S7} & /E123/245/45/466/6/E6// \\ 
\texttt{111-M7} & /EE12/234/56/456/56/6// & 
\texttt{229-S7} & /E112/34/E35/45/66/66// & 
\texttt{347-S7} & /E123/245/46/455/6/6/E/ \\ 
\texttt{112-M7} & /EE12/234/56/555/666/// & 
\texttt{230-S7} & /E112/34/E35/46/56/66// & 
\texttt{348-S7} & /E123/245/46/456/E/66// \\ 
\texttt{113-M7} & /EE12/234/56/556/566/// & 
\texttt{231-S7} & /E112/34/E35/55/666/6// & 
\texttt{349-S7} & /E123/245/46/456/5/6/E/ \\ 
\texttt{114-M7} & /EE12/333/444/5/6/666// & 
\texttt{232-S7} & /E112/34/E35/56/566/6// & 
\texttt{350-S7} & /E123/245/46/556/56//E/ \\ 
\texttt{115-M7} & /EE12/333/445/4/6/666// & 
\texttt{233-S7} & /E112/34/E35/66/556/6// & 
\texttt{351-S7} & /E123/444/556/556/6//E/ \\ 
\texttt{116-M7} & /EE12/333/445/5/66/66// & 
\texttt{234-S7} & /E112/34/E55/456/66/6// & 
\texttt{352-S7} & /E123/445/456/566/6/E// \\ 
\texttt{117-M7} & /EE12/333/445/6/56/66// & 
\texttt{235-S7} & /E112/34/E55/566/566/// & 
\texttt{353-S7} & /E123/456/456/456/E/6// \\ 
\texttt{118-M7} & /EE12/333/456/4/56/66// & 
\texttt{236-S7} & /E112/34/E56/445/6/66// & 
 &  \\ 
\hline
\elt

\nc{\allresultscaption}
{Weights $w_n$, group factors $g_n$ and numerical
results for diagrams through seven loops ($^*=$ probable
error in the final digit;
$^{**}=$ probable error in the two final digits;
$( )=$  estimated statistical error in final digits).
The explicit group factors are listed in Table~\ref{groupfactors}.
The numerical data were provided by B.~Nickel and have
only partially been checked by the author.
The numbering proceeds according to \cite{NiMeBa}.
Since the momentum derivatives have not been published
before, we also provide them here, although they are not
needed for the purposes of this work.
The numerical values are taken at $r=1$ and $k=0$, and $l$ is the
number of loops of the corresponding diagram, indicated by the last
digit of the diagram name $n$.}
\blt[c]{|r|c|c|l|l||r|c|c|l|l|}
\caption{\label{allresults}\allresultscaption}
\\\hline\rule{0pt}{1pt}&&&&&&&&&\\
\multicolumn{1}{|c|}{$n$} & $w_n$ & $g_n$
& \multicolumn{1}{|c|}{$\ds-(8\pi)^l\f{\p I_n}{\p r}$}
& \multicolumn{1}{|c||}{$\ds-(8\pi)^l\f{\p I_n}{\p k^2}$}
& \multicolumn{1}{|c|}{$n$} & $w_n$ & $g_n$
& \multicolumn{1}{|c|}{$\ds-(8\pi)^l\f{\p I_n}{\p r}$}
& \multicolumn{1}{|c|}{$\ds-(8\pi)^l\f{\p I_n}{\p k^2}$}
\\\rule{0pt}{1pt}&&&&&&&&&\\\hline
\texttt{1-S0} &  1 &  1 &  $-1.000000000000$ &  $-1.000000000000$& 
\texttt{178-S7} & 1/16 & $g_{46}$ & $\phantom{+}1.112198567717$ & $\phantom{+}0.059426697530$\\
\texttt{2-M1} & 1/2 & $g_{1}$ & $\phantom{+}1.000000000000$ & $\phantom{+}0$& 
\texttt{179-S7} & 1/12 & $g_{15}$ &  $-0.099385135580$ &  $-0.003196017365$\\
\texttt{3-S2} & 1/6 & $g_{1}$ & $\phantom{+}2.000000000000$ & $\phantom{+}0.074074074074$& 
\texttt{180-S7} & 1/8 & $g_{47}$ & $\phantom{+}0.613041462344$ & $\phantom{+}0.025955207773$\\
\texttt{4-M3} & 1/12 & $g_{2}$ &  $-0.575364144904$ & $\phantom{+}0$& 
\texttt{181-S7} & 1/2 & $g_{44}$ & $\phantom{+}0.477782071483$ & $\phantom{+}0.020153442432$\\
\texttt{5-S3} & 1/4 & $g_{3}$ & $\phantom{+}2.053735627745$ & $\phantom{+}0.094651431944$& 
\texttt{182-S7} & 1/8 & $g_{46}$ & $\phantom{+}0.822000194835$ & $\phantom{+}0.039293990833$\\
\texttt{6-M4} & 1/8 & $g_{4}$ &  $-0.817603121794$ & $\phantom{+}0$& 
\texttt{183-S7} & 1/4 & $g_{41}$ & $\phantom{+}0.62652529607$ & $\phantom{+}0.03117089736$\\
\texttt{7-S4} & 1/12 & $g_{2}$ &  $-0.296452722985$ &  $-0.003098338156$& 
\texttt{184-S7} & 1/12 & $g_{16}$ &  $-0.088398198119$ &  $-0.002799922292$\\
\texttt{8-S4} & 1/4 & $g_{5}$ & $\phantom{+}1.723490549736$ & $\phantom{+}0.086595023215$& 
\texttt{185-S7} & 1/4 & $g_{48}$ & $\phantom{+}0.633014398591$ & $\phantom{+}0.031431108546$\\
\texttt{9-S4} & 1/8 & $g_{6}$ & $\phantom{+}2.065719357141$ & $\phantom{+}0.101317290347$& 
\texttt{186-S7} & 1/4 & $g_{44}$ & $\phantom{+}0.69594974685$ & $\phantom{+}0.03412961347$\\
\texttt{10-S4} & 1/4 & $g_{5}$ & $\phantom{+}1.240596097829$ & $\phantom{+}0.050040931841$& 
\texttt{187-S7} & 1/2 & $g_{41}$ & $\phantom{+}0.44224379962$ & $\phantom{+}0.01909934773$\\
\texttt{11-M5} & 1/24 & $g_{7}$ & $\phantom{+}0.063289032026$ & $\phantom{+}0$& 
\texttt{188-S7} & 1/2 & $g_{49}$ & $\phantom{+}0.2738240559^{**}$ & $\phantom{+}0.0117842659^*$\\
\texttt{12-M5} & 1/16 & $g_{8}$ &  $-0.950830436253$ & $\phantom{+}0$& 
\texttt{189-S7} & 1/4 & $g_{42}$ & $\phantom{+}0.587039378427$ & $\phantom{+}0.025512006666$\\
\texttt{13-M5} & 1/8 & $g_{9}$ &  $-0.522389299127$ & $\phantom{+}0$& 
\texttt{190-S7} & 1/4 & $g_{50}$ & $\phantom{+}0.71485758848$ & $\phantom{+}0.03528193562$\\
\texttt{14-M5} & 1/72 & $g_{7}$ & $\phantom{+}0.034976834929$ & $\phantom{+}0$& 
\texttt{191-S7} & 1/2 & $g_{41}$ & $\phantom{+}0.45315198278$ & $\phantom{+}0.01965173249$\\
\texttt{15-M5} & 1/8 & $g_{9}$ &  $-0.810810317465$ & $\phantom{+}0$& 
\texttt{192-S7} & 1/24 & $g_{16}$ &  $-0.091382042321$ &  $-0.002633887708$\\
\texttt{16-S5} & 1/8 & $g_{4}$ &  $-0.432936983596$ &  $-0.003699002180$& 
\texttt{193-S7} & 1/8 & $g_{40}$ &  $-0.200696579093$ &  $-0.004500005469$\\
\texttt{17-S5} & 1/6 & $g_{4}$ &  $-0.184366712188$ &  $-0.003701938896$& 
\texttt{194-S7} & 1/24 & $g_{14}$ &  $-0.128843563571$ &  $-0.003942067525$\\
\texttt{18-S5} & 1/4 & $g_{10}$ & $\phantom{+}1.280344835327$ & $\phantom{+}0.067020792129$& 
\texttt{195-S7} & 1/64 & $g_{51}$ & $\phantom{+}2.024860402022$ & $\phantom{+}0.099350544942$\\
\texttt{19-S5} & 1/4 & $g_{11}$ & $\phantom{+}1.608449036612$ & $\phantom{+}0.083298877665$& 
\texttt{196-S7} & 1/16 & $g_{45}$ & $\phantom{+}1.106144793719$ & $\phantom{+}0.050277183440$\\
\texttt{20-S5} & 1/16 & $g_{12}$ & $\phantom{+}2.059638006775$ & $\phantom{+}0.102643784933$& 
\texttt{197-S7} & 1/24 & $g_{14}$ &  $-0.128843563571$ &  $-0.003942067525$\\
\texttt{21-S5} & 1/24 & $g_{4}$ &  $-0.213590327764$ &  $-0.003731682713$& 
\texttt{198-S7} & 1/16 & $g_{52}$ & $\phantom{+}0.815707802129$ & $\phantom{+}0.032877730545$\\
\texttt{22-S5} & 1/4 & $g_{11}$ & $\phantom{+}1.177687787771$ & $\phantom{+}0.051418670702$& 
\texttt{199-S7} & 1/4 & $g_{43}$ & $\phantom{+}0.635702829750$ & $\phantom{+}0.025518064643$\\
\texttt{23-S5} & 1/2 & $g_{10}$ & $\phantom{+}1.093576383486$ & $\phantom{+}0.049451372165$& 
\texttt{200-S7} & 1/16 & $g_{45}$ & $\phantom{+}1.106144793719$ & $\phantom{+}0.050277183440$\\
\texttt{24-S5} & 1/8 & $g_{11}$ & $\phantom{+}0.944371175125$ & $\phantom{+}0.032955416087$& 
\texttt{201-S7} & 1/24 & $g_{15}$ &  $-0.113126023891$ &  $-0.003290535327$\\
\texttt{25-S5} & 1/2 & $g_{10}$ & $\phantom{+}0.734587686279$ & $\phantom{+}0.025567072553$& 
\texttt{202-S7} & 1/16 & $g_{52}$ & $\phantom{+}1.342508904936$ & $\phantom{+}0.071390284829$\\
\texttt{26-S5} & 1/6 & $g_{5}$ & $\phantom{+}0.47723573065^*$ & $\phantom{+}0.01715756028$& 
\texttt{203-S7} & 1/12 & $g_{15}$ &  $-0.096261106599$ &  $-0.003155179167$\\
\texttt{27-M6} & 1/16 & $g_{13}$ & $\phantom{+}0.092703296625$ & $\phantom{+}0$& 
\texttt{204-S7} & 1/16 & $g_{53}$ & $\phantom{+}0.942042143809$ & $\phantom{+}0.050485695523$\\
\texttt{28-M6} & 1/12 & $g_{13}$ & $\phantom{+}0.062064397151$ & $\phantom{+}0$& 
\texttt{205-S7} & 1/4 & $g_{42}$ & $\phantom{+}0.664812550858$ & $\phantom{+}0.032801749909$\\
\texttt{29-M6} & 1/32 & $g_{14}$ &  $-1.032433915322$ & $\phantom{+}0$& 
\texttt{206-S7} & 1/8 & $g_{47}$ & $\phantom{+}0.746247163126$ & $\phantom{+}0.036654897006$\\
\texttt{30-M6} & 1/8 & $g_{15}$ &  $-0.558960605344$ & $\phantom{+}0$& 
\texttt{207-S7} & 1/144 & $g_{13}$ & $\phantom{+}0.015400329826$ & $\phantom{+}0.000164624707$\\
\texttt{31-M6} & 1/16 & $g_{15}$ &  $-0.406156736719$ & $\phantom{+}0$& 
\texttt{208-S7} & 1/96 & $g_{14}$ &  $-0.148081924590$ &  $-0.003305011557$\\
\texttt{32-M6} & 1/4 & $g_{16}$ &  $-0.316465247271$ & $\phantom{+}0$& 
\texttt{209-S7} & 1/48 & $g_{13}$ & $\phantom{+}0.015548592044$ & $\phantom{+}0.000160182577$\\
\texttt{33-M6} & 1/24 & $g_{13}$ & $\phantom{+}0.045271743432$ & $\phantom{+}0$& 
\texttt{210-S7} & 1/32 & $g_{40}$ &  $-0.363469627011$ &  $-0.004727828507$\\
\texttt{34-M6} & 1/48 & $g_{13}$ & $\phantom{+}0.070805748949$ & $\phantom{+}0$& 
\texttt{211-S7} & 1/16 & $g_{38}$ &  $-0.31134215536$ &  $-0.00404674919$\\
\texttt{35-M6} & 1/4 & $g_{16}$ &  $-0.535115380809$ & $\phantom{+}0$& 
\texttt{212-S7} & 1/8 & $g_{42}$ & $\phantom{+}0.81924717224$ & $\phantom{+}0.04055898550$\\
\texttt{36-M6} & 1/8 & $g_{15}$ &  $-0.831785654370$ & $\phantom{+}0$& 
\texttt{213-S7} & 1/24 & $g_{15}$ &  $-0.112542632056$ &  $-0.003478053745$\\
\texttt{37-M6} & 1/8 & $g_{16}$ &  $-0.667694545359$ & $\phantom{+}0$& 
\texttt{214-S7} & 1/8 & $g_{47}$ & $\phantom{+}0.91405093408$ & $\phantom{+}0.04486160082$\\
\texttt{38-M6} & 1/12 & $g_{9}$ &  $-0.21440337147$ & $\phantom{+}0$& 
\texttt{215-S7} & 1/8 & $g_{54}$ & $\phantom{+}0.828736871089$ & $\phantom{+}0.040880626848$\\
\texttt{39-S6} & 1/24 & $g_{7}$ & $\phantom{+}0.025334411533$ & $\phantom{+}0.000134151236$& 
\texttt{216-S7} & 1/8 & $g_{46}$ & $\phantom{+}0.95215768567$ & $\phantom{+}0.04559036202$\\
\texttt{40-S6} & 1/16 & $g_{8}$ &  $-0.512249210261$ &  $-0.003730210466$& 
\texttt{217-S7} & 1/4 & $g_{42}$ & $\phantom{+}0.59759777279$ & $\phantom{+}0.02595026422$\\
\texttt{41-S6} & 1/8 & $g_{9}$ &  $-0.271221697673$ &  $-0.001890072089$& 
\texttt{218-S7} & 1/8 & $g_{53}$ & $\phantom{+}0.774920557284$ & $\phantom{+}0.033638518217$\\
\texttt{42-S6} & 1/72 & $g_{7}$ & $\phantom{+}0.021623868322$ & $\phantom{+}0.000141394973$& 
\texttt{219-S7} & 1/32 & $g_{40}$ &  $-0.246180375270$ &  $-0.004550733992$\\
\texttt{43-S6} & 1/8 & $g_{9}$ &  $-0.43846576066$ &  $-0.00319405093$& 
\texttt{220-S7} & 1/8 & $g_{43}$ & $\phantom{+}0.961197181072$ & $\phantom{+}0.046855489196$\\
\texttt{44-S6} & 1/4 & $g_{17}$ &  $-0.259325943995$ &  $-0.004300120594$& 
\texttt{221-S7} & 1/16 & $g_{38}$ &  $-0.188790327282$ &  $-0.002374789360$\\
\texttt{45-S6} & 1/6 & $g_{9}$ &  $-0.119009992648$ &  $-0.003308367453$& 
\texttt{222-S7} & 1/4 & $g_{42}$ & $\phantom{+}0.58123025452$ & $\phantom{+}0.02495949785$\\
\texttt{46-S6} & 1/4 & $g_{18}$ & $\phantom{+}0.879291241521$ & $\phantom{+}0.047113097898$& 
\texttt{223-S7} & 1/4 & $g_{55}$ & $\phantom{+}0.35733921887$ & $\phantom{+}0.01511183749$\\
\texttt{47-S6} & 1/4 & $g_{19}$ & $\phantom{+}1.120968561756$ & $\phantom{+}0.059552411259$& 
\texttt{224-S7} & 1/16 & $g_{40}$ &  $-0.200696579093$ &  $-0.004500005469$\\
\texttt{48-S6} & 1/8 & $g_{20}$ & $\phantom{+}1.542237652030$ & $\phantom{+}0.080253512938$& 
\texttt{225-S7} & 1/12 & $g_{15}$ &  $-0.065174820497$ &  $-0.001839165015$\\
\texttt{49-S6} & 1/8 & $g_{21}$ & $\phantom{+}1.176947197512$ & $\phantom{+}0.062822045769$& 
\texttt{226-S7} & 1/16 & $g_{52}$ & $\phantom{+}0.702185354197$ & $\phantom{+}0.023943701788$\\
\texttt{50-S6} & 1/12 & $g_{9}$ &  $-0.140199070719$ &  $-0.003403733352$& 
\texttt{227-S7} & 1/8 & $g_{47}$ & $\phantom{+}0.407663379492$ & $\phantom{+}0.013494432296$\\
\texttt{51-S6} & 1/4 & $g_{21}$ & $\phantom{+}0.865925201214$ & $\phantom{+}0.040831704607$& 
\texttt{228-S7} & 1/12 & $g_{15}$ &  $-0.08195863410$ &  $-0.00211512505$\\
\texttt{52-S6} & 1/2 & $g_{18}$ & $\phantom{+}0.782731311849$ & $\phantom{+}0.037539120424$& 
\texttt{229-S7} & 1/4 & $g_{53}$ & $\phantom{+}0.490372577726$ & $\phantom{+}0.016658394455$\\
\texttt{53-S6} & 1/12 & $g_{8}$ &  $-0.147488212902$ &  $-0.003907983810$& 
\texttt{230-S7} & 1/2 & $g_{42}$ & $\phantom{+}0.36993498100$ & $\phantom{+}0.01245375453$\\
\texttt{54-S6} & 1/32 & $g_{22}$ & $\phantom{+}2.044600225603$ & $\phantom{+}0.101571794444$& 
\texttt{231-S7} & 1/12 & $g_{15}$ &  $-0.08195863410$ &  $-0.00211512505$\\
\texttt{55-S6} & 1/8 & $g_{20}$ & $\phantom{+}1.137071072414$ & $\phantom{+}0.051222861796$& 
\texttt{232-S7} & 1/2 & $g_{42}$ & $\phantom{+}0.36993498100$ & $\phantom{+}0.01245375453$\\
\texttt{56-S6} & 1/16 & $g_{23}$ & $\phantom{+}1.438346940100$ & $\phantom{+}0.076201517483$& 
\texttt{233-S7} & 1/4 & $g_{47}$ & $\phantom{+}0.52108161787$ & $\phantom{+}0.01794555506$\\
\texttt{57-S6} & 1/48 & $g_{8}$ &  $-0.172986991913$ &  $-0.003634710346$& 
\texttt{234-S7} & 1/8 & $g_{43}$ & $\phantom{+}0.635702829750$ & $\phantom{+}0.025518064643$\\
\texttt{58-S6} & 1/16 & $g_{17}$ &  $-0.308545719756$ &  $-0.004585599956$& 
\texttt{235-S7} & 1/32 & $g_{52}$ & $\phantom{+}0.815707802129$ & $\phantom{+}0.032877730545$\\
\texttt{59-S6} & 1/4 & $g_{19}$ & $\phantom{+}1.013873425897$ & $\phantom{+}0.048344744022$& 
\texttt{236-S7} & 1/8 & $g_{46}$ & $\phantom{+}0.68905771535$ & $\phantom{+}0.02848129895$\\
\texttt{60-S6} & 1/24 & $g_{8}$ &  $-0.147488212902$ &  $-0.003907983810$& 
\texttt{237-S7} & 1/8 & $g_{54}$ & $\phantom{+}0.478222481915$ & $\phantom{+}0.016667771417$\\
\texttt{61-S6} & 1/8 & $g_{23}$ & $\phantom{+}0.865346183548$ & $\phantom{+}0.033272188246$& 
\texttt{238-S7} & 1/4 & $g_{55}$ & $\phantom{+}0.2289777693^{**}$ & $\phantom{+}0.0077046092^*$\\
\texttt{62-S6} & 1/2 & $g_{19}$ & $\phantom{+}0.674258545727$ & $\phantom{+}0.025819821639$& 
\texttt{239-S7} & 1/72 & $g_{13}$ & $\phantom{+}0.012011538097$ & $\phantom{+}0.000163713616$\\
\texttt{63-S6} & 1/16 & $g_{20}$ & $\phantom{+}1.137071072414$ & $\phantom{+}0.051222861796$& 
\texttt{240-S7} & 1/8 & $g_{42}$ & $\phantom{+}0.46370717475^*$ & $\phantom{+}0.01606754321$\\
\texttt{64-S6} & 1/8 & $g_{18}$ & $\phantom{+}0.91444037725$ & $\phantom{+}0.04390592970$& 
\texttt{241-S7} & 1/12 & $g_{16}$ &  $-0.07540859139$ &  $-0.00204387664$\\
\texttt{65-S6} & 1/24 & $g_{9}$ &  $-0.135779338809$ &  $-0.003596057150$& 
\texttt{242-S7} & 1/16 & $g_{38}$ &  $-0.187536113779$ &  $-0.004224294792$\\
\texttt{66-S6} & 1/8 & $g_{24}$ & $\phantom{+}0.921487251154$ & $\phantom{+}0.044136661234$& 
\texttt{243-S7} & 1/12 & $g_{16}$ &  $-0.075738914634$ &  $-0.002043899165$\\
\texttt{67-S6} & 1/4 & $g_{21}$ & $\phantom{+}1.00712450325$ & $\phantom{+}0.04748442921$& 
\texttt{244-S7} & 1/12 & $g_{16}$ &  $-0.07818970216$ &  $-0.00189022267$\\
\texttt{68-S6} & 1/2 & $g_{18}$ & $\phantom{+}0.63804938664$ & $\phantom{+}0.02549067586$& 
\texttt{245-S7} & 1/4 & $g_{50}$ & $\phantom{+}0.60353187598$ & $\phantom{+}0.02624560199$\\
\texttt{69-S6} & 1/2 & $g_{25}$ & $\phantom{+}0.40041775626$ & $\phantom{+}0.01597778408$& 
\texttt{246-S7} & 1/4 & $g_{41}$ & $\phantom{+}0.56993372746$ & $\phantom{+}0.02564664969$\\
\texttt{70-S6} & 1/4 & $g_{19}$ & $\phantom{+}0.839686306795$ & $\phantom{+}0.033838333483$& 
\texttt{247-S7} & 1/4 & $g_{48}$ & $\phantom{+}0.57239203577$ & $\phantom{+}0.02569079046$\\
\texttt{71-S6} & 1/2 & $g_{18}$ & $\phantom{+}0.64945908709$ & $\phantom{+}0.02613967761$& 
\texttt{248-S7} & 1/6 & $g_{16}$ &  $-0.062067310900$ &  $-0.001813515421$\\
\texttt{72-S6} & 1/12 & $g_{9}$ &  $-0.078329154158$ &  $-0.001805644050$& 
\texttt{249-S7} & 1/4 & $g_{42}$ & $\phantom{+}0.46691192976$ & $\phantom{+}0.01661222202$\\
\texttt{73-S6} & 1/16 & $g_{20}$ & $\phantom{+}0.784299477998$ & $\phantom{+}0.023940846450$& 
\texttt{250-S7} & 1/2 & $g_{41}$ & $\phantom{+}0.35089341850$ & $\phantom{+}0.01238298317$\\
\texttt{74-S6} & 1/8 & $g_{21}$ & $\phantom{+}0.459957864776$ & $\phantom{+}0.013458616281$& 
\texttt{251-S7} & 1/4 & $g_{56}$ & $\phantom{+}0.32526887363$ & $\phantom{+}0.01381319757$\\
\texttt{75-S6} & 1/6 & $g_{9}$ &  $-0.098563025016$ &  $-0.002060047146$& 
\texttt{252-S7} & 1/8 & $g_{57}$ & $\phantom{+}0.32687895181$ & $\phantom{+}0.01384306584$\\
\texttt{76-S6} & 1/4 & $g_{19}$ & $\phantom{+}0.545976922937$ & $\phantom{+}0.016655215393$& 
\texttt{253-S7} & 1/4 & $g_{41}$ & $\phantom{+}0.54819221041$ & $\phantom{+}0.02383525773$\\
\texttt{77-S6} & 1/1 & $g_{18}$ & $\phantom{+}0.41299428301$ & $\phantom{+}0.01244181957$& 
\texttt{254-S7} & 1/2 & $g_{58}$ & $\phantom{+}0.19468385257$ & $\phantom{+}0.00712440090$\\
\texttt{78-S6} & 1/4 & $g_{21}$ & $\phantom{+}0.57773667234$ & $\phantom{+}0.01796147180$& 
\texttt{255-S7} & 1/2 & $g_{56}$ & $\phantom{+}0.28701805127$ & $\phantom{+}0.01064946492$\\
\texttt{79-S6} & 1/8 & $g_{21}$ & $\phantom{+}0.72502789578$ & $\phantom{+}0.02886965480$& 
\texttt{256-S7} & 1/8 & $g_{43}$ & $\phantom{+}0.697364736943$ & $\phantom{+}0.025166949580$\\
\texttt{80-S6} & 1/8 & $g_{24}$ & $\phantom{+}0.525569264321$ & $\phantom{+}0.016702878464$& 
\texttt{257-S7} & 1/12 & $g_{16}$ &  $-0.07780036312$ &  $-0.00206345134$\\
\texttt{81-S6} & 1/4 & $g_{25}$ & $\phantom{+}0.2537676580^{**}$ & $\phantom{+}0.0076995725^*$& 
\texttt{258-S7} & 1/4 & $g_{44}$ & $\phantom{+}0.50747236144$ & $\phantom{+}0.01842390147$\\
\texttt{82-S6} & 1/72 & $g_{7}$ & $\phantom{+}0.018045754289$ & $\phantom{+}0.000141297494$& 
\texttt{259-S7} & 1/8 & $g_{48}$ & $\phantom{+}0.55986740563$ & $\phantom{+}0.02443161047$\\
\texttt{83-S6} & 1/8 & $g_{18}$ & $\phantom{+}0.51101057979$ & $\phantom{+}0.01609272071$& 
\texttt{260-S7} & 1/4 & $g_{42}$ & $\phantom{+}0.48161327069$ & $\phantom{+}0.01740560635$\\
\texttt{84-S6} & 1/24 & $g_{9}$ &  $-0.102061722421$ &  $-0.001882627644$& 
\texttt{261-S7} & 1/2 & $g_{56}$ & $\phantom{+}0.29353728091$ & $\phantom{+}0.01099247480$\\
\texttt{85-S6} & 1/4 & $g_{18}$ & $\phantom{+}0.69473734752$ & $\phantom{+}0.02934001606$& 
\texttt{262-S7} & 1/4 & $g_{44}$ & $\phantom{+}0.58759041059$ & $\phantom{+}0.02495250774$\\
\texttt{86-S6} & 1/2 & $g_{26}$ & $\phantom{+}0.23847649789$ & $\phantom{+}0.00778731763$& 
\texttt{263-S7} & 1/4 & $g_{48}$ & $\phantom{+}0.45969626338$ & $\phantom{+}0.01695129072$\\
\texttt{87-S6} & 1/2 & $g_{25}$ & $\phantom{+}0.34480051893$ & $\phantom{+}0.01152224582$& 
\texttt{264-S7} & 1/16 & $g_{42}$ & $\phantom{+}0.72541357539$ & $\phantom{+}0.03174350610$\\
\texttt{88-S6} & 1/4 & $g_{25}$ & $\phantom{+}0.33720102751$ & $\phantom{+}0.01113173461$& 
\texttt{265-S7} & 1/8 & $g_{47}$ & $\phantom{+}0.76145735178$ & $\phantom{+}0.03241477483$\\
\texttt{89-M7} & 1/48 & $g_{27}$ &  $-0.004609529361$ & $\phantom{+}0$& 
\texttt{266-S7} & 1/4 & $g_{44}$ & $\phantom{+}0.402492934390$ & $\phantom{+}0.014050307068$\\
\texttt{90-M7} & 1/32 & $g_{28}$ & $\phantom{+}0.110191210376$ & $\phantom{+}0$& 
\texttt{267-S7} & 1/2 & $g_{41}$ & $\phantom{+}0.35576134348$ & $\phantom{+}0.01253183085$\\
\texttt{91-M7} & 1/16 & $g_{29}$ & $\phantom{+}0.057057476340$ & $\phantom{+}0$& 
\texttt{268-S7} & 1/2 & $g_{41}$ & $\phantom{+}0.35875548318$ & $\phantom{+}0.01278571172$\\
\texttt{92-M7} & 1/144 & $g_{27}$ &  $-0.004646692505$ & $\phantom{+}0$& 
\texttt{269-S7} & 1/1 & $g_{59}$ & $\phantom{+}0.19941886961$ & $\phantom{+}0.00725142437$\\
\texttt{93-M7} & 1/16 & $g_{29}$ & $\phantom{+}0.09439061787$ & $\phantom{+}0$& 
\texttt{270-S7} & 1/2 & $g_{57}$ & $\phantom{+}0.29058013586$ & $\phantom{+}0.01075914097$\\
\texttt{94-M7} & 1/8 & $g_{30}$ & $\phantom{+}0.086255985441$ & $\phantom{+}0$& 
\texttt{271-S7} & 1/4 & $g_{44}$ & $\phantom{+}0.58151786330$ & $\phantom{+}0.02454625495$\\
\texttt{95-M7} & 1/48 & $g_{28}$ & $\phantom{+}0.060934949636$ & $\phantom{+}0$& 
\texttt{272-S7} & 1/8 & $g_{48}$ & $\phantom{+}0.54469915316$ & $\phantom{+}0.02347973464$\\
\texttt{96-M7} & 1/64 & $g_{31}$ &  $-1.085167325405$ & $\phantom{+}0$& 
\texttt{273-S7} & 1/2 & $g_{49}$ & $\phantom{+}0.2148706018^{**}$ & $\phantom{+}0.0076033420^*$\\
\texttt{97-M7} & 1/32 & $g_{32}$ &  $-0.581418054856$ & $\phantom{+}0$& 
\texttt{274-S7} & 1/2 & $g_{56}$ & $\phantom{+}0.28321821960$ & $\phantom{+}0.01041920621$\\
\texttt{98-M7} & 1/24 & $g_{28}$ & $\phantom{+}0.060934949636$ & $\phantom{+}0$& 
\texttt{275-S7} & 1/12 & $g_{16}$ &  $-0.078531030620$ &  $-0.001926078277$\\
\texttt{99-M7} & 1/16 & $g_{33}$ &  $-0.415203240084$ & $\phantom{+}0$& 
\texttt{276-S7} & 1/36 & $g_{13}$ & $\phantom{+}0.012465829906$ & $\phantom{+}0.000153902490$\\
\texttt{100-M7} & 1/4 & $g_{34}$ &  $-0.323424309472$ & $\phantom{+}0$& 
\texttt{277-S7} & 1/24 & $g_{15}$ &  $-0.084798571521$ &  $-0.001740909795$\\
\texttt{101-M7} & 1/16 & $g_{32}$ &  $-0.581418054856$ & $\phantom{+}0$& 
\texttt{278-S7} & 1/4 & $g_{44}$ & $\phantom{+}0.51605751592$ & $\phantom{+}0.01887196029$\\
\texttt{102-M7} & 1/24 & $g_{29}$ & $\phantom{+}0.029781711859$ & $\phantom{+}0$& 
\texttt{279-S7} & 1/2 & $g_{41}$ & $\phantom{+}0.36766342213$ & $\phantom{+}0.01317373214$\\
\texttt{103-M7} & 1/32 & $g_{32}$ &  $-0.340755240980$ & $\phantom{+}0$& 
\texttt{280-S7} & 1/4 & $g_{41}$ & $\phantom{+}0.44945925264$ & $\phantom{+}0.01646917857$\\
\texttt{104-M7} & 1/16 & $g_{35}$ &  $-0.196062158014$ & $\phantom{+}0$& 
\texttt{281-S7} & 1/4 & $g_{44}$ & $\phantom{+}0.63993915178$ & $\phantom{+}0.02847857167$\\
\texttt{105-M7} & 1/12 & $g_{29}$ & $\phantom{+}0.037007306205$ & $\phantom{+}0$& 
\texttt{282-S7} & 1/16 & $g_{54}$ & $\phantom{+}0.733996157205$ & $\phantom{+}0.031992253791$\\
\texttt{106-M7} & 1/8 & $g_{34}$ &  $-0.238176055425$ & $\phantom{+}0$& 
\texttt{283-S7} & 1/4 & $g_{57}$ & $\phantom{+}0.29124465780$ & $\phantom{+}0.01078593684$\\
\texttt{107-M7} & 1/2 & $g_{36}$ &  $-0.17920527706$ & $\phantom{+}0$& 
\texttt{284-S7} & 1/8 & $g_{38}$ &  $-0.102831822363$ &  $-0.001963907242$\\
\texttt{108-M7} & 1/8 & $g_{35}$ &  $-0.25405258372$ & $\phantom{+}0$& 
\texttt{285-S7} & 1/12 & $g_{15}$ &  $-0.055793955510$ &  $-0.001239522763$\\
\texttt{109-M7} & 1/16 & $g_{35}$ &  $-0.35337393251$ & $\phantom{+}0$& 
\texttt{286-S7} & 1/32 & $g_{45}$ & $\phantom{+}0.681842034844$ & $\phantom{+}0.018439194095$\\
\texttt{110-M7} & 1/16 & $g_{37}$ &  $-0.234481729265$ & $\phantom{+}0$& 
\texttt{287-S7} & 1/8 & $g_{46}$ & $\phantom{+}0.388596044366$ & $\phantom{+}0.010188219302$\\
\texttt{111-M7} & 1/8 & $g_{38}$ &  $-0.1112555703^{**}$ & $\phantom{+}0$& 
\texttt{288-S7} & 1/16 & $g_{47}$ & $\phantom{+}0.306680788759$ & $\phantom{+}0.007632976875$\\
\texttt{112-M7} & 1/144 & $g_{27}$ &  $-0.004476954704$ & $\phantom{+}0$& 
\texttt{289-S7} & 1/4 & $g_{44}$ & $\phantom{+}0.239104577809$ & $\phantom{+}0.005952085972$\\
\texttt{113-M7} & 1/16 & $g_{36}$ &  $-0.22686994808$ & $\phantom{+}0$& 
\texttt{290-S7} & 1/4 & $g_{38}$ &  $-0.136445344241$ &  $-0.002439738284$\\
\texttt{114-M7} & 1/432 & $g_{27}$ &  $-0.002157445026$ & $\phantom{+}0$& 
\texttt{291-S7} & 1/6 & $g_{16}$ &  $-0.040988108137$ &  $-0.000916369833$\\
\texttt{115-M7} & 1/72 & $g_{27}$ &  $-0.002517936782$ & $\phantom{+}0$& 
\texttt{292-S7} & 1/8 & $g_{43}$ & $\phantom{+}0.444684697062$ & $\phantom{+}0.012048266739$\\
\texttt{116-M7} & 1/48 & $g_{28}$ & $\phantom{+}0.049619035559$ & $\phantom{+}0$& 
\texttt{293-S7} & 1/4 & $g_{44}$ & $\phantom{+}0.258027818279$ & $\phantom{+}0.006779105555$\\
\texttt{117-M7} & 1/24 & $g_{29}$ & $\phantom{+}0.04248235153$ & $\phantom{+}0$& 
\texttt{294-S7} & 1/6 & $g_{16}$ &  $-0.04924750611$ &  $-0.00108567543$\\
\texttt{118-M7} & 1/24 & $g_{29}$ & $\phantom{+}0.026414905961$ & $\phantom{+}0$& 
\texttt{295-S7} & 1/2 & $g_{42}$ & $\phantom{+}0.29509588412$ & $\phantom{+}0.00780329148$\\
\texttt{119-M7} & 1/32 & $g_{30}$ & $\phantom{+}0.102504909593$ & $\phantom{+}0$& 
\texttt{296-S7} & 1/1 & $g_{41}$ & $\phantom{+}0.21984447334$ & $\phantom{+}0.00571613352$\\
\texttt{120-M7} & 1/48 & $g_{29}$ & $\phantom{+}0.05689249345$ & $\phantom{+}0$& 
\texttt{297-S7} & 1/6 & $g_{16}$ &  $-0.05226800912$ &  $-0.00105297682$\\
\texttt{121-M7} & 1/12 & $g_{29}$ & $\phantom{+}0.050866270685$ & $\phantom{+}0$& 
\texttt{298-S7} & 1/1 & $g_{41}$ & $\phantom{+}0.22784166071$ & $\phantom{+}0.00604295950$\\
\texttt{122-M7} & 1/8 & $g_{34}$ &  $-0.418158972550$ & $\phantom{+}0$& 
\texttt{299-S7} & 1/2 & $g_{44}$ & $\phantom{+}0.32103472773$ & $\phantom{+}0.00869741529$\\
\texttt{123-M7} & 1/4 & $g_{36}$ &  $-0.31676596355$ & $\phantom{+}0$& 
\texttt{300-S7} & 1/8 & $g_{46}$ & $\phantom{+}0.48492854629$ & $\phantom{+}0.01352273240$\\
\texttt{124-M7} & 1/16 & $g_{32}$ &  $-0.847760133200$ & $\phantom{+}0$& 
\texttt{301-S7} & 1/2 & $g_{44}$ & $\phantom{+}0.32792589489$ & $\phantom{+}0.00904574305$\\
\texttt{125-M7} & 1/24 & $g_{29}$ & $\phantom{+}0.05799468900$ & $\phantom{+}0$& 
\texttt{302-S7} & 1/12 & $g_{15}$ &  $-0.067370543631$ &  $-0.001354748387$\\
\texttt{126-M7} & 1/8 & $g_{35}$ &  $-0.53390645444$ & $\phantom{+}0$& 
\texttt{303-S7} & 1/8 & $g_{47}$ & $\phantom{+}0.43217408742$ & $\phantom{+}0.01209244215$\\
\texttt{127-M7} & 1/8 & $g_{35}$ &  $-0.458493427150$ & $\phantom{+}0$& 
\texttt{304-S7} & 1/4 & $g_{50}$ & $\phantom{+}0.32318954732$ & $\phantom{+}0.00871680902$\\
\texttt{128-M7} & 1/4 & $g_{36}$ &  $-0.32400906607$ & $\phantom{+}0$& 
\texttt{305-S7} & 1/4 & $g_{48}$ & $\phantom{+}0.28070537197$ & $\phantom{+}0.00756061771$\\
\texttt{129-M7} & 1/8 & $g_{34}$ &  $-0.625069918658$ & $\phantom{+}0$& 
\texttt{306-S7} & 1/2 & $g_{49}$ & $\phantom{+}0.1299127414^{**}$ & $\phantom{+}0.0033287282^*$\\
\texttt{130-M7} & 1/16 & $g_{37}$ &  $-0.494415425222$ & $\phantom{+}0$& 
\texttt{307-S7} & 1/36 & $g_{13}$ & $\phantom{+}0.010485980621$ & $\phantom{+}0.000143949321$\\
\texttt{131-M7} & 1/4 & $g_{36}$ &  $-0.41989663865$ & $\phantom{+}0$& 
\texttt{308-S7} & 1/4 & $g_{41}$ & $\phantom{+}0.27709244190^*$ & $\phantom{+}0.00746928985$\\
\texttt{132-M7} & 1/8 & $g_{34}$ &  $-0.542017150796$ & $\phantom{+}0$& 
\texttt{309-S7} & 1/12 & $g_{15}$ &  $-0.072993670101$ &  $-0.001976781189$\\
\texttt{133-M7} & 1/4 & $g_{38}$ &  $-0.2010875739^{**}$ & $\phantom{+}0$& 
\texttt{310-S7} & 1/8 & $g_{47}$ & $\phantom{+}0.54154753559$ & $\phantom{+}0.01945547104$\\
\texttt{134-M7} & 1/32 & $g_{30}$ & $\phantom{+}0.057209654200$ & $\phantom{+}0$& 
\texttt{311-S7} & 1/2 & $g_{44}$ & $\phantom{+}0.41728030634$ & $\phantom{+}0.01494610501$\\
\texttt{135-M7} & 1/16 & $g_{35}$ &  $-0.65804614052$ & $\phantom{+}0$& 
\texttt{312-S7} & 1/16 & $g_{46}$ & $\phantom{+}0.607654217590$ & $\phantom{+}0.025416236198$\\
\texttt{136-M7} & 1/8 & $g_{36}$ &  $-0.49327378592$ & $\phantom{+}0$& 
\texttt{313-S7} & 1/6 & $g_{16}$ &  $-0.05189612468$ &  $-0.00105446532$\\
\texttt{137-M7} & 1/96 & $g_{28}$ & $\phantom{+}0.069176047800$ & $\phantom{+}0$& 
\texttt{314-S7} & 1/2 & $g_{48}$ & $\phantom{+}0.30611471184$ & $\phantom{+}0.00872378220$\\
\texttt{138-M7} & 1/16 & $g_{36}$ &  $-0.491439771^{**}$ & $\phantom{+}0$& 
\texttt{315-S7} & 1/8 & $g_{54}$ & $\phantom{+}0.416000342949$ & $\phantom{+}0.012096422783$\\
\texttt{139-M7} & 1/32 & $g_{33}$ &  $-0.800600314818$ & $\phantom{+}0$& 
\texttt{316-S7} & 1/1 & $g_{56}$ & $\phantom{+}0.17349841526$ & $\phantom{+}0.00475514760$\\
\texttt{140-M7} & 1/8 & $g_{38}$ &  $-0.15678153586$ & $\phantom{+}0$& 
\texttt{317-S7} & 1/2 & $g_{57}$ & $\phantom{+}0.17914799102$ & $\phantom{+}0.00497574659$\\
\texttt{141-M7} & 1/48 & $g_{29}$ & $\phantom{+}0.038380451088$ & $\phantom{+}0$& 
\texttt{318-S7} & 1/2 & $g_{59}$ & $\phantom{+}0.12223230107$ & $\phantom{+}0.00332646474$\\
\texttt{142-M7} & 1/4 & $g_{38}$ &  $-0.16148138890$ & $\phantom{+}0$& 
\texttt{319-S7} & 1/24 & $g_{13}$ & $\phantom{+}0.023539759080$ & $\phantom{+}0.000168392224$\\
\texttt{143-M7} & 1/8 & $g_{36}$ &  $-0.35422809760$ & $\phantom{+}0$& 
\texttt{320-S7} & 1/8 & $g_{42}$ & $\phantom{+}0.39745790655$ & $\phantom{+}0.01134906488$\\
\texttt{144-M7} & 1/4 & $g_{39}$ &  $-0.11081591890$ & $\phantom{+}0$& 
\texttt{321-S7} & 1/2 & $g_{41}$ & $\phantom{+}0.29746997541$ & $\phantom{+}0.00837896648$\\
\texttt{145-S7} & 1/16 & $g_{13}$ & $\phantom{+}0.036278976466$ & $\phantom{+}0.000158781404$& 
\texttt{322-S7} & 1/8 & $g_{50}$ & $\phantom{+}0.35097358861$ & $\phantom{+}0.01003156199$\\
\texttt{146-S7} & 1/12 & $g_{13}$ & $\phantom{+}0.028592826473$ & $\phantom{+}0.000147520009$& 
\texttt{323-S7} & 1/4 & $g_{58}$ & $\phantom{+}0.11200684(65)$ & $\phantom{+}0.002934044(22)$\\
\texttt{147-S7} & 1/32 & $g_{14}$ &  $-0.562933518492$ &  $-0.003584909865$& 
\texttt{324-S7} & 1/6 & $g_{16}$ &  $-0.066786319580$ &  $-0.001901654937$\\
\texttt{148-S7} & 1/8 & $g_{15}$ &  $-0.297654173362$ &  $-0.001836208735$& 
\texttt{325-S7} & 1/4 & $g_{44}$ & $\phantom{+}0.532227035472$ & $\phantom{+}0.023800456805$\\
\texttt{149-S7} & 1/16 & $g_{15}$ &  $-0.207555228859$ &  $-0.001215771810$& 
\texttt{326-S7} & 1/4 & $g_{41}$ & $\phantom{+}0.471398155593$ & $\phantom{+}0.021205531919$\\
\texttt{150-S7} & 1/4 & $g_{16}$ &  $-0.161644383198$ &  $-0.000945042482$& 
\texttt{327-S7} & 1/16 & $g_{38}$ &  $-0.143600626236$ &  $-0.002341515026$\\
\texttt{151-S7} & 1/24 & $g_{13}$ & $\phantom{+}0.028711924535$ & $\phantom{+}0.000168339233$& 
\texttt{328-S7} & 1/2 & $g_{41}$ & $\phantom{+}0.40256253226$ & $\phantom{+}0.01541675510$\\
\texttt{152-S7} & 1/48 & $g_{13}$ & $\phantom{+}0.031801483472$ & $\phantom{+}0.000155670095$& 
\texttt{329-S7} & 1/2 & $g_{49}$ & $\phantom{+}0.2408938124^{**}$ & $\phantom{+}0.0091028154^*$\\
\texttt{153-S7} & 1/4 & $g_{16}$ &  $-0.28664763500$ &  $-0.00177064771$& 
\texttt{330-S7} & 1/4 & $g_{42}$ & $\phantom{+}0.53011891513$ & $\phantom{+}0.02052915892$\\
\texttt{154-S7} & 1/8 & $g_{15}$ &  $-0.45651280619$ &  $-0.00291467186$& 
\texttt{331-S7} & 1/2 & $g_{41}$ & $\phantom{+}0.40699861908$ & $\phantom{+}0.01578213619$\\
\texttt{155-S7} & 1/8 & $g_{16}$ &  $-0.36721882223$ &  $-0.00234684076$& 
\texttt{332-S7} & 1/6 & $g_{9}$ &  $-0.03327128907^*$ &  $-0.00071484464$\\
\texttt{156-S7} & 1/12 & $g_{9}$ &  $-0.1101540890^{**}$ &  $-0.00064080576$& 
\texttt{333-S7} & 1/12 & $g_{9}$ &  $-0.03233713444$ &  $-0.00074619804$\\
\texttt{157-S7} & 1/12 & $g_{13}$ & $\phantom{+}0.013391408702$ & $\phantom{+}0.000160623702$& 
\texttt{334-S7} & 1/1 & $g_{49}$ & $\phantom{+}0.19856030714$ & $\phantom{+}0.00601526961$\\
\texttt{158-S7} & 1/8 & $g_{40}$ &  $-0.300396192752$ &  $-0.004260505313$& 
\texttt{335-S7} & 1/1 & $g_{60}$ & $\phantom{+}0.11669921848$ & $\phantom{+}0.00353726638$\\
\texttt{159-S7} & 1/4 & $g_{38}$ &  $-0.156922345580$ &  $-0.002172374372$& 
\texttt{336-S7} & 1/1 & $g_{59}$ & $\phantom{+}0.17040027980$ & $\phantom{+}0.00516337873$\\
\texttt{160-S7} & 1/36 & $g_{13}$ & $\phantom{+}0.012817474926$ & $\phantom{+}0.000165556889$& 
\texttt{337-S7} & 1/4 & $g_{55}$ & $\phantom{+}0.27313504687$ & $\phantom{+}0.00842786560$\\
\texttt{161-S7} & 1/4 & $g_{38}$ &  $-0.25729167687$ &  $-0.00364837451$& 
\texttt{338-S7} & 1/2 & $g_{59}$ & $\phantom{+}0.17869340447$ & $\phantom{+}0.00559074588$\\
\texttt{162-S7} & 1/4 & $g_{38}$ &  $-0.160550866158$ &  $-0.003782129544$& 
\texttt{339-S7} & 1/4 & $g_{57}$ & $\phantom{+}0.26110837786$ & $\phantom{+}0.00835316544$\\
\texttt{163-S7} & 1/6 & $g_{16}$ &  $-0.075881836191$ &  $-0.002524110547$& 
\texttt{340-S7} & 1/2 & $g_{25}$ & $\phantom{+}0.11163367834$ & $\phantom{+}0.00333921861$\\
\texttt{164-S7} & 1/4 & $g_{41}$ & $\phantom{+}0.572724291583$ & $\phantom{+}0.031135889252$& 
\texttt{341-S7} & 1/2 & $g_{58}$ & $\phantom{+}0.16178590989$ & $\phantom{+}0.00485764945$\\
\texttt{165-S7} & 1/4 & $g_{42}$ & $\phantom{+}0.733711248698$ & $\phantom{+}0.039635195293$& 
\texttt{342-S7} & 1/12 & $g_{16}$ &  $-0.052565424208$ &  $-0.000929056468$\\
\texttt{166-S7} & 1/8 & $g_{43}$ & $\phantom{+}1.033924815474$ & $\phantom{+}0.054881037579$& 
\texttt{343-S7} & 1/4 & $g_{56}$ & $\phantom{+}0.25493509530$ & $\phantom{+}0.00810369020$\\
\texttt{167-S7} & 1/4 & $g_{44}$ & $\phantom{+}0.782547139439$ & $\phantom{+}0.042410572712$& 
\texttt{344-S7} & 1/2 & $g_{56}$ & $\phantom{+}0.24870245765$ & $\phantom{+}0.00779831998$\\
\texttt{168-S7} & 1/12 & $g_{16}$ &  $-0.089380432289$ &  $-0.002665803419$& 
\texttt{345-S7} & 1/8 & $g_{55}$ & $\phantom{+}0.26307959329$ & $\phantom{+}0.00791848968$\\
\texttt{169-S7} & 1/4 & $g_{44}$ & $\phantom{+}0.576651527776$ & $\phantom{+}0.028328751343$& 
\texttt{346-S7} & 1/4 & $g_{49}$ & $\phantom{+}0.19864889227$ & $\phantom{+}0.00603498255$\\
\texttt{170-S7} & 1/2 & $g_{41}$ & $\phantom{+}0.516195727537$ & $\phantom{+}0.025716040464$& 
\texttt{347-S7} & 1/2 & $g_{58}$ & $\phantom{+}0.16748684(75)$ & $\phantom{+}0.005145002(32)$\\
\texttt{171-S7} & 1/12 & $g_{15}$ &  $-0.099385135580$ &  $-0.003196017365$& 
\texttt{348-S7} & 1/4 & $g_{49}$ & $\phantom{+}0.18925230589$ & $\phantom{+}0.00553447373$\\
\texttt{172-S7} & 1/16 & $g_{45}$ & $\phantom{+}1.494759772685$ & $\phantom{+}0.077174394769$& 
\texttt{349-S7} & 1/1 & $g_{60}$ & $\phantom{+}0.11070332(35)$ & $\phantom{+}0.003243273(14)$\\
\texttt{173-S7} & 1/8 & $g_{46}$ & $\phantom{+}0.822000194835$ & $\phantom{+}0.039293990833$& 
\texttt{350-S7} & 1/2 & $g_{59}$ & $\phantom{+}0.17193262172$ & $\phantom{+}0.00525283507$\\
\texttt{174-S7} & 1/8 & $g_{47}$ & $\phantom{+}1.011972832768$ & $\phantom{+}0.054702376849$& 
\texttt{351-S7} & 1/48 & $g_{15}$ &  $-0.067427158448$ &  $-0.001192271595$\\
\texttt{175-S7} & 1/24 & $g_{15}$ &  $-0.115595929722$ &  $-0.003026809384$& 
\texttt{352-S7} & 1/8 & $g_{57}$ & $\phantom{+}0.24582805220$ & $\phantom{+}0.00758817302$\\
\texttt{176-S7} & 1/8 & $g_{38}$ &  $-0.196626012265$ &  $-0.004116199520$& 
\texttt{353-S7} & 1/12 & $g_{25}$ & $\phantom{+}0.10201073910$ & $\phantom{+}0.00287693468$\\
\texttt{177-S7} & 1/4 & $g_{44}$ & $\phantom{+}0.721067216848$ & $\phantom{+}0.035832522650$& 
&&&&\\
\hline
\elt

\nc{\groupfactorscaption}
{Group factors $g_n$ for Table~\ref{allresults}.}
\blt[c]{|r|c||r|c|}
\caption{\label{groupfactors}\groupfactorscaption}
\\\hline\rule{0pt}{1pt}&&&\\
\multicolumn{1}{|c|}{$n$} & $g_n$
& \multicolumn{1}{|c|}{$n$} & $g_n$
\\\rule{0pt}{1pt}&&&\\\hline
1 & $(2{+}N)/3$&
31 & ${(2{+}N)}^2(112{+}80N{+}40N^2{+}10N^3{+}N^4)/2187$ \\
2 & ${(2{+}N)}^2/9$&
32 & ${(2{+}N)}^2(136{+}80N{+}24N^2{+}3N^3)/2187$ \\
3 & $(2{+}N)(8{+}N)/27$&
33 & ${(2{+}N)}^2(144{+}80N{+}18N^2{+}N^3)/2187$ \\
4 & ${(2{+}N)}^2(8{+}N)/81$&
34 & ${(2{+}N)}^2(152{+}76N{+}14N^2{+}N^3)/2187$ \\
5 & $(2{+}N)(22{+}5N)/81$&
35 & ${(2{+}N)}^2(156{+}76N{+}11N^2)/2187$ \\
6 & $(2{+}N)(20{+}6N{+}N^2)/81$&
36 & ${(2{+}N)}^2(164{+}72N{+}7N^2)/2187$ \\
7 & ${(2{+}N)}^3/27$&
37 & ${(2{+}N)}^2(160{+}72N{+}10N^2{+}N^3)/2187$ \\
8 & ${(2{+}N)}^2(20{+}6N{+}N^2)/243$&
38 & ${(2{+}N)}^2(8{+}N)(22{+}5N)/2187$ \\
9 & ${(2{+}N)}^2(22{+}5N)/243$&
39 & ${(2{+}N)}^2(186{+}55N{+}2N^2)/2187$ \\
10 & $(2{+}N)(60{+}20N{+}N^2)/243$&
40 & ${(2{+}N)}^2(8{+}N)(20{+}6N{+}N^2)/2187$ \\
11 & $(2{+}N)(56{+}22N{+}3N^2)/243$&
41 & $(2{+}N)(448{+}244N{+}36N^2{+}N^3)/2187$ \\
12 & $(2{+}N)(48{+}24N{+}8N^2{+}N^3)/243$&
42 & $(2{+}N)(416{+}252N{+}56N^2{+}5N^3)/2187$ \\
13 & ${(2{+}N)}^3(8{+}N)/243$&
43 & $(2{+}N)(368{+}256N{+}88N^2{+}16N^3{+}N^4)/2187$ \\
14 & ${(2{+}N)}^2(48{+}24N{+}8N^2{+}N^3)/729$&
44 & $(2{+}N)(424{+}252N{+}50N^2{+}3N^3)/2187$ \\
15 & ${(2{+}N)}^2(56{+}22N{+}3N^2)/729$&
45 & $(2{+}N)(320{+}256N{+}120N^2{+}30N^3{+}3N^4)/2187$ \\
16 & ${(2{+}N)}^2(60{+}20N{+}N^2)/729$&
46 & $(2{+}N)(384{+}260N{+}76N^2{+}9N^3)/2187$ \\
17 & ${(2{+}N)}^2{(8{+}N)}^2/729$&
47 & $(2{+}N)(400{+}260N{+}64N^2{+}5N^3)/2187$ \\
18 & $(2{+}N)(164{+}72N{+}7N^2)/729$&
48 & $(2{+}N)(440{+}244N{+}42N^2{+}3N^3)/2187$ \\
19 & $(2{+}N)(152{+}76N{+}14N^2{+}N^3)/729$&
49 & $(2{+}N){(22{+}5N)}^2/2187$ \\
20 & $(2{+}N)(136{+}80N{+}24N^2{+}3N^3)/729$&
50 & $(2{+}N)(432{+}252N{+}44N^2{+}N^3)/2187$ \\
21 & $(2{+}N)(156{+}76N{+}11N^2)/729$&
51 & $(2{+}N)(256{+}240N{+}160N^2{+}60N^3{+}12N^4{+}N^5)/2187$ \\
22 & $(2{+}N)(112{+}80N{+}40N^2{+}10N^3{+}N^4)/729$&
52 & $(2{+}N)(352{+}264N{+}96N^2{+}16N^3{+}N^4)/2187$ \\
23 & $(2{+}N)(144{+}80N{+}18N^2{+}N^3)/729$&
53 & $(2{+}N)(384{+}256N{+}76N^2{+}12N^3{+}N^4)/2187$ \\
24 & $(2{+}N)(160{+}72N{+}10N^2{+}N^3)/729$&
54 & $(2{+}N)(400{+}248N{+}68N^2{+}12N^3{+}N^4)/2187$ \\
25 & $(2{+}N)(8{+}N)(22{+}5N)/729$&
55 & $(2{+}N)(22{+}5N)(20{+}6N{+}N^2)/2187$ \\
26 & $(2{+}N)(186{+}55N{+}2N^2)/729$&
56 & $(2{+}N)(472{+}224N{+}32N^2{+}N^3)/2187$ \\
27 & ${(2{+}N)}^4/81$&
57 & $(2{+}N)(464{+}224N{+}38N^2{+}3N^3)/2187$ \\
28 & ${(2{+}N)}^3(20{+}6N{+}N^2)/729$&
58 & $(2{+}N)(504{+}206N{+}19N^2)/2187$ \\
29 & ${(2{+}N)}^3(22{+}5N)/729$&
59 & $(2{+}N)(492{+}210N{+}26N^2{+}N^3)/2187$ \\
30 & ${(2{+}N)}^3{(8{+}N)}^2/2187$&
60 & $(2{+}N)(526{+}189N{+}14N^2)/2187$ \\
\hline
\elt


\begin{thebibliography}{99}

\bibitem{expbec}
M.H.~Anderson, J.R.~Ensher, M.R.~Matthews, C.~Wieman, and E.A.~Cornell,
\Sc{269}{198}{1995};
K.B.~Davis, M.O.~Mewes, M.R.~Andrews, N.J.~van Druten, D.S.~Durfee,
D.M.~Kurn, and W.~Ketterle,
\PRL{75}{3969}{1995};
C.C.~Bradley, C.A.~Sackett, J.J.~Tollett, and R.G.~Hulet,
\PRL{75}{1687}{1995}.

\bibitem{BaBlHoLaVa1}
G.~Baym, J.-P.~Blaizot, M.~Holzmann, F.~Lalo\"e, and D.~Vautherin,
\PRL{83}{1703}{1999}%
\ [\eprint{cond-mat}{9905430}]%
.

\bibitem{BaBlZi}
G.~Baym, J.-P.~Blaizot, and J.~Zinn-Justin,
\EPL{49}{150}{2000}%
\ [\eprint{cond-mat}{9907241}]%
.

\bibitem{HoBaBlLa}
M.~Holzmann, G.~Baym, J.-P.~Blaizot, and F.~Lalo\"e,
\PRL{87}{120403}{2001}%
\ [\eprint{cond-mat}{0103595}]%
.

\bibitem{An}
J.O.~Andersen,
Rev.\ Mod.\ Phys., to be published%
\ [\eprint{cond-mat}{0305138v2}]%
.

\bibitem{GiPiSt}
S.~Giorgini, L.P.~Pitaevskii, and S.~Stringari,
\PRA{54}{R4633}{1996}.


\bibitem{ArTo}
P.~Arnold and B.~Tom\'a\v{s}ik,
\PRA{64}{053609}{2001}%
\ [\eprint{cond-mat}{0105147}]%
.

\bibitem{ArMoTo}
P.~Arnold, G.~Moore, and B.~Tom\'a\v{s}ik,
\PRA{65}{013606}{2001}%
\ [\eprint{cond-mat}{0107124}]%
.

\bibitem{HuYaLu}
K.~Huang and C.N.~Yang,
\PR{105}{767}{1957};
K.~Huang, C.N.~Yang, and J.M.~Luttinger,
\PR{105}{776}{1957}.

\bibitem{Su}
X.~Sun,
\PRE{67}{066702}{2003}%
\ [\eprint{hep-lat}{0209144}]%
.

\bibitem{Ka6}
B.~Kastening,
\PRA{68}{061601(R)}{2003}%
\ [\eprint{cond-mat}{0303486}]%
.

\bibitem{Ka7}
B.~Kastening,
\PRA{69}{043613}{2004}%
\ [\eprint{cond-mat}{0309060}]%
.

\bibitem{KaLPHYS03}
B.~Kastening,
\LP{14}{586}{2004}%
\ [\eprint{cond-mat}{0404354}]%
.

\bibitem{c1kaN}
The quantity actually considered in \cite{Ka7,KaLPHYS03} is
$c_1=\al\ka_N$, where $\al=-256\pi^3/[\ze(3/2)]^{4/3}\approx-2206.19$.

\bibitem{KaPrSv}
V.A.~Kashurnikov, N.V.~Prokof'ev, and B.V.~Svistunov,
\PRL{87}{120402}{2001}%
\ [\eprint{cond-mat}{0103149}]%
;
N.V.~Prokof'ev and B.V.~Svistunov,
\PRL{87}{160601}{2001}%
\ [\eprint{cond-mat}{0103146}]%
.

\bibitem{ArMoc1}
P.~Arnold and G.~Moore,
\PRL{87}{120401}{2001}%
\ [\eprint{cond-mat}{0103228}]%
.

\bibitem{ArMoMC}
P.~Arnold and G.~Moore,
\PRE{64}{066113}{2001}%
\ [\eprint{cond-mat}{0103227}]%
.

\bibitem{phi4book}
H.~Kleinert and V.~Schulte-Frohlinde,
{\em Critical Properties of $\phi^4$-Theories},
1st ed. (World Scientific, Singapore, 2001).

\bibitem{ArToN}
P.~Arnold and B.~Tom\'a\v{s}ik,
\PRA{62}{063604}{2000}%
\ [\eprint{cond-mat}{0005197}]%
.

\bibitem{recrel}
H.~Kleinert, A.~Pelster, B.~Kastening, and M.~Bachmann,
\PRE{62}{1537}{2000}%
\ [\eprint{hep-th}{9907168}]%
;
B.~Kastening,
\PRE{61}{3501}{2000}%
\ [\eprint{hep-th}{9908172}]%
.

\bibitem{MurNi}
D.B.~Murray and B.G.~Nickel, 
Univ.\ of Guelph Report, 1991, unpublished.

\bibitem{MutNi}
M.~Muthukumar and B.~Nickel,
\JChP{80}{5839}{1984}.

\bibitem{We}
F.~Wegner,
\PRB{5}{4529}{1972}.

\bibitem{ZiPeVi}
J.~Zinn-Justin,
{\em Quantum Field Theory and Critical Phenomena}, 4th ed.\
(Clarendon, Oxford, 2002);
A.~Pelissetto and E.~Vicari,
\PR{368}{549}{2002}%
\ [\eprint{cond-mat}{0012164}]%
. 

\bibitem{HaKl}
B.~Hamprecht and H.~Kleinert,
\PRD{68}{065001}{2003}%
\ [\eprint{hep-th}{0302116}]%
.

\bibitem{Kl2}
H.~Kleinert,
\PLA{207}{133}{1995}%
\ [\eprint{quant-ph}{9507005}]%
.

\bibitem{Kl3}
H.~Kleinert,
\PRD{57}{2264}{1998};
\PRD{58}{107702}{1998}%
\ [\eprint{cond-mat}{9803268}]%
.

\bibitem{Kl4}
H.~Kleinert,
\PRD{60}{085001}{1999}%
\ [\eprint{hep-th}{9812197}]%
;
\PLA{277}{205}{2000}%
\ [\eprint{cond-mat}{9906107}]%
.

\bibitem{pibook}
H.~Kleinert,
{\em Path Integrals in Quantum Mechanics, Statistics and Polymer Physics},
3rd ed. (World Scientific, Singapore, 2004).

\bibitem{Yu}
V.I.~Yukalov,
\MUPB{31}{10}{1976}.

\bibitem{KlvdB}
H.~Kleinert and B.~Van den Bossche,
\PRE{63}{056113}{2001}%
\ [\eprint{cond-mat}{0011329}]%
.

\bibitem{Ste}
P.M.~Stevenson,
\PRD{23}{2916}{1981}.

\bibitem{SCPiRaSe}
F.F.~de Souza Cruz, M.B.~Pinto, R.O.~Ramos, and P.~Sena,
\PRA{65}{053613}{2002}%
\ [\eprint{cond-mat}{0112306}]%
.

\bibitem{KnNePi}
J.-L.~Kneur, A.~Neveu, and M.B.~Pinto,
\PRA{69}{053624}{2004}%
\ [\eprint{cond-mat}{0401324}]%
.

\bibitem{ldeerror}
The connection between the coefficients $A_l$ of \cite{KnNePi} and our
coefficients for $N=2$, given in Eqs.~(\ref{bcoeffs}) and
Table~\ref{pertcoeffs}, is $A_2=4b_2'$, $A_3=-16b_3$, $A_4=64b_4$,
$A_5=-256b_5$.
Their values $A_4=3.57259\times10^{-6}$ and $A_5=2.25332\times10^{-7}$
should, according to our results, read $A_4=3.46941\times10^{-6}$
and $A_5=2.23296\times10^{-7}$.
With the conventions of \cite{KnNePi}, the constant under the logarithm,
quoted in \cite{KnNePi} as $-0.59775$, should read
$\f{1}{2}-\ln3\approx-0.598612$.

\bibitem{NiMeBa}
B.G.~Nickel, D.I.~Meiron, and G.A.~Baker, Jr.,
Univ.\ of Guelph Report, 1977, unpublished.

\bibitem{Na}
J.F.~Nagle,
\JMP{7}{1588}{1966}.


\end{thebibliography}
\end{document}